\documentclass{aa_old}

\pdfoutput=1

\usepackage[varg]{txfonts}
\usepackage{graphicx}
\usepackage{microtype}
\usepackage[english]{babel}

\usepackage{colortbl}
\usepackage{arydshln}
\usepackage{siunitx}

\bibpunct{(}{)}{;}{a}{}{,} 

\let\stdsection\section
\newcommand\appsection{\clearpage\stdsection}

\newcommand{\mycomment}[1]{}

\usepackage{hyperref}
\hypersetup{
    colorlinks = true,
    linkcolor = blue,
    citecolor = blue,
    urlcolor = blue,
    linktocpage,
}

\graphicspath{{figures/}}
\newcommand\mc[1]{\multicolumn{1}{c}{#1}}

\newcommand{\ocfigure}[4][tbp]{
    \begin{figure}[#1]
        \resizebox{\hsize}{!}{\includegraphics{#2}}
        \caption{#3}
        \label{#4}
    \end{figure}
}

\newcommand{\tcfigure}[4][tbp]{
    \begin{figure*}[#1]
        \centering
        \includegraphics[width=17cm]{#2}
        \caption{#3}
        \label{#4}
    \end{figure*}
}

\newcommand{\wfigure}[4][tbp]{
    \begin{figure*}[#1]
        \sidecaption
        \includegraphics[width=12cm]{#2}
        \caption{#3}
        \label{#4}
    \end{figure*}
}

\newcommand{\resnum}[1]{#1}

\begin{document}

\title{
    (433)~Eros and (25143)~Itokawa surface properties from reflectance spectra
    }

\author{
    David Korda\inst{1}
    \and
    Tom{\' a}{\v s} Kohout\inst{1, 2}
    \and
    Kate{\v r}ina Flanderov{\' a}\inst{1, 2, 3}
    \and
    Jean-Baptiste Vincent\inst{4}
    \and
    Antti Penttil{\" a}\inst{5}
}

\offprints{David Korda, \\ \email{david.korda@helsinki.fi}}

\institute{
    Department of Geosciences and Geography, University of Helsinki, P.O.~Box~64, FI-00014, Helsinki, Finland
    \and
    Institute of Geology, Czech Academy of Sciences, Rozvojov{\' a}~269, CZ-16500, Prague, Czech Republic
    \and
    Astronomical Institute, Charles University, V Hole{\v s}ovi{\v c}k{\' a}ch 2, CZ-18000, Prague, Czech Republic
    \and
    DLR Institute of Planetary Research, Rutherfordstra{\ss}e 2, D-12489 Berlin, Germany
    \and
    Department of Physics, University of Helsinki P.O.~Box~64, FI-00014, Helsinki, Finland
} 

\abstract
{Our knowledge of near-Earth asteroid~(NEA) composition is important for planetary research, planetary defence, and future in-space resource utilisation. Upcoming space missions, for example, Hera, M-ARGO, or missions to the asteroid (99942)~Apophis, will provide us with surface-resolved NEA reflectance spectra. Neural networks are useful tools for analysing reflectance spectra and determining material composition with high precision and low processing time.}
{We applied neural-network models on disk-resolved spectra of the Eros and Itokawa asteroids observed by the NEAR Shoemaker and Hayabusa spacecraft. With this approach, the mineral variations or intensity of space weathering can be mapped.}
{We built and tested two types of convolutional neural networks~(CNNs). The first one was trained using asteroid reflectance spectra with known taxonomy classes. The other one used silicate reflectance spectra with assigned mineral abundances and compositions.}
{The reliability of the classification model depends on the resolution of reflectance spectra. Typical $F_1$ score and Cohen's $\kappa_C$ values decrease from about \resnum{0.90} for high-resolution spectra to about \resnum{0.70} for low-resolution spectra. 
The predicted silicate composition does not strongly depend on spectrum resolution and coverage of the 2\textmu{}m band of pyroxene. The typical root mean square error is between \resnum{6} and \resnum{10}~percentage points. 
For the Eros and Itokawa asteroids, the predicted taxonomy classes favour the S-type and the predicted surface compositions are homogeneous and correspond to the composition of L/LL and LL ordinary chondrites, respectively. On the Itokawa surface, the model identified fresh spots that were connected with craters or coarse-grain areas.}
{The neural network models trained with measured spectra of asteroids and silicate samples are suitable for deriving surface silicate mineralogy with a reasonable level of accuracy. The predicted surface mineralogy is comparable to the mineralogy of returned samples measured in the laboratory. Moreover, the taxonomical predictions can point out locations of fresher areas.}

\keywords{
    Minor planets, asteroids: general -- Methods: numerical -- Methods: data analysis -- Techniques: spectroscopic
}

\authorrunning{D. Korda et al.}
\maketitle


\section{Introduction}
Near-Earth asteroids~(NEAs) are relatively easy-to-reach and compositionally diverse targets for spacecraft exploration. The dominant NEA fraction is of S and Q spectral types \citep{Binzel_2004, Dunn_2013, Binzel_2019} predominantly composed of dry silicates such as olivine and pyroxene. These two minerals are easily identified in the near-infrared reflectance spectra through multiple characteristic absorption bands around 1 and 2~\textmu{}m. These bands are due to the presence of Fe$^{2+}$ cations in the crystalline structure. In olivine, three strong overlapping absorption bands are present close to 1-\textmu{}m wavelength. Similarly, in pyroxene, there are also three absorption bands observed:\ two strong bands are located around 1 and 2~\textmu{}m and one weak band is around 1.25~\textmu{}m \citep{Gaffey_1979, Cloutis_1986}. 

The specific position and depth of these absorption bands depend on mineral composition, texture, or material grain size. In a mixture of olivine and pyroxene, the overall shape of a spectrum depends on the olivine-to-pyroxene mixture ratio. The spectral mixing is, however, non-linear. Previous studies \citep[e.g.][]{Adams_1974, Cloutis_1986, Gaffey_2002, Burbine_2007, Dunn_2010} have found mainly empirical quantitative composition relations using the areas and central positions of the absorption bands. This has opened up one way of studying the quantitative mineralogy of silicate asteroids. However, the determination of the band positions and their areas is sensitive to the quality of a spectrum (with asteroid spectra being often of low signal-to-noise ratios) and spectral continuum subtraction (often not easy to reliably determine due to the limited spectral range). 
Together with quantitative mineralogy, several asteroid classification schemes (taxonomies) were introduced \citep[e.g.][]{Tholen_1984, Bus_1999, DeMeo_2009, Mahlke_2022}. The taxonomies are based on the reflectance spectra of asteroids, usually taken in visible and near-infrared wavelengths. Different taxonomy classes provide us with qualitative information about the surface composition of asteroids.

In this work, we focus on the surface composition of (433)~Eros and (25143)~Itokawa asteroids. Their surface-resolution reflectance spectra of the asteroids have been studied in the past. Taxonomically, both asteroids belong into the S or Sw-types \citep{Fujiwara_2006, DeMeo_2009, Mahlke_2022}.
Eros is a near-Earth elongated asteroid with dimensions of the order of tens of kilometres. The asteroid was studied in detail by the NEAR Shoemaker, the first dedicated asteroid rendezvous mission \citep{Cheng_1997}. The surface of Eros seemed to be compositionally homogeneous with extremely weak spectral variations across the northern hemisphere \citep{Nittler_2001, McFadden_2001, Bell_2002}, and with a good match to L/LL ordinary chondrite composition \citep{McCoy_2001, McFadden_2001}. Small spectral variability detected on Eros in 1-\textmu{}m and 2-\textmu{}m band absorptions is associated with large topographic slopes in and near large craters and along one nose of the asteroid. The weak variations in the band-area ratio are observed between the large impact craters Psyche and Himeros \citep{Bell_2002}.

Itokawa is a near-Earth asteroid resembling a contact binary \citep{Fujiwara_2006}. Compared to Eros, Itokawa is smaller, with dimensions on the order of hundreds of metres. The asteroid was studied in detail by the JAXA Hayabusa mission \citep{Kawaguchi_2003}. Most works related to surface composition and space weathering are based on the AMICA multi-band imaging camera \citep{Ishiguro_2007, Koga_2018}. Fresher areas are often correlated with younger impact craters or with coarser areas associated with tomographic highs \citep{Tancredi_2015}. The more mature areas correspond to accumulation areas such as Sagamihara, Muses Sea, or the bottom of prominent older craters. The laboratory analyses of returned dust particles from the Itokawa Muses Sea area showed uniform composition similar to LL chondrites \citep{Nakamura_2011, Nakamura_2014}.

Machine learning uses arbitrary mathematically complex empirical models. Neural networks \citep{Goodfellow_2016} represent a specific family of machine learning models. Neural networks are formed by layers of neurons which perform simple non-linear operations. The nonlinearity, optional complexity, and data-driven basis of the neural networks make them suitable for searching for complicated empirical relations. 
Recently, authors have used neural networks to extract information from reflectance spectra. \citet{Korda_2023} demonstrated the usefulness of neural networks in determining the mineralogical composition of olivine and pyroxene and \citet{Penttila_2021} and \citet{Klimczak_2021} used neural networks to perform a categorical classification of asteroid spectra, to name a few. The main advantage of neural networks compared to traditional statistical methods or empirical relations is that the flexible model finds the empirical relation itself and such a relation is versatile over a large range of material compositions. Reliability metrics used by neural networks are well-defined but often less intuitive. The main disadvantages of neural networks are a requirement for a large and diverse training dataset and a less apparent understanding of how individual reflectance values contribute to final model predictions.

On the other hand, the empirical relations using band-centre and band-area-ratio parameters are very intuitive but with a limited range of validity. A possible trade-off between intuitive parameterisation and versatility is statistical methods, for example, principal component analysis~(PCA). The high flexibility makes neural networks useful also in related fields of astronomy \citep[e.g.][]{Miller_1993, Ciaramella_2005} or asteroid studies \citep[e.g.][]{deLeon_2010, Hefele_2020, Carruba_2021, Viavattene_2022}.

In this work, we combine the outputs of two different neural network models. The first model classifies reflectance spectra to the Bus--DeMeo taxonomy classes \citep{DeMeo_2009} and is referred to as the `classification model'. The second model is the adaptation of previously published work by \citet{Korda_2023} and quantifies relative abundances of olivine, orthopyroxene, clinopyroxene, and their chemical compositions and this is referred to as the `composition model'. 
Subsequently, we evaluate error metrics on test data to get the error estimate of the two models. Finally, we apply the models to the disk-resolved reflectance spectral data of the Eros and Itokawa asteroids and discuss correlations between the outputs of both models. 
Both neural network models were built using the Keras library \citep{keras} in Python.

The presented results are based on an improved version of the Python scripts introduced in \citet{Korda_2023}. The improvements include performance optimisation, additional functionality, and, most importantly,  the possibility to easily switch between classification and composition models. The scripts are driven by user-friendly configuration files. Compared to works by other authors, this work has refined output in terms of asteroid classes (16 in this work compared to 12 in \citealt{Klimczak_2021} and \citealt{Penttila_2022}).
All Python scripts, data files, and metadata used in this article can be downloaded from GitHub repository\footnote{\url{https://github.com/Sirrah91/Asteroid-spectra}} as the  version `v2.0'.


\section{Data}

\tcfigure{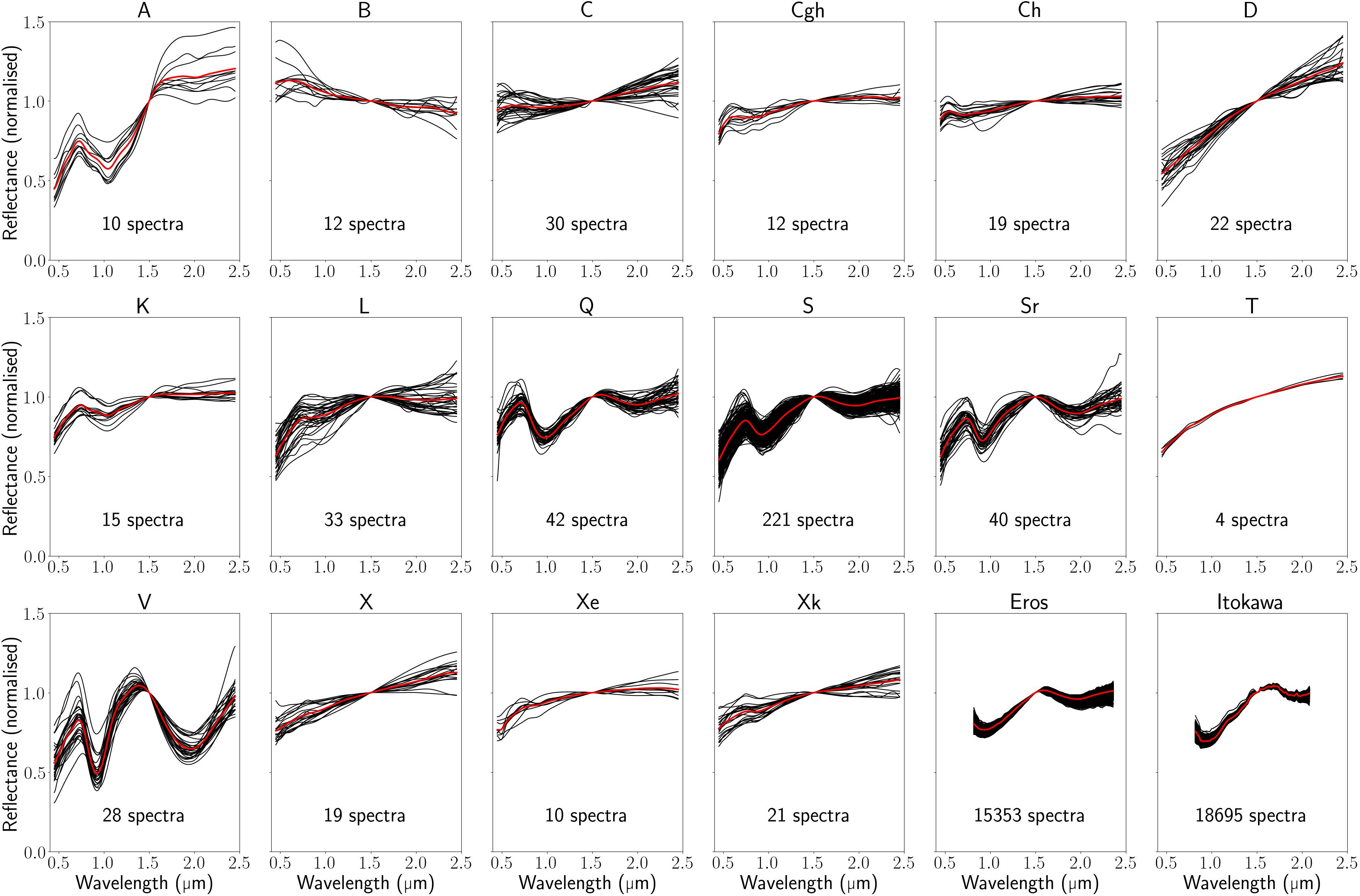}
{Individual reflectance spectra of the reduced taxonomy classes and of Eros and Itokawa. The thick red spectrum on top of each subplot represents the average spectrum of the type. For visualisation purposes, all spectra were re-normalised at \resnum{1500}~nm.}
{fig:spectra}

For the analysis of Eros and Itokawa surfaces, we utilised two different types of labelled reflectance spectra. The taxonomic classification is based on the dataset of asteroids with a known Bus--DeMeo class. The estimate of the surface mineral chemical composition is based on reflectance spectra of pure olivine, orthopyroxene, and clinopyroxene, their laboratory mixtures, and `natural' mixtures (meteorites).


\subsection{Taxonomic classification}

The basic dataset for training our classification model is a combined dataset drawn from \citet{DeMeo_2009} and \citet{Binzel_2019}. This dataset originally contains 591 reflectance spectra of asteroids in 33 taxonomy classes. Some classes include only a few examples. For this reason, we combined or deleted some classes (see Table~\ref{tab:taxonomy_data} for details). In the reduced taxonomy system, we have 538 reflectance spectra in 16 classes and an additional 41 reflectance spectra of Sq asteroids, which we set aside for additional model validation only, and we did not include these for the neural network training. 

The reflectance spectra were pre-processed in three steps. Firstly, we interpolated the spectra to a uniform wavelength grid from 450~nm to 2450~nm with a resolution of 5~nm (the grid is referred to as `full'). Secondly, we removed high-frequency noise from the re-sampled data using a Gaussian convolutional kernel with $\sigma = 7$~nm. Thirdly, the re-sampled and denoised spectra were normalised to unity at 550~nm, which highlighted differences among spectra. The final spectra are plotted in Fig.~\ref{fig:spectra}.

\begin{table}
    \sisetup{table-format = 3}
    \caption{Original taxonomy dataset and our reduced version.}
    \label{tab:taxonomy_data}
    \centering
    \begin{tabular}{l S S l}
        \hline\hline
        & \multicolumn{2}{c}{Counts} & \\
        \hline
        \mc{Class} & \mc{Original} & \mc{Reduced} & \mc{Notes}\\
        \hline
        A & 7 & 10\\
        B & 12 & 12\\
        C & 26 & 30\\
        Cb & 4 & 0 & Cb $\rightarrow$ C\\
        Cg & 2 & 0 & Cg $\rightarrow$ Cgh\\
        Cgh & 10 & 12\\
        Ch & 19 & 19\\
        D & 22 & 22\\
        K & 15 & 15\\
        L & 33 & 33\\
        O & 1 & 0 & not used\\
        Q & 42 & 42\\
        Qw & 1 & 0 & not used\\
        R & 1 & 0 & R $\rightarrow$ Sr\\
        S & 147 & 221\\
        Sa & 3 & 0 & Sa $\rightarrow$ A\\
        Sq & 41 & 41 & validation only\\
        Sq: & 1 & 0 & not used\\
        Sqw & 17 & 0 & Sqw $\rightarrow$ S\\
        Sr & 31 & 40\\
        Srw & 8 & 0 & Srw $\rightarrow$ Sr\\
        Sv & 3 & 0 & not used\\
        Svw & 2 & 0 & not used\\
        Sw & 57 & 0 & Sw $\rightarrow$ S\\
        T & 4 & 4\\
        U & 3 & 0 & not used\\
        V & 26 & 28\\
        Vw & 2 & 0 & Vw $\rightarrow$ V\\
        X & 16 & 19\\
        Xc & 3 & 0 & Xc $\rightarrow$ X\\
        Xe & 10 & 10\\
        Xk & 21 & 21\\
        Xn & 1 & 0 & not used\\
        \hline
    \end{tabular}
\end{table}


\subsection{Composition regression}

For the training of the composition model, we used reflectance spectra of pure olivine~(OL; 100~samples), orthopyroxene~(OPX; 102~samples), and clinopyroxene~(CPX; 108~samples), their laboratory silicate mixtures~(137~samples), and meteorites~(45 ordinary chondrites [OC],  7 HEDs, 5 lodranites, four SNC achondrites, and 2 brachinites; total:\ 63 samples). Except for seven reflectance spectra measured at the University of Helsinki, all other reflectance spectra were sourced from the RELAB\footnote{\url{https://pds-geosciences.wustl.edu/spectrallibrary/}} and C-Tape\footnote{\url{https://uwinnipeg.ca/c-tape/sample-database.html}} databases. Relative volumetric modal abundances and mineral chemical compositions represented by end-members are known for all the samples. 
The pre-processing is the same as in the case of reflectance spectra used in the classification model. The same wavelength grid and normalisation enable further validations, for example,  assigning taxonomy classes to reflectance spectra of olivine, pyroxene, ordinary chondrites, or HED meteorites (see Sect.~\ref{sect:xtest}). For more information about the data, we refer to Sect.~2 and Appendix~A, both in \citet{Korda_2023}.


\subsection{Eros and Itokawa}
\label{sect:E_I_data}

The spectra of Eros were collected by the Near-Infrared Spectrometer~(NIS) instrument on board the NEAR Shoemaker \citep{Warren_1997} spacecraft, while the Itokawa data are based on the Near Infrared Spectrometer \citep[NIRS;][]{Abe_2011} measurements by the Hayabusa spacecraft. 
Both datasets were obtained from the NASA Planetary Data System~(PDS) archive. In the case of Eros, the NEAR\_A\_NIS\_5\_EDR\_ALL\_PHASES\_PDSREV\_V1\_0 dataset was used. In the case of Itokawa the Home phase data including footprint information from HAY\_A\_NIRS\_3\_NIRSCAL\_V1\_0 dataset was used. Both datasets include basic dark frame and/or dark current, as well as lighting geometry corrections necessary to convert raw data to reflectance ($\mathrm{radiance} / \mathrm{irradiance}$, $I / F$) units of reflectance.

Comparing the NIS Eros spectra to those obtained by ground telescopes there is a good match at lower wavelengths, but increasing excessive reddening above 1500~nm \citep[e.g.][]{Veverka_2000, Bell_2002} in the InGaAs detector segment \citep{Warren_1997}. While still matching ground-based spectra within the error bar, the NIS red slope is not fully removed during calibration \citep{Izenberg_2000} and its origin is not fully understood \citep{McFadden_2001}. Therefore, we calculated a correction coefficient for each wavelength as a fraction of the normalised spectrum of Eros used in \citet{DeMeo_2009} over the mean normalised NIS spectrum. This correction was applied to all NIS spectra (see Fig.~\ref{fig:Eros_correction}). No other photometric or thermal corrections were utilised.

To filter a coherent and reliable dataset, we applied a set of criteria on the reflectance geometry and data quality specified in Table~\ref{tab:I_E_criteria}, verified that the field of view is filled with the target asteroid, interpolated the original wavelength grid to uniform spacing, and normalised the spectra. Among the selection criteria, we also filtered out all untrustworthy spectra (e.g. with too low or even negative reflectance) and in the case of Itokawa, we also filtered out reflectance values at wavelengths that were biased by the instrument (low sensitivity of the detector below 800~nm, Abe 2020, personal communication). A total of \resnum{1685} Eros and \resnum{8358} Itokawa spectra passed our criteria and were used in this work.

The observed spectra 
cover 
overlapping areas of the surface depending on the exact observation geometry. To scale the spectral measurements to a uniform grid, we divided the asteroid surface into `target regions' defined as areas of 1~deg $\times$ 1~deg in latitude $\times$ longitude. To recover the spectrum of a desired target region, we computed a weighted average of all spectra covering that region. The averaging procedure has three steps. 
Firstly, we collected all measured spectra $\vec{R}^{\mathrm{raw}}$ overlapping with the desired target region. Secondly, for all collected spectra $\vec{R}^{\mathrm{raw}}$, we weighted their contribution to the target region based on their overlap area $\vec{S}^o$ with the region divided by their total area $\vec{S}^g$. Thirdly, the final spectrum $R$ of the target region is a weighted sum of all collected spectra, namely,
\begin{equation}
    R = \frac{\sum \limits_i \vec{R}^{\mathrm{raw}}_i \left(\vec{S}^o_i / \vec{S}^g_i \right)}{\sum \limits_i \vec{S}^o_i / \vec{S}^g_i},
\end{equation}
where $i$ indexes individual spectra and areas. In total, 35\% of Eros and 45\% of Itokawa surfaces are covered with target regions containing spectral information. After the averaging procedure, we denoised the spectra using the Gaussian convolutional kernel and normalised them again. The final spectra are plotted in Fig.~\ref{fig:spectra}. We note that the normalisation wavelength was set to 1500~nm, which is roughly in the middle between the two absorption bands. This must have been modified in the case of Eros due to the excessive reddening and was set to 1300~nm, where the germanium detector provides more reliable data \citep{Veverka_2000}.

\begin{table}
    \caption{Selection criteria, number of spectra and their average size, and final wavelength grid for Eros and Itokawa spectra.}
    \label{tab:I_E_criteria}
    \centering
    \begin{tabular}{l S[table-format=3.2] S[table-format=4]}
        \hline\hline
        \mc{Parameter} & \mc{Min} & \mc{Max} \\
        \hline
        \multicolumn{3}{c}{Eros}\\
        \hline
        phase angle (deg) & 0 & 40\\
        incidence angle (deg) & 0 & 60\\
        emission angle (deg) & 0 & 60\\
        observed area (deg$^2$) & 0 & 750\\
        mean$\left(R^{\mathrm{raw}} \right)$ & 0.01 & \mc{-}\\
        \hdashline
        $\#R^{\mathrm{raw}}$ & \multicolumn{2}{c}{\phantom{0}1685}\\
        $\#R$ & \multicolumn{2}{c}{15353}\\
        $S^{R^{\mathrm{raw}}}$ (deg$^2$) & \multicolumn{2}{c}{$199 \pm 111$} \\
        \hdashline
        wavelength grid (nm) & 820 & 2360\\
        wavelength spacing (nm) & \multicolumn{2}{c}{\phantom{000}20}\\
        normalised at (nm) & \multicolumn{2}{c}{\phantom{0}1300}\\
        \hline
        \multicolumn{3}{c}{Itokawa}\\
        \hline
        phase angle (deg) & 0 & 30\\
        incidence angle (deg) & 0 & 50\\
        emission angle (deg) & 0 & 50\\
        distance from the surface (km) & 0 & 5\\
        mean$\left(R^{\mathrm{raw}} \right)$ & 0.01 & \mc{-}\\
        \hdashline
        $\#R^{\mathrm{raw}}$ & \multicolumn{2}{c}{\phantom{0}8358}\\
        $\#R$ & \multicolumn{2}{c}{18695}\\
        $S^{R^{\mathrm{raw}}}$ (deg$^2$) & \multicolumn{2}{c}{$6.0 \pm 5.2$} \\
        \hdashline
        wavelength grid (nm) & 820 & 2080\\
        wavelength spacing (nm) & \multicolumn{2}{c}{\phantom{000}20}\\
        normalised at (nm) & \multicolumn{2}{c}{\phantom{0}1500}\\
        \hline
    \end{tabular}
\end{table}


\section{Neural network}

We utilised artificial neural networks to find a relation between the reflectance spectra and taxonomic classes or mineral composition. In both cases, the input for the neural network is a reflectance spectrum. The layers of neurons in the network extract diagnostic patterns from the reflectance values, combine them, and finally predict the taxonomic class or the composition.

Every layer of the neural network performs a linear combination of the layer inputs $\vec{h}$ followed by a simple non-linear operation that takes the form $\vec{o} = \mathrm{f}\!\left( \tens{W} \vec{h} + \vec{b} \right)$, where $\vec{o}$ is the layer output, $\tens{W}$ are the coefficients of the linear combination (weights), $\vec{b}$ are the bias, and $\mathrm{f}$ is the non-linear (activation) function. The activation function is fixed for each layer, while weights and biases are trained to produce the mapping between the inputs and outputs. 


\subsection{Model architecture}

Model architecture defines the structure of a neural network. The structure is composed of an input layer, hidden layers, and an output layer. The widths of the input and output layers are fixed and depend on the dimension of the input data and the number of outputs. The input data of both models is a reflectance spectrum with a number of reflectance values. The output values of our classification models are probabilities of individual taxonomic classes (16 classes in our reduced list), while the composition model returns predicted modal abundances and chemical compositions of minerals (10 values).

The number and widths of hidden layers as well as their interconnections were chosen using model evaluation and parameter search as described in Sect.~\ref{sect:hp}. The preferred architecture of the classification model consists of \resnum{one} convolutional hidden layer with \resnum{24} different kernels (filters) of width \resnum{20~nm (5~values)}. The composition model includes an additional convolutional layer with \resnum{eight} filters of width \resnum{20~nm}. Each of the filters learns a different pattern from the data, for instance, a position of an absorption band. The output of the last convolution layer is flattened and fully connected with the output layer.

The parameter search favoured models with the \resnum{exponential linear unit~(ELU)} or \resnum{rectified linear unit~(ReLU)} activation functions between the input and the hidden layers, and the \resnum{softmax} or \resnum{sigmoid} activation functions between the hidden and the output layers. These are defined as:
\begin{align}
     \mathrm{ELU}\!\left( x \right) &=
     \begin{cases}
        x,& \text{if } x\geq 0\\
        \exp \left( x \right) - 1,& \text{if } x < 0,
    \end{cases}\\
    \mathrm{ReLU}\!\left( x \right) &= \max \left(0, x \right),\\
    \mathrm{softmax}\!\left( \vec{x} \right) &= \frac{\exp \left( \vec{x} \right)}{\sum \limits_i \exp \left( \vec{x}_i \right)},\\
    \mathrm{sigmoid}\!\left( x \right) &= \left(1 + \exp \left( -x \right)\right)^{-1}.
\end{align}
Because of additional sigmoid normalisation \citep[see Sect.~3.1 in][]{Korda_2023}, both softmax and sigmoid activation functions return a vector of non-negative values that (by parts) sum up to one. Therefore, the vector can be directly interpreted as a probability (match score) of individual taxonomic classes or as silicate composition in percent.


\subsection{Loss function and accuracy metrics}

Before the training of the neural network model, we need to define a loss function that is minimised and metrics to measure the accuracy of the model. The classification and composition models solve different tasks, namely, classification and regression. We used standard loss functions and accuracy metrics for these. For the classification, we used the categorical cross-entropy loss function and categorical accuracy, $F_1$ score, and Cohen's $\kappa_C$ as the metrics. The regression loss and metrics (root-mean-square error~[RMSE], coefficient of determination~[$R^2$], and spectral angle mapper~[SAM]) are described in Sect.~3.2 in \citet{Korda_2023}. Categorical cross-entropy $\mathcal{L}_{{\rm tax}}$, categorical accuracy $CA$, $F_1$ score, and Cohen's $\kappa_C$ are defined as:
\begin{align}
    \mathcal{L}_{{\rm tax}} &= - \frac{1}{N} \sum \limits_{i = 1}^N \sum \limits_{c = 1}^M \vec{1}_{y_i \in C_c} \log \left( p_{\mathrm{model}}\left[ y_i \in C_c \right]\right),\\
    CA &= \frac{\text{true positives}}{N}, \label{eq:CA}\\
    F_1 &= 2 \, \frac{\text{precision} \cdot \text{recall}}{\text{precision} + \text{recall}},\\
    \kappa_C &= \frac{CA - p_e}{1 - p_e}, \\
    \text{precision} &= \frac{\text{true positives}}{\text{true positives} + \text{false positives}},\\
    \text{recall} &= \frac{\text{true positives}}{\text{true positives} + \text{false negatives}}, \label{eq:recall}\\
    p_e &= N^{-2} \sum \limits_{c = 1}^M n_{\mathrm{actual}}^c n_{\mathrm{model}}^c,
\end{align}
where $N$ is the number of observations (reflectance spectra), $M$ is the number of classes, $C_c$ is the individual class, $\vec{1}_{y_i \in C_c}$ is the indicator function (equals to 1 if $y_i \in C_c$ and 0 otherwise), $p_{\mathrm{model}}\left[ y_i \in C_c \right]$ is the predicted probability of the observation $i$ to belong to class $C_c$, and $n^c$ is the number of observations belonging to class $C_c$. We note that the predicted class is the one with the maximum predicted probability.

The categorical accuracy simply tells us in what fraction of observations the model prediction is correct. Sometimes, even a few misclassifications can make the model useless. An example is a situation with one of the classes being significantly over-represented in the number of cases compared to other classes. Then a simple model which always `predicts' the observation belonging to the dominant class has very high categorical accuracy but is useless. 
For this reason, it is also useful to introduce other metrics. Cohen's $\kappa_C$ is similar to categorical accuracy but takes into account a baseline of random chance. Precision measures the probability that if the model predicts the observation with a certain class, it will be correct. Recall measures the probability that if the observation is from a certain class, the model will correctly predict it. $F_1$ score (harmonic average of precision and recall) is useful if the model does not need to concentrate on either precision or recall, which is the case we are examining here.


\subsection{Training and regularisation}
\label{sect:training}

The training of a neural network is an iterative process of weights and bias optimisation. The optimisation is based on gradient descent algorithms. The gradients of the loss function are computed using the backpropagation algorithm. We solved the optimisation problem using the Keras library. The parameter search favoured the \resnum{Adam} optimisation scheme \citep{adam}. 
We note that the training of a single model with optimised hyperparameters using 4 Intel 1.6~GHz cores takes about 2~minutes for a low-resolution classification model and about 25~minutes for a high-resolution composition model. Subsequent evaluation of over 18500 Itokawa spectra with the trained models takes about 1.3~seconds per model.

The available data can be split into three parts: training, validation, and test data. Training data are only used in the training algorithm. Validation data are used to find the optimal model architecture and other hyperparameters (see Sect.~\ref{sect:hp}) and to avoid overfitting during training. Test data only enters the trained model, where its purpose is to evaluate the model performance.

To avoid overfitting, we utilised three types of regularisation techniques: $L_1$ (absolute values) and $L_2$ (Euclidean norm) regularisations of weights and biases and dropout regularisation \citep{Srivastava_2014}. The dropout randomly cuts connections between neurons and effectively makes the model smaller. The $L_1$ and $L_2$ trade-off parameters and dropout rates between individual layers were selected via the parameter search. Among these, we used a batch normalisation between the layers, applied after the activation function. The batch normalisation leads to a faster convergence of the optimisation algorithm.


\subsection{Hyperparameters}
\label{sect:hp}

The specific configuration and behaviour of a neural network are driven by a set of hyperparameters. The hyperparameters contain the model architecture (number and widths of hidden layers, interlayer connections, activation functions), optimisation routine (optimisation algorithm and its hyperparameters), or regularisation techniques (trade-off parameters, dropout rates). The hyperparameters are selected via parameter search and accuracy evaluation on the validation data. First, we split the data to train (80\%) and validation (20\%) parts. The splitting was done separately for each taxonomy class and each type of mineral mixture to keep relative numbers in both sets. Second, we used random and Bayes search \citep{Kandasamy_2018} algorithms and computed about 3500 models with unique combinations of hyperparameters \citep[additionally, over 2500 more models were computed for the purposes of our previous study][]{Korda_2023}. The selected set of hyperparameters fulfils two conditions: (1)~the accuracy metrics computed from validation and training data are optimal and comparable (to prevent overfitting) and (2)~the loss function computed from validation and training data are low and comparable. The final parameters and ranges used in the parameter search are listed in Table~\ref{tab:hp}. We note that for the composition models, we adopted hyperparameters from \citet{Korda_2023}.


\subsection{Model evaluation}

The accuracy evaluation of the models was done using a $k$-fold procedure. The $k$-fold procedure splits the data into $k$ different parts. All parts except one are used as a training dataset and the single remaining part as a test dataset. The training dataset is utilised to train a model with the previously set hyperparameters, and the test dataset is used to evaluate accuracy metrics. This is repeated $k$ times so all data parts are finally evaluated in the test part. For this work, we selected $k = 100$, a compromise between computational complexity and the reliable estimation of the accuracy metrics.

To further test the classification model, we used spectra of Sq-type asteroids, meteorites (OC and HED), and pure mineral samples (olivine and pyroxene) to verify that they are evaluated within expected classification categories (i.e. Sq-type asteroid and ordinary chondrite meteorite spectra should be evaluated as S and Q-types; pyroxenes and HEDs as V-type; and olivines as A-type). We note that similar tests on the composition model on asteroid spectra were carried out in \citet{Korda_2023}.


\section{Results}

Both the classification and composition networks were trained three times:\ once with the full available spectral span and resolution and twice to match the spectral datasets of Eros and Itokawa, respectively. The full resolution covers both the 1-\textmu{}m and 2-\textmu{}m absorption bands with the high 5-nm resolution. Eros resolution contains both bands too, but with a coarser 20-nm spacing. Itokawa resolution does not fully cover the 2-\textmu{}m band and is of the same 20-nm spacing. First, we validated the individual classification and composition models. Second, we used them to make predictions about the surfaces of the Eros and Itokawa asteroids.


\tcfigure{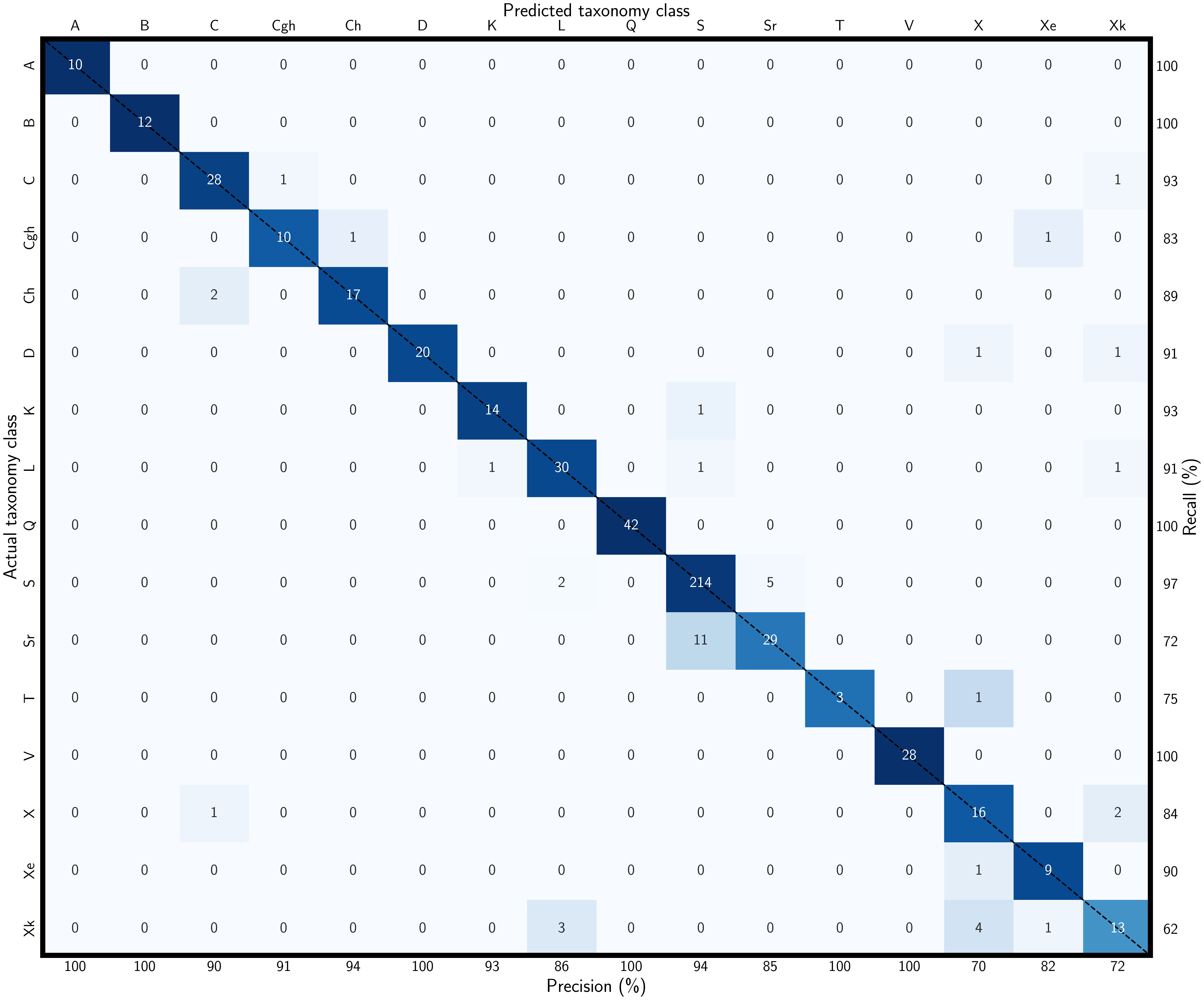}
{Confusion matrix of the full-resolution classification model.}
{fig:full_conf_mat}

\subsection{Classification model}

The results of the classification models are presented via confusion matrices. The confusion matrix summarises predictions of individual classes. Rows represent true classes and columns predicted classes. For example, in the full-resolution model represented in Fig.~\ref{fig:full_conf_mat}, one Cgh-type asteroid was predicted as Ch-type, one as Xe-type, and 10 correctly as Cgh-type (line 4). From the predicted Cgh-type (column 4) one belongs to the C-type and 10 belong to Cgh-type. The accuracy metrics calculated using Eqs.~\ref{eq:CA}--\ref{eq:recall} are listed in Table~\ref{tab:metrics_tax}. Confusion matrices for the Eros and Itokawa-resolution models can be found in Appendix~\ref{app:taxonomical}.

Among asteroids rich in dry silicates, the A-type and V-type asteroids have special positions. Both of them are thought to be predominantly composed of pure minerals, namely, the A-type is connected with olivine and the V-type is connected with pyroxene \citep{deSanctis_2011}. Even though there are only 10 A-type and 28 V-type asteroids in our dataset, all classification models classed all these samples correctly. In addition, no other sample was classed in these two categories.


\subsubsection{Model with full resolution}
\label{sect:xtest}

The model predictions of the taxonomy classes are visualised via a confusion matrix in Fig.~\ref{fig:full_conf_mat}. A total of \resnum{495} out of 538 asteroids were correctly classified in our reduced taxonomy system. Most of the mismatches are due to unsharp boundaries among the taxonomy classes, for instance, within the S-complex (among S, Sr, and Q-types) and within the C$+$X-complex (among B, C, Cgh, Ch, X, Xe, and Xk-types). The most common mismatch exists between S and Sr-types (\resnum{16} cases) and between X and Xk-types (\resnum{six} cases). These led to lower $F_1$ scores for Sr, X, and Xk-types, compared to other types. An important performance indicator is that no mismatch is observed between S-complex and C$+$X-complex. The precision, recall, $F_1$ score, and categorical accuracy values computed from all the data resulted in \resnum{92\%} reliability of the model. Slightly lower accuracy is indicated by Cohen's $\kappa_C$.

\begin{table}
    \sisetup{table-format = 2.1}
    \caption{Mean match scores and predicted classes of the Sq-type asteroids and meteorite and mineral samples.}
    \label{tab:met_ast_tax}
    \centering
    \begin{tabular}{l S S S S S}
        \hline\hline
        \mc{} & \mc{Sq-type} & \mc{OC} & \mc{HED} & \mc{PX} & \mc{OL}\\ 
        \hline
        \mc{} & \multicolumn{5}{c}{Match score (\%)} \\
        \hline
        A & 0.0 & 0.0 & 0.0 & 7.2 & 46.2 \\
        K & 0.0 & 0.1 & 10.6 & 0.4 & 39.3 \\
        Q & 15.6 & 61.5 & 32.3 & 22.8 & 13.8 \\
        S & 80.6 & 23.7 & 0.0 & 2.0 & 0.0 \\
        Sr & 3.7 & 14.1 & 0.0 & 9.1 & 0.0 \\
        V & 0.0 & 0.0 & 57.1 & 56.8 & 0.0 \\
        \hline
        \mc{} & \multicolumn{5}{c}{Predicted class (\%)} \\
        \hline
        A & 0.0 & 0.0 & 0.0 & 8.0 & 49.0 \\
        K & 0.0 & 0.0 & 14.3 & 0.4 & 40.0 \\
        Q & 12.2 & 62.2 & 28.6 & 21.7 & 11.0 \\
        S & 85.4 & 24.4 & 0.0 & 1.9 & 0.0 \\
        Sr & 2.4 & 13.3 & 0.0 & 9.5 & 0.0 \\
        V & 0.0 & 0.0 & 57.1 & 57.0 & 0.0 \\
        \hline
    \end{tabular}
\end{table}

The reliability test on Sq-type asteroids, ordinary chondrites and HED meteorites, and pure pyroxene and olivine samples are summarised in Table~\ref{tab:met_ast_tax}. 
The most probable classes of the 41 Sq-type asteroids are S-type with mean match score \resnum{80.6}\%, Q-type (mean match score \resnum{15.6}\%), and Sr-type (mean match score \resnum{3.7}\%). The preference for S-type over Q-type is probably caused by an imbalanced training set in which S-type asteroids dominate. 
Most of the Sq-type asteroids are predicted as very similar to S or Q-types. 

Ordinary chondrites are spectrally similar to Q and S-type asteroids \citep{Chapman_1996}. This is correctly predicted by the classification model. Overall, the mean match score with Q-type asteroids is above \resnum{60\%} and the remaining almost \resnum{40\%} tends to S and Sr-types. Specifically, from 45 ordinary chondrites, the model assigns Q-type to \resnum{28} samples, S-type to \resnum{11} samples, and Sr-type to \resnum{6} samples. All these types contain the 1-\textmu{}m and 2-\textmu{}m absorption bands that are formed by olivine and pyroxene. We note that the mean olivine fraction, mean Fa, and mean Fs (OPX) are about 5~pp higher for the ordinary chondrites that were predicted as `almost' Q-type (match score over 90\%) compared to those predicted as `almost' S-type.
HED meteorites are usually rich in pyroxene. The taxonomy class related to pyroxene-rich asteroids is the V-type. From 7 HED meteorites, \resnum{4} are predicted as V-type, \resnum{2} as Q-type, and \resnum{1} as K-type. The mean match scores of these classes are 57.1\%, 32.3\%, and 10.6\% for V, Q, and K, respectively. We identified the misclassifications as fragments of MIL-03443 meteorite. This meteorite is olivine-rich (94~vol\%) dunite-like HED \citep{Beck_2011}. Its spectra show a dominant 1-\textmu{}m band with a broach, a shallow 2-\textmu{}m band, and no reddening. For these reasons, the classifications as K or Q-types are expected and correctly predicted by the model.

The predictions for the pyroxene samples (PX; pure OPX, pure CPX, and binary mixtures of OPX and CPX) should be similar to the predictions for HED meteorites. The classes with the highest mean match score are V-type with the match score of \resnum{56.8\%}, Q-type (22.8\%), and Sr-type (9.1\%). From 263 pyroxene samples, V-type is the most probable for \resnum{150} of them, Q-type for \resnum{57} samples, Sr-type for \resnum{25} samples, and A-type for \resnum{21} samples. Other types contain only a minor fraction of samples, namely S (\resnum{$5\times$}), L (\resnum{$2\times$}), Ch (\resnum{$2\times$}), and K (\resnum{$1\times$}). 
Olivine spectrum is dominated by a strong 1-\textmu{}m absorption band. The taxonomy classes whose only strong absorption band is around 1~\textmu{}m are A and K-types. The mean match scores of these are \resnum{46.2\%} and \resnum{39.3\%} for A and K, respectively. Among the 100 olivine samples, \resnum{49} are predicted as A-type, \resnum{40} as K-type, and \resnum{11} as Q-type.


\subsubsection{Model with Eros resolution}

Taxonomy predictions based on low-resolution reflectance spectra are significantly worse than in the case of the full-resolution model. The limited wavelength range that starts at 820~nm significantly affects predictions on the individual types in C$+$X-complex, particularly Xe, Xk, and Cgh-types, because these types are defined via features located around 500 and 700~nm. In general, the wavelength coverage is crucial for the accuracy of taxonomy predictions (see the left panel in Fig.~\ref{fig:error_all}). The categorical accuracy, $F_1$ score, and Cohen's $\kappa_C$ of the model are 0.77, 0.75, and 0.71, respectively. That means that there are more than twice that many incorrectly predicted classes. The misclassifications are predominantly found within the S-complex and within the C$+$X-complex (see Fig.~\ref{fig:E_conf_mat}).


\subsubsection{Model with Itokawa resolution}

The limited wavelength range of the Itokawa spectra does not affect the accuracy metrics compared to the Eros-resolution model. The overall predictions of the Itokawa-resolution model are comparable to the Eros-resolution model. The categorical accuracy, $F_1$ score, and Cohen's $\kappa_C$ are \resnum{0.78}, \resnum{0.77}, and \resnum{0.72}, respectively, and misclassifications are mostly within S-complex and C$+$X-complex as can be seen in Fig.~\ref{fig:I_conf_mat}.


\subsection{Composition model}

In the following subsections, we present the results of our composition models. The results are visualised via scatter and error plots and summarised in tables. In the scatter plots, the true abundances and compositions are on the horizontal axes, while the predicted ones are on the vertical axes. The error bars were estimated as RMSE between the predicted and actual values in bins with a width of 10~percentage points (pp, the difference between two percentage values); for details, see \citet{Korda_2023}. The diagonal lines delimit different accuracy intervals. 
Table~\ref{tab:metrics_comp} contains the evaluation of accuracy metrics, Table~\ref{tab:quantile} lists what part of predictions are within a given accuracy interval, and Table~\ref{tab:one_sigma} summarises the 1-$\sigma$ error estimates. All tables include models with different resolutions. The result plots for the Eros and Itokawa-resolution models can be found in Appendix~\ref{app:compositional}. We note that the absolute errors of individual predicted quantities behave similarly in all the models (see the right panel in Fig.~\ref{fig:error_all} for overall error statistics).


\wfigure{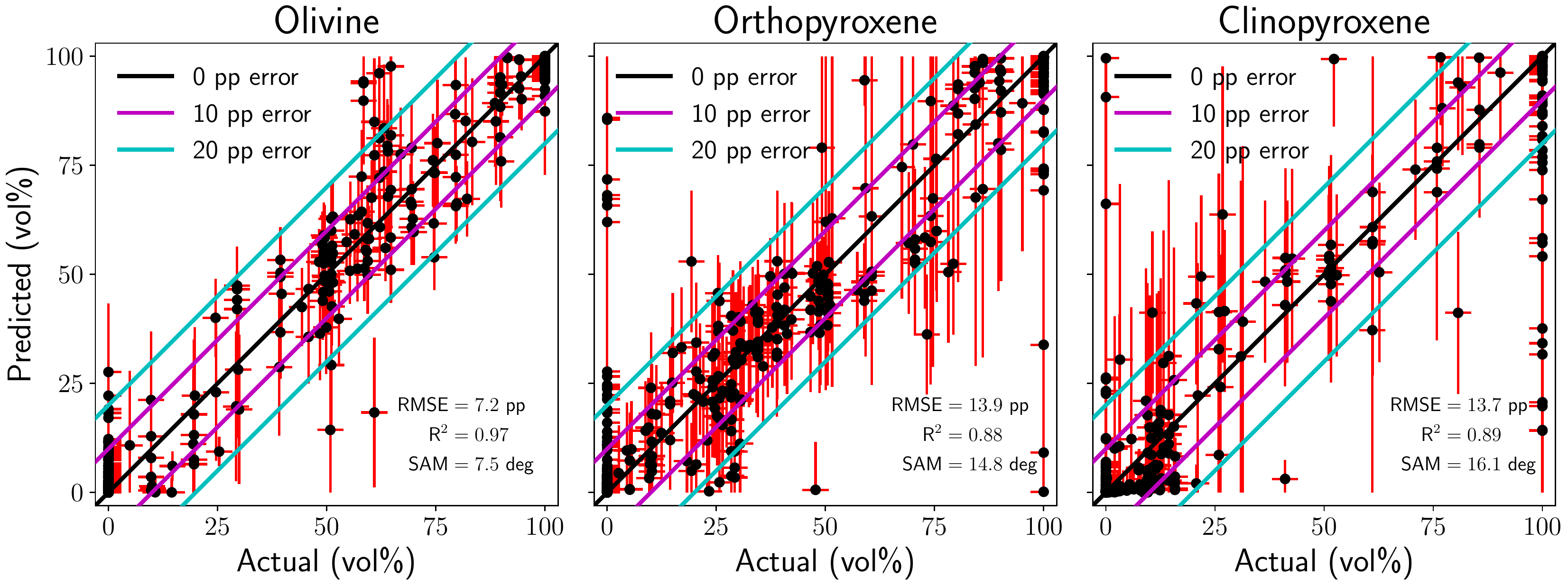}
{Comparison of the true and predicted (by the full-resolution model) modal abundances.}
{fig:full_mineral}

\begin{figure*}
    \centering
    \begin{minipage}[b]{.48\textwidth}
        \resizebox{\hsize}{!}{\includegraphics{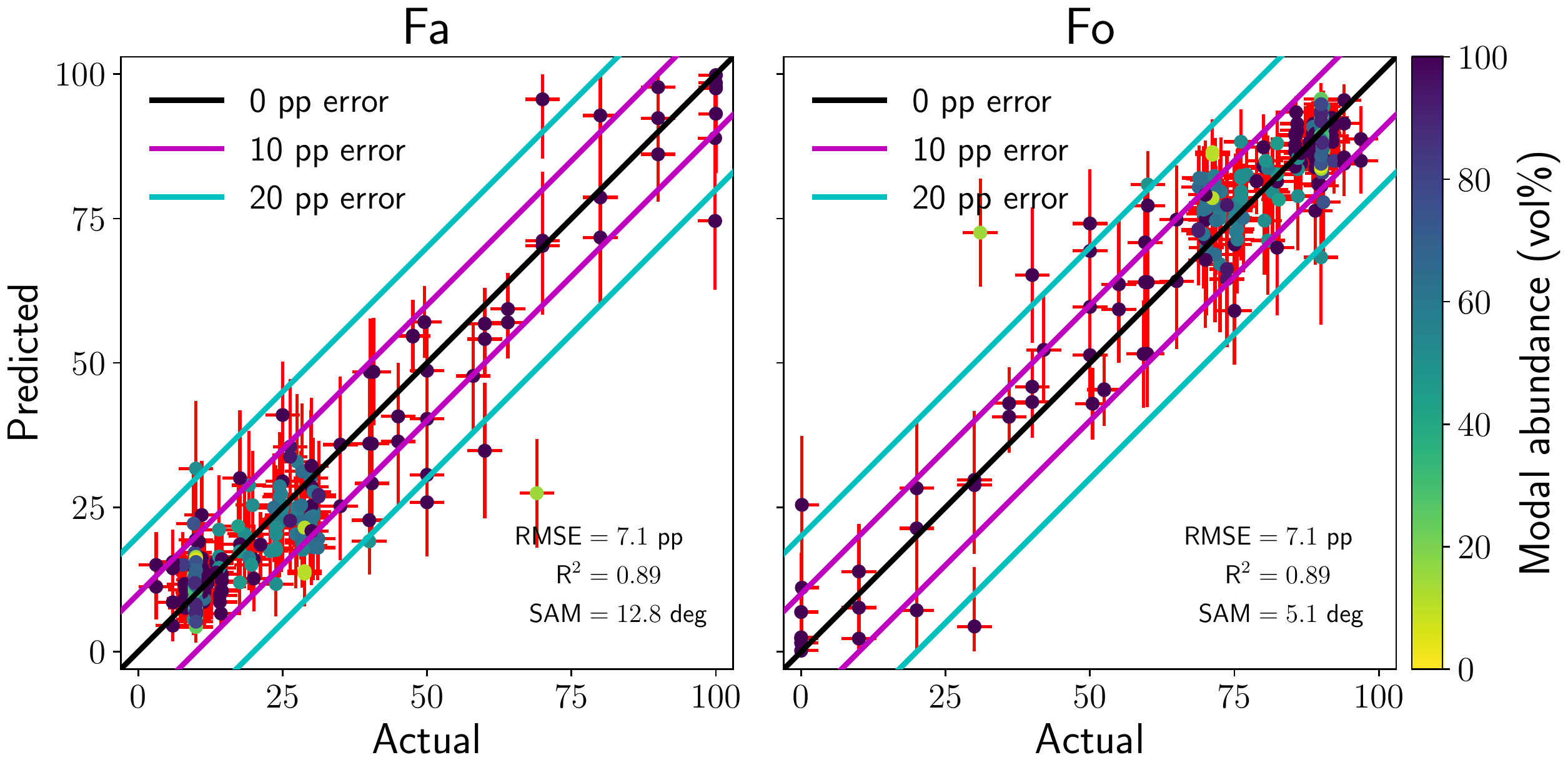}}
        \caption{Comparison of the true and predicted (by the full-resolution model) olivine composition. \textit{Left}: Iron content. \textit{Right}: Magnesium content. The point colours indicate the actual modal abundance of olivine in the samples.}
        \label{fig:full_ol}
    \end{minipage}
    \hfill
    \begin{minipage}[b]{.48\textwidth}
        \resizebox{\hsize}{!}{\includegraphics{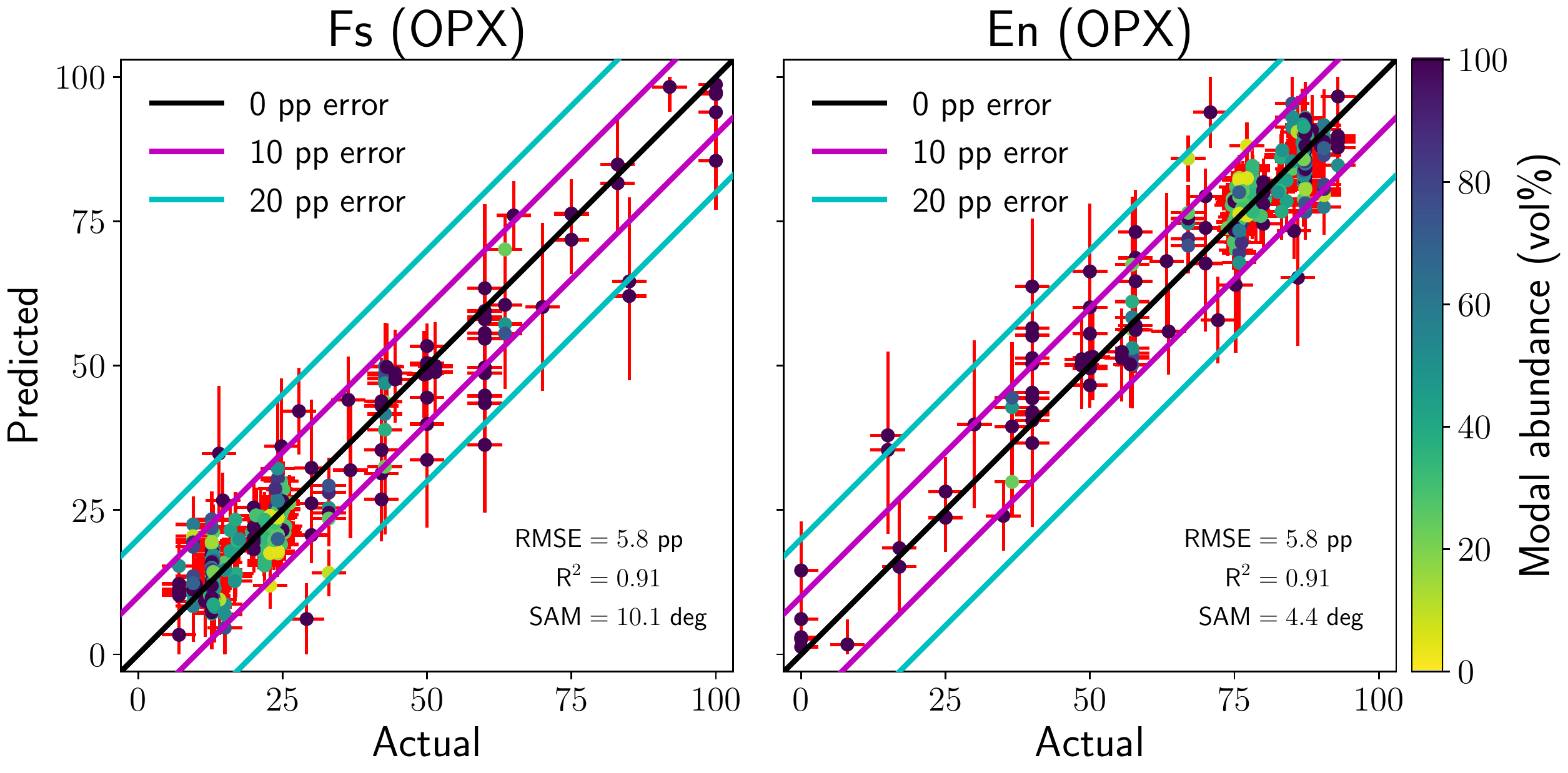}}
        \caption{Comparison of the true and predicted (by the full-resolution model) orthopyroxene composition. \textit{Left}: Iron content. \textit{Right}: Magnesium content. The point colours indicate the actual modal abundance of orthopyroxene in the samples.}
        \label{fig:full_opx}
    \end{minipage}
\end{figure*}

\wfigure{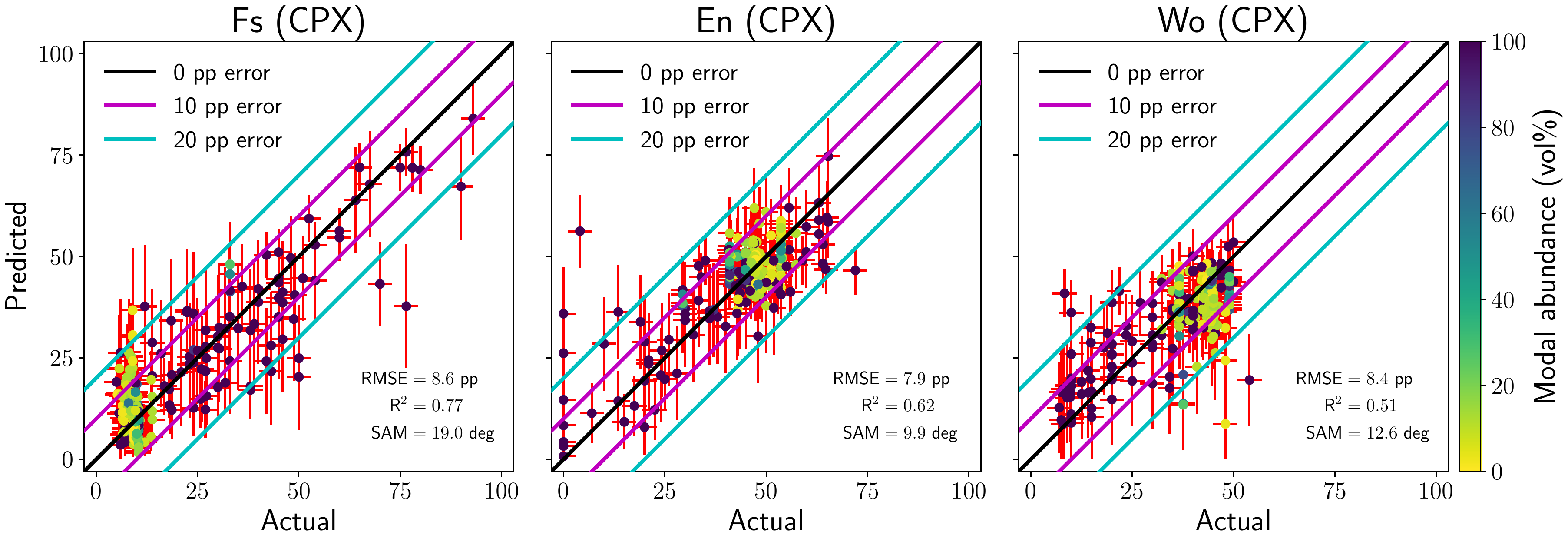}
{Comparison of the true and predicted (by the full-resolution model) clinopyroxene composition. \textit{Left}: Iron content. \textit{Middle}: Magnesium content. \textit{Right}: Calcium content. The point colours indicate the actual modal abundance of clinopyroxene in the samples.}
{fig:full_cpx}


\subsubsection{Model with full resolution}

The chemical mineral composition in full resolution was evaluated and discussed in \citet{Korda_2023}. Here, we used a more advanced $k$-fold procedure to get a better estimate of model precision. In Figs.~\ref{fig:full_mineral}--\ref{fig:full_cpx}, there are the predictions of the neural network model.

The results are mostly comparable with those of the previous study. The best results are obtained for the modal abundance of olivine and olivine and orthopyroxene chemical compositions. Compared to \citet{Korda_2023}, the modal orthopyroxene and clinopyroxene composition error estimates are lower. The lower error is mostly attributed to a lower fraction of orthopyroxene--clinopyroxene mismatches that highly contribute to the RMSE estimate. The orthopyroxene--clinopyroxene mismatches are naturally presented due to the smooth spectral transition between these two pyroxene phases. We define orthopyroxene--clinopyroxene outlier as orthopyroxene--clinopyroxene mixture (volume fraction of both pyroxene phases is 95\% or higher) for which absolute error in prediction of orthopyroxene or clinopyroxene modal abundance is higher than 40~vol\%. In this model, we found \resnum{15} such outliers. None of them contain olivine, \resnum{3} of them are pure orthopyroxene, \resnum{11} are pure clinopyroxene, and \resnum{1} is an OPX--CPX binary mixture. Significant improvement is seen in the case of the chemical composition of clinopyroxene, where the increased amount of training data led to a reduction of biases, and therefore better estimates of composition in a wider compositional range. This is best seen from the $R^2$ metric. For the clinopyroxene enstatite and wollastonite, $R^2$ is between \resnum{0.5} and \resnum{0.6}, while in the previous study, these were zeros in both cases. In total, \resnum{86\%} of absolute error values are within 10~pp. Assuming the normal distribution of prediction errors, 1-$\sigma$ errors are between \resnum{3.1}~pp and \resnum{6.8}~pp.


\subsubsection{Model with Eros resolution}

Unlike in the classification model, the lower resolution of the Eros spectra does not worsen the predictions of the mineral composition. Most of the computed metrics indicate even better results than for the full-resolution model. It may be because the reflectance spectra of silicates are well-separated in mineral composition space and the loss of fine details is not important. Also, for the Eros-resolution model, the most precise are predictions of olivine modal abundance and olivine and orthopyroxene chemical compositions, all with the RMSE of about 5~pp. In this model, we find 15 orthopyroxene--clinopyroxene mismatches (five pure orthopyroxene, eight pure clinopyroxene, and \resnum{two} OPX--CPX binary mixtures). Overall, 86\% of all predictions are within 10~pp error. The 1-$\sigma$ error estimates are in a range from 4.0~pp to 7.6~pp. The predictions are plotted in Figs.~\ref{fig:E_mineral}--\ref{fig:E_cpx}.

\begin{table}
    \sisetup{separate-uncertainty, table-align-uncertainty = true, table-format = 2.1(2)}
    \caption{Mean and standard deviations of Eros and Itokawa composition predictions together with Itokawa particle laboratory analyses.}
    \label{tab:E_I_predictions}
    \centering
    \begin{tabular}{l S S S}
        \hline\hline
        \mc{Label} & 
        \mc{Eros} & 
        \mc{Itokawa} &
        \mc{Itokawa lab.} \\
        \hline
        OL (vol\%) & 57.6(45) & 69.0(15) & 75.2 \\
        OPX (vol\%) & 28.9(18) & 27.0(17) & 21.5 \\
        CPX (vol\%) & 13.5(41) & 4.0(19) & 3.3 \\
        Fa & 22.8(11) & 23.1(7) & 28.4(12) \\
        Fs (OPX) & 21.9(36) & 22.8(4) & 23.4(20) \\
        Fs (CPX) & 10.1(6) & 21.3(14) & 9.0(15) \\
        En (CPX) & 49.6(22) & 35.8(15) & 43.2(43) \\
        Wo (CPX) & 40.4(22) & 43.0(8) & 47.9(29) \\
        \hdashline
        %
        Q (\%) & 8.5(55) & 19.6(91) & \mc{-} \\
        S (\%) & 89.5(53) & 80.0(90) & \mc{-} \\
        \hline
    \end{tabular}
    \tablefoot{Itokawa particle analyses were adopted from \citet{Nakamura_2014} and \citet{Tsuchiyama_2014}. Modal mineral compositions are always normalised to $\mathrm{OL} + \mathrm{OPX} + \mathrm{CPX} = 100~\mathrm{vol}\%$.}
\end{table}


\subsubsection{Model with Itokawa resolution}

Similarly as for the Eros-resolution classification model, the incompletely covered 2-\textmu{}m band of pyroxene and the lower resolution of the Itokawa spectra do not worsen the predictions of the pyroxene abundances and their compositions. Therefore, not only the spectra are well-separated in the compositional space, but also the whole 2-\textmu{}m band is not necessary to derive proper pyroxene composition. We find \resnum{11} orthopyroxene--clinopyroxene mismatches (\resnum{1} pure orthopyroxene, \resnum{7} pure clinopyroxene, and \resnum{3} OPX--CPX binary mixtures). In sum, \resnum{86\%} of the predictions are within the 10~pp error interval. The typical 1-$\sigma$ error is about \resnum{5}~pp with a range from \resnum{2.3}~pp to \resnum{7.0}~pp. The predictions
are plotted in Figs.~\ref{fig:I_mineral}--\ref{fig:I_cpx}.


\subsection{Evaluation of the Eros and Itokawa spectra}

We applied the trained models to Eros and Itokawa spectra pre-processed by the routine indicated in Sect.~\ref{sect:E_I_data} to derive the mean taxonomic class and composition of these asteroids as well as to map local variations. The overall results (averaged over the predictions) are summarised in Table~\ref{tab:E_I_predictions}. We present only taxonomy classes which have mean match scores above \resnum{5\%}. We note that the only additional taxonomy class with the mean match score above 1\% is the L-type in the case of Eros predictions. The 3D projection maps of Eros and Itokawa were obtained using shapeViewer\footnote{\url{https://www.comet-toolbox.com/shapeViewer.html}} software \citep{Vincent_2018}. The classification and composition maps for the asteroids can be found in Figs.~\ref{fig:Eros_S}--\ref{fig:Itokawa_OL} and in Appendices~\ref{app:Eros} and \ref{app:Itokawa}.


\subsubsection{Eros}

{The taxonomical prediction presented in Table~\ref{tab:E_I_predictions} shows that the dominant predicted taxonomy type on the Eros surface is S-type with \resnum{89.5\%} mean match score. Moreover, the S-type is the most probable type in \resnum{100\%} of cases. The second and third most probable types are the Q-type and L-type with the mean match scores \resnum{8.5\%} and \resnum{1.6\%}, respectively. The predicted surface mineral modal and chemical compositions indicate a very homogeneous surface. The highest standard deviations of predictions are in olivine and clinopyroxene abundances (\resnum{4.5}~pp and \resnum{4.1}~pp, respectively) and in orthopyroxene ferrosilite number (\resnum{3.6}~pp). The mean predicted mineral modal abundances (normalised to 100~vol\%) of olivine, orthopyroxene, and clinopyroxene are \resnum{57.6~vol\%}, \resnum{28.9~vol\%}, and \resnum{13.5~vol\%}, respectively. The predicted mineral chemical composition of olivine is $\mathrm{Fa} = \resnum{22.8}$ and of orthopyroxene is $\mathrm{Fs} = \resnum{21.9}$.


\subsubsection{Itokawa}

The taxonomical prediction (see Table~\ref{tab:E_I_predictions}) indicates that the most likely spectral match to Itokawa surface in \resnum{99\%} of all cases is S-type spectrum with the mean match score of \resnum{80.0\%}. The second likely match is the Q-type spectrum with the mean score of \resnum{19.6\%}. The predictions of the surface mineral abundance and compositions are very homogeneous with standard deviations lower than \resnum{2.0}~pp. The mean predicted mineral modal abundances of olivine, orthopyroxene, and clinopyroxene are \resnum{69.0~vol\%}, \resnum{27.0~vol\%}, and \resnum{4.0~vol\%}, respectively. The predicted chemical composition of olivine is $\mathrm{Fa} = \resnum{23.1}$ and of orthopyroxene is $\mathrm{Fs} = \resnum{22.8}$.

\tcfigure{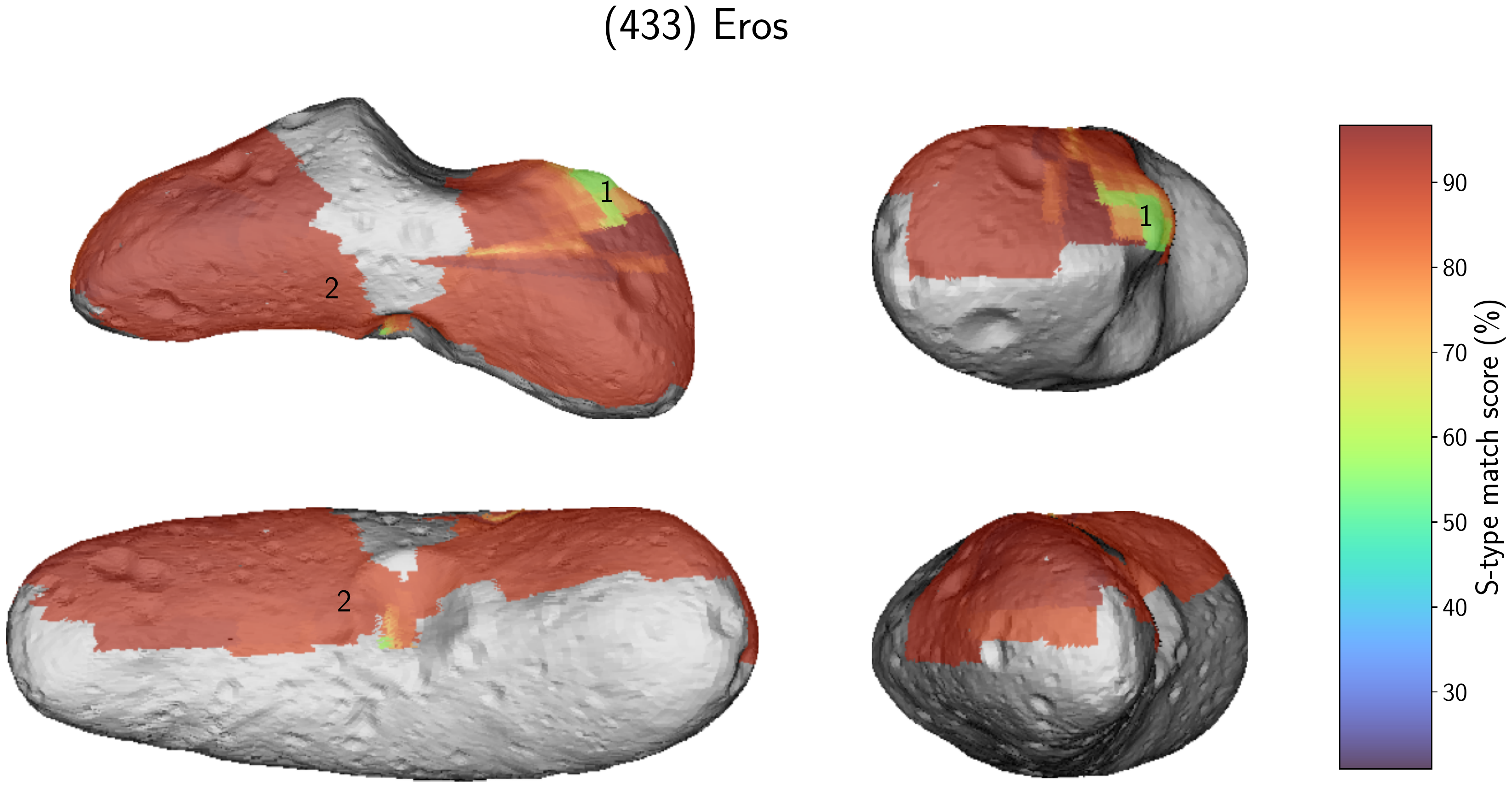}
{Predicted match score of the S-type asteroids on the surface of Eros. The numbers designate areas discussed in the text.}
{fig:Eros_S}

\tcfigure{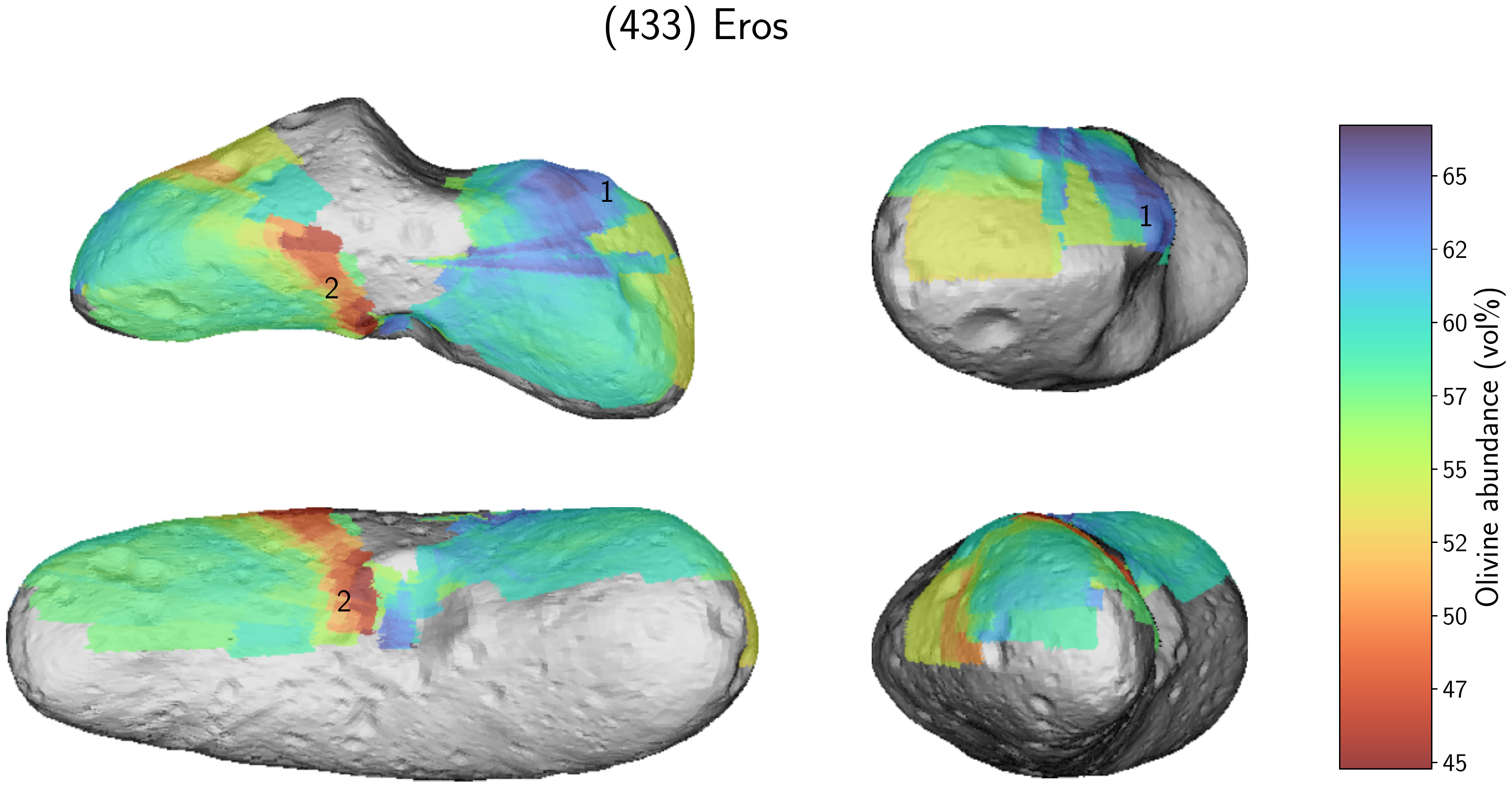}
{Predictions of olivine abundance on the surface of Eros. The numbers designate areas discussed in the text.}
{fig:Eros_OL}

\tcfigure{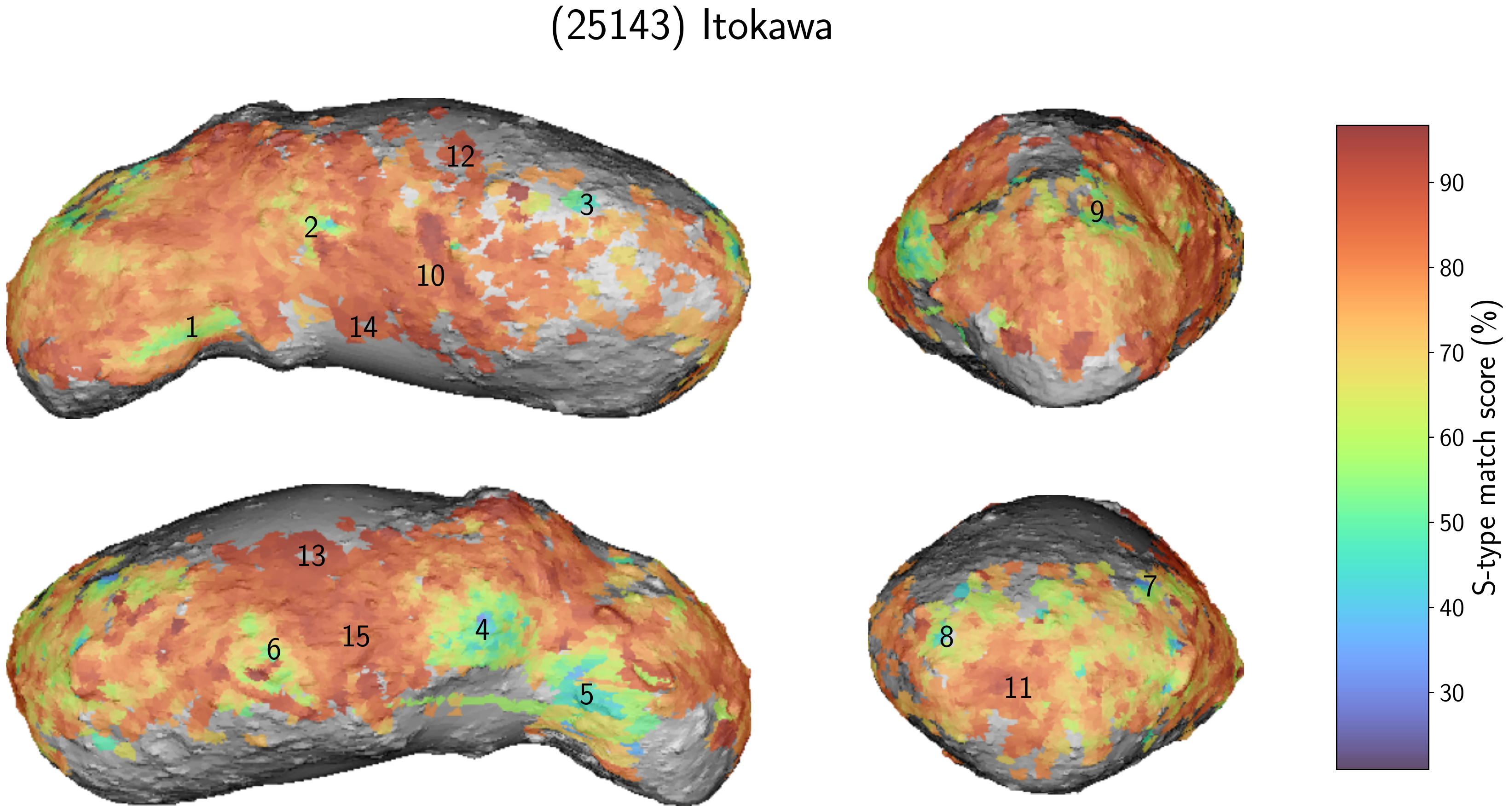}
{Predicted match score of the S-type asteroids on the surface of Itokawa. The numbers designate fresh and mature areas.}
{fig:Itokawa_S}

\tcfigure{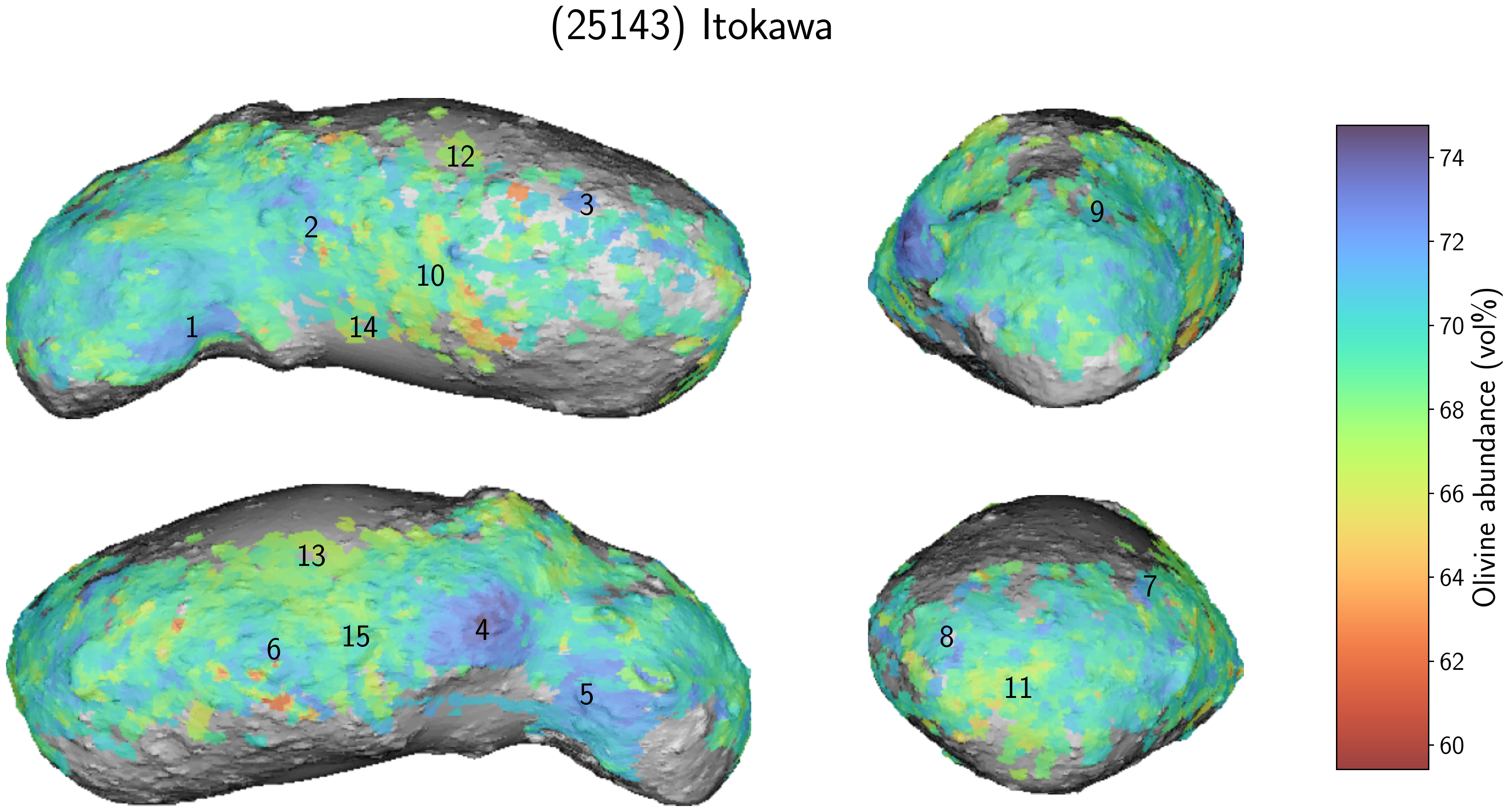}
{Predictions of olivine abundance on the surface of Itokawa. The numbers designate fresh and mature areas.}
{fig:Itokawa_OL}


\section{Discussion}

\subsection{Biases}

The performance of a neural network depends on the data used to train it. Our training data were selected manually based on visual quality screening. The manual screening lowers the total number of samples by about 15\%. The reduced amount of samples can make the model training less stable and make the predictions potentially biased because the number of spectra per individual taxonomy classes are not comparable or not all combinations of silicate abundances and chemistries are present.

The estimation of bias for classification is not straightforward. Based on Fig.~\ref{fig:full_conf_mat}, the biases among the asteroid classes are rare and amount to a few cases at most. Weak biases are possible between S and Sr-types (biased towards S-type) and higher between X and Xk-types (biased towards X-type). The higher bias can be due to the spectral variability within Xk-type asteroids.
Using spectra with lower resolution, about one-third to one-half of Q and Sr-types were predicted as S-type (see Figs.~\ref{fig:E_conf_mat} and \ref{fig:I_conf_mat}). Therefore, the predicted probabilities of the S-type can be artificially higher, while predicted probabilities of Q and Sr-types can be lowered for low-resolution spectra. 

Our data for composition models cover most of the solid solutions of olivine and pyroxene, and a wide range of mixture modal abundances \citep[see Fig.~1 in][]{Korda_2023}. The regression biases can be studied via prediction errors (e.g. computed as predicted value minus actual value). The statistics of the prediction errors are plotted in Fig.~\ref{fig:full_density}. The biases are present when the mean value of the prediction error is significantly non-zero, is dependent on the composition, or when the prediction error distributions are asymmetric. The predictions of modal abundances as well as predictions of olivine and orthopyroxene compositions do not indicate any biases. In the clinopyroxene chemical composition, the enstatite and wollastonite contents are biased. Especially the enstatite is significantly biased because of the trend between the composition and the prediction error. This decreases the interval in which the enstatite predictions are valid. We estimate the enstatite validity interval to $\left< 40; 60 \right>$.

\begin{figure*}
    \centering
    \resizebox{0.48\textwidth}{!}{\includegraphics{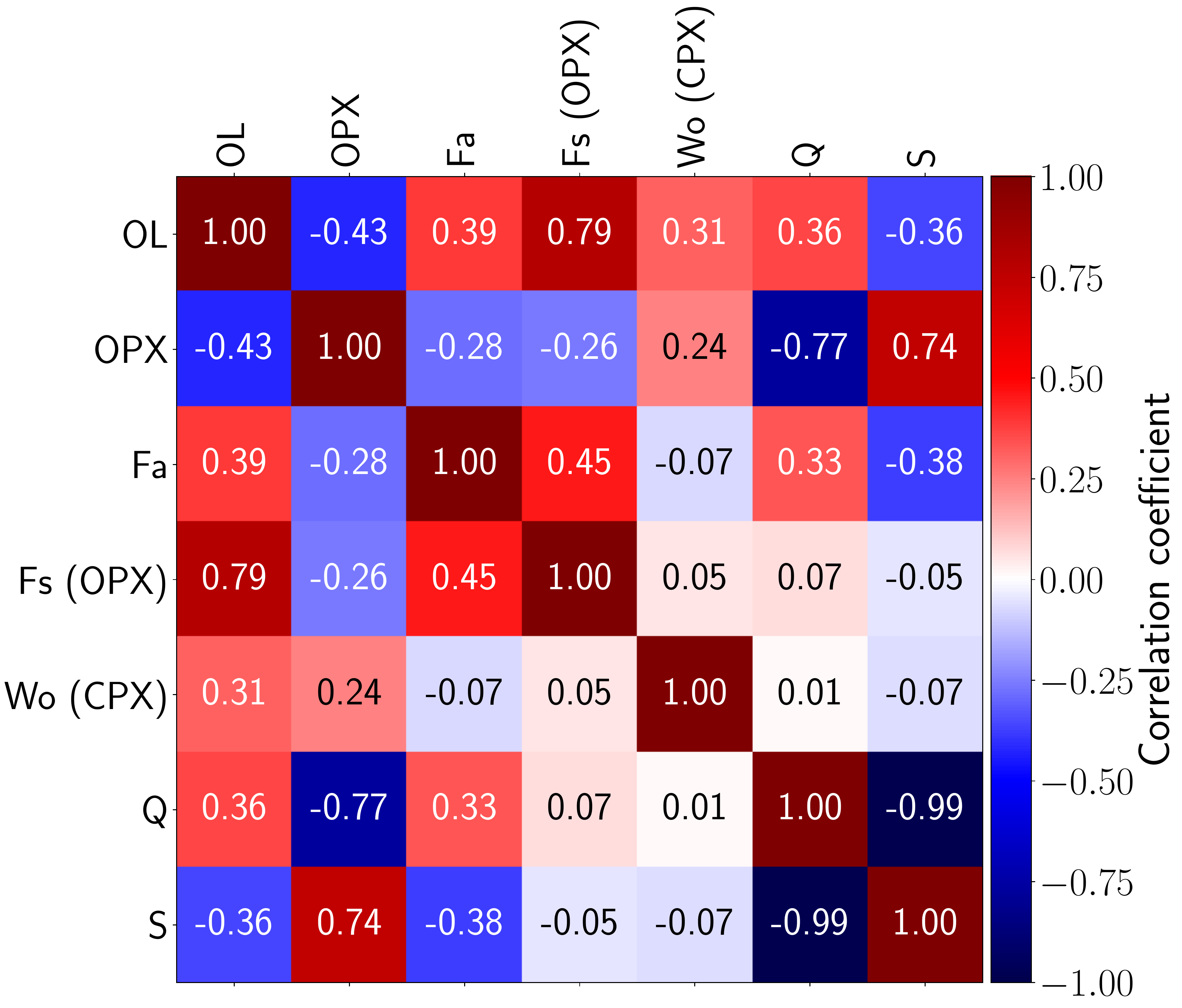}}
    \hfill
    \resizebox{0.48\textwidth}{!}{\includegraphics{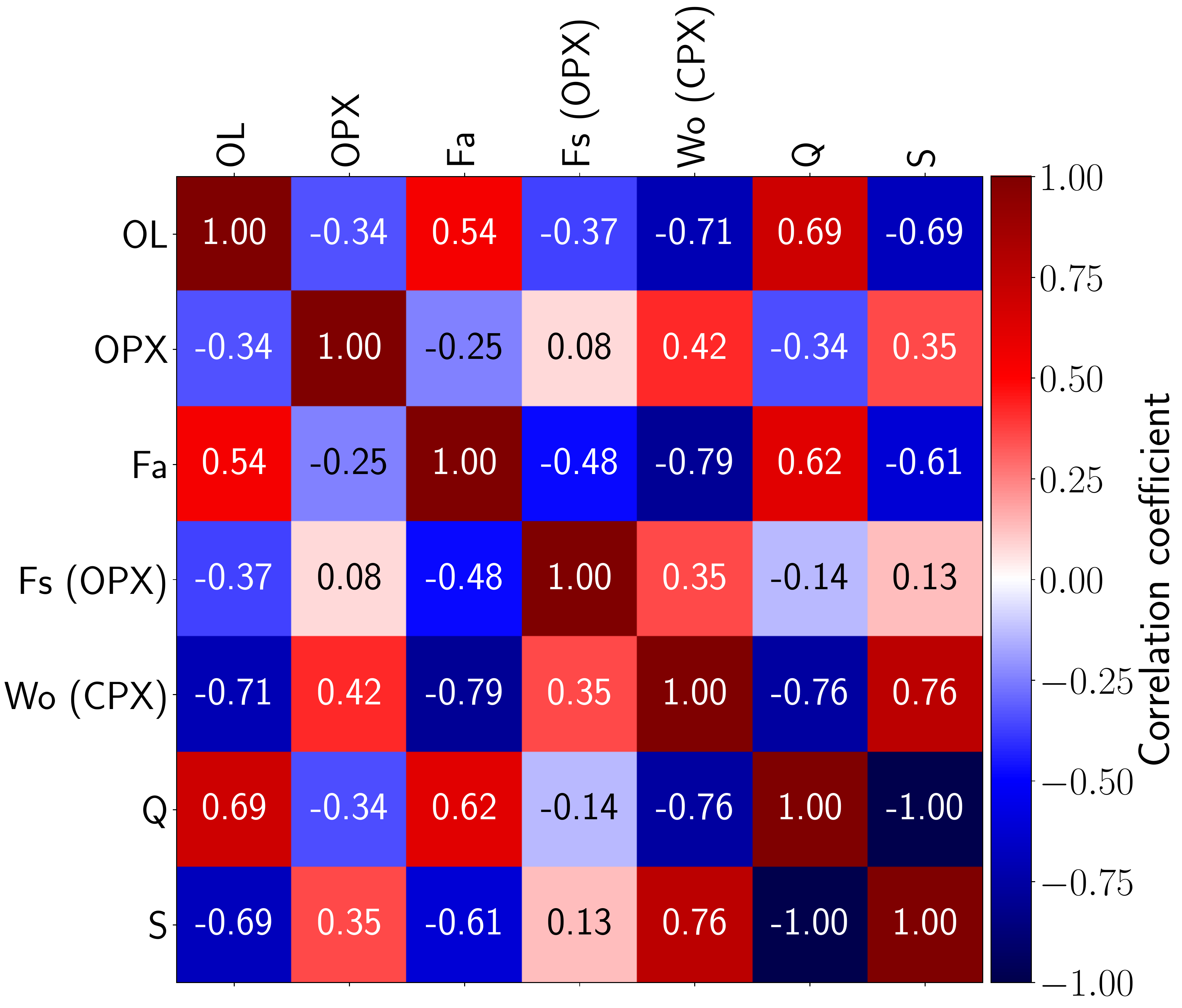}}
    \caption{Correlation matrices for combined outputs of composition and classification models. \textit{Left}: Eros. \textit{Right}: Itokawa.}
    \label{fig:corr_mat}
\end{figure*}

\ocfigure{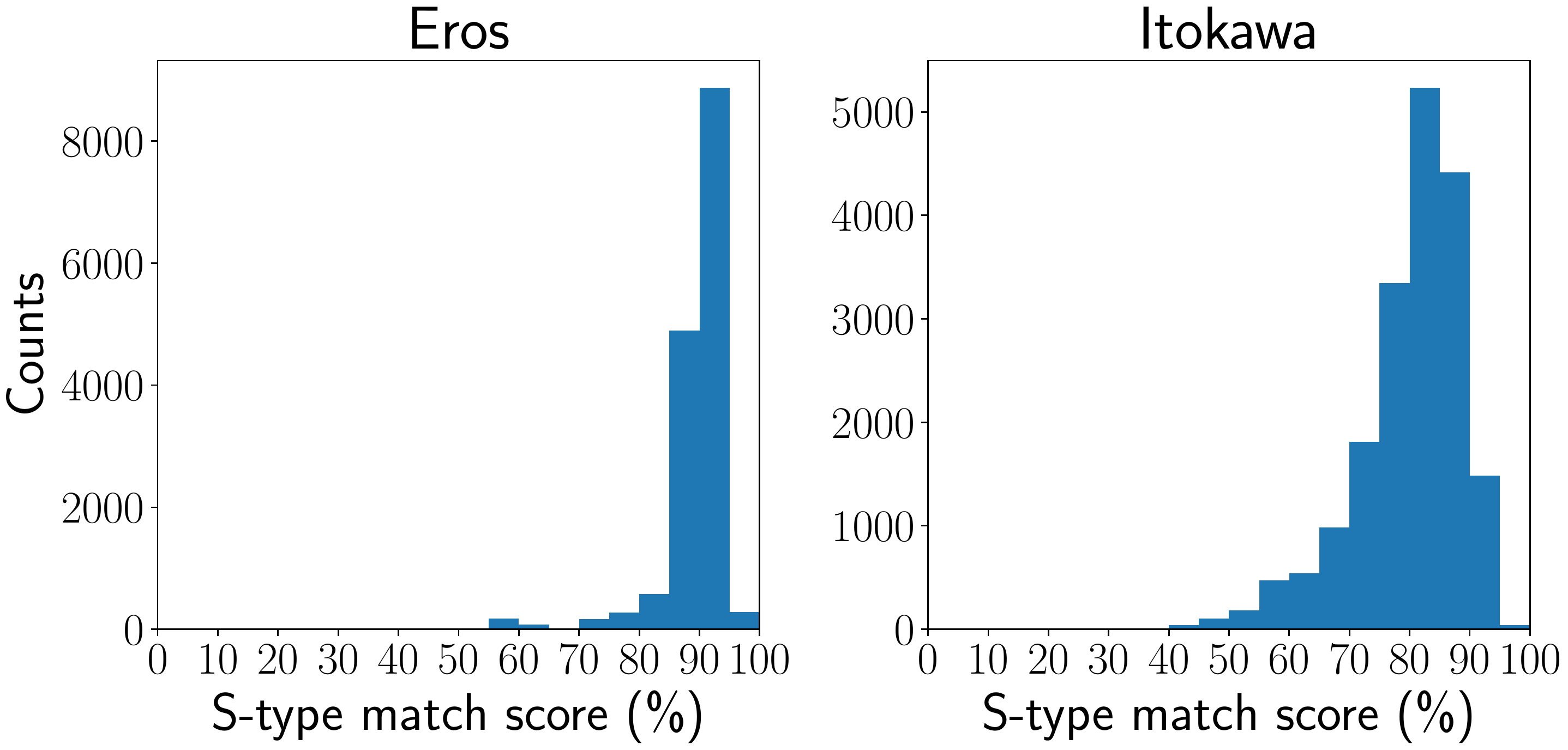}
{Histograms of the S-type match score. \textit{Left}: Eros. \textit{Right}: Itokawa.}
{fig:EI_S_hist}


\subsection{Asteroid surfaces}

\subsubsection{Eros}

Eros is classified as an S-type asteroid with its surface predominantly composed of matured dry silicates \citep{DeMeo_2009, Mahlke_2022}. This is consistent with our classification model results indicating the highest mean match score of almost \resnum{90}\% for S-type asteroids (see Table~\ref{tab:E_I_predictions}), with the second likely spectral type being Q-type with a mean match score of about \resnum{9}\% and the third being L-type with a mean match score of about \resnum{2}\%.

The S-type match score of the covered Eros surface is relatively constant with the exception being a fresher (lower S-type score) area (1)~north of Charlois Regio (see Figs.~\ref{fig:Eros_S} and \ref{fig:Eros_Q}). This region also corresponds to relatively higher olivine and lower orthopyroxene abundance predicted by the composition model (see Figs.~\ref{fig:Eros_OL} and \ref{fig:Eros_OPX}). We attribute this anti-correlation between S-type score and olivine abundance to the apparent attenuation of olivine spectral signature (relative to pyroxene) in the asteroid spectra due to olivine's faster response to space weathering \citep{Chrbolkova_2021}. We discuss this in more detail in Sect.~\ref{sect:I_disc} for Itokawa, where this phenomenon is more prominent than in the case of Eros (compare olivine--S-type correlations in Fig.~\ref{fig:corr_mat}). The area marked (1) is in the vicinity of the local minimum in 1000~nm band depth (more weathered) in \citet{Bell_2002} despite being fresher and more olivine-rich in our model. 

Given the normalisation to the sum of the three evaluated silicates, olivine abundance always anti-correlates with the sum of the two pyroxenes. We consider the anomalous area~(2) manifested as a low-olivine, high-clinopyroxene stripe (see Figs.~\ref{fig:Eros_OL} and \ref{fig:Eros_CPX}) west of crater Psyche to be an artefact. It is not correlated with any prominent geological feature, but it roughly coincides with higher 2000-nm band depth and 900-nm albedo values in \citet{Bell_2002}. The olivine, orthopyroxene, and clinopyroxene composition indicators (Fa, Fs, and Wo) show rather small variations (see figures in Appendix~\ref{app:Eros}), with the exception being the above-mentioned anomalous areas (1) and (2). The interpretation of the area (1) being variation in space weathering and area (2) being rather a composition artefact is supported by \citet{Nittler_2001} who reported a homogeneous surface composition from the NEAR Shoemaker X-ray spectrometer data. The overall lack of spectral variation in our S-type match score map in Fig.~\ref{fig:Eros_S} also agrees with the finding of \citet{McFadden_2001}, who reported a high level of spectral homogeneity of the Eros surface reflectance spectra recorded by the NIS instrument. The distribution of S-type scores is slightly asymmetric with a tail towards smaller values, but the asymmetry is less prominent than in the case of Itokawa (see Fig.~\ref{fig:EI_S_hist}).

Table~\ref{tab:E_I_predictions} shows our predicted Eros mineral composition. Mineral modal abundances (normalised to the sum of evaluated silicates) predicted by our composition model show mean values of \resnum{58~vol\%} olivine, \resnum{29~vol\%} orthopyroxene, and \resnum{13~vol\%} clinopyroxene. This is within the (normalised) L chondrite range \citep[see, e.g. Table~3 in][]{McCoy_2001} and also consistent with three-component ($\mathrm{olivine} + \mathrm{orthopyroxene} + \mathrm{clinopyroxene}$) composition detected by \citet{McFadden_2001}. The model-predicted mean chemical composition of olivine \mbox{$\mathrm{Fa} \approx \resnum{23}$} corresponds to L ordinary chondrite value \citep{McCoy_2001} and orthopyroxene \mbox{$\mathrm{Fs} \approx \resnum{22}$} is at the boundary between L and LL chondrites. These values are by a few pp lower than the values determined by \citet{Dunn_2013} from the ground-based asteroid spectrum. As discussed more in detail in Sect.~\ref{sect:I_disc}, our olivine prediction values may be lowered by space-weathering effects and are thus less reliable than these or orthopyroxene.


\subsubsection{Itokawa}
\label{sect:I_disc}

Itokawa is classified as an S-type asteroid with its surface predominantly composed of matured dry silicates \citep{Fujiwara_2006}. This is consistent with our classification model results indicating the highest mean match score of \resnum{80}\% for S-type spectra (see Table~\ref{tab:E_I_predictions}), with the second likely spectral type being Q-type with a mean match score of \resnum{20}\%. 

The exceptions are in certain regions of, e.g. rough terrains, steep slopes, and craters \citep[subsequent crater numbering follows designation in][]{Hirata_2009}, where the surface is fresher and, therefore, more Q-type-like. To detect such regions, we evaluated the spatially-resolved Itokawa reflectance spectra using the classification model and obtained maps of S-type (see Fig.~\ref{fig:Itokawa_S}) and Q-type match scores (see Fig.~\ref{fig:Itokawa_Q}). We note that in the case of Itokawa, all other spectral types besides S and Q have match scores below 1\%,  therefore, the S and Q-type maps are almost perfectly anti-correlated. A few regions have relatively suppressed S-type and elevated Q-type match scores. In some areas, the Q-type match score reaches as high values as \resnum{60}\%. The high Q-type match scores are indicators of relatively fresh terrain and their spatial distribution is in quite a good match to regions of lower space weathering as indicated in Fig.~2 in \citet{Koga_2018} based on PCA of AMICA images. They also generally agree with areas characterised by a lower space weathering index, based on the spectral inflexion method, as defined in \citet{Ishiguro_2007}, especially seen in  their Fig.~7.
In particular, the fresher areas were identified as (1)~Shirakami, (2)~areas around fresh craters no.~24 and 29, (3)~area south-east of crater no.~9, (4) and (5)~rough areas on both sides of the neck corresponding to areas of locally higher total (gravity) potential \citep[see Fig.~2 in][]{Tancredi_2015}, (6)~fresh crater no.~35 Kamoi and Ohsumi boulder area south of it, (7)~area around crater no.~12, (8)~area around crater no.~22, and (9)~area corresponding with no.~6 Miyabaru crater. Surprisingly, (10)~crater no.~15 Komaba indicated as being fresher by \citet{Koga_2018} shows only a small increase in Q-type match, and (11)~no.~2 Arcoona and (12)~no.~5 Uchinoura craters show an even more mature interior compared to its surrounding. The Arcoona and Uchinoura case may be explained by its locally lower total potential \citep[see Fig.~2 in][]{Tancredi_2015} relative to its surroundings and associated material accumulation in its interior. The areas (1), (3), and (5) also correspond to higher local slopes (see Fig.~\ref{fig:Itokawa_slope}) calculated after \citet{Werner_1997} and \citet{Cheng_2012} assuming the constant density of 1950~kg\,m$^{-3}$ \citep{Abe_2006}.
The prominent mature areas correspond to fine material accumulation in association with locally low total potential such as (13)~Sagamihara and (14)~Muses Sea. The areas (8) and (12)--(14) also correspond to local slope minima (see Fig.~\ref{fig:Itokawa_slope}). Area~(15) roughly coincides with no.~3 Ohsumi crater, which could  also be a material accumulation area, even though the correlation with total gravity potential or slope is not as evident as it is in the cases of Arcoona and Uchinoura.
The histogram of S-type match scores (see Fig.~\ref{fig:EI_S_hist}) shows asymmetric distribution with a tail towards lower scores (or more Q-like) and rather a sharp boundary at high S-type score values. A similar trend was observed in PCA values presented in Fig.~4 in \citet{Koga_2018}. We interpret this as a manifestation of the temporal non-linearity of space weathering with larger spectral changes at lower exposure times and an approach to spectral saturation at higher exposure times \citep[e.g.][]{Vernazza_2009, Kohout_2014, Chrbolkova_2021}.

Table~\ref{tab:E_I_predictions} shows our predicted Itokawa mineral composition. Mineral modal abundances (normalised to the sum of evaluated silicates) show mean values of \resnum{69~vol\%} olivine, \resnum{27~vol\%} orthopyroxene, and \resnum{4~vol\%} clinopyroxene. This is within the normalised LL chondrite range \citep{McCoy_2001}. From Fig.~\ref{fig:Itokawa_OL} one can see that variation in olivine content is rather small in the order of percentage points. Comparing the map of olivine abundance with that of the S-type match score in Fig.~\ref{fig:Itokawa_S}, we can notice relatively good spatial anti-correlation between the predicted abundance of olivine from the composition model and intensity of space weathering (represented by an S-type match score) from the classification model. This is also indicated by the correlation matrix in Fig.~\ref{fig:corr_mat}. A similar trend was reported in our previous work \citep{Korda_2023}, where we observed similar anti-correlation between neural network-predicted asteroid olivine content and its position within the S-Q cloud of the \citet{DeMeo_2009} PCA taxonomy. We attribute this to the apparent attenuation of olivine spectral signature (relative to pyroxene) in the asteroid spectra due to olivine's faster response to space weathering \citep[see][for detailed discussion and references]{Korda_2023}. Thus, in the case of Itokawa, we interpret the variation in the olivine modal abundance as a spatial manifestation of space weathering rather than a real change in surface composition. Due to the normalisation of the total sum of detected silicates, the olivine abundance is a complement to the sum of both pyroxenes. The observed variations between orthopyroxene and clinopyroxene fractions plotted in Figs.~\ref{fig:Itokawa_OPX} and \ref{fig:Itokawa_CPX} are rather a fluctuation of the model due to lower accuracy metrics of pyroxenes (see Tables~\ref{tab:metrics_comp} and \ref{tab:quantile}), as well as their lower mean abundances (see Table~\ref{tab:E_I_predictions}), compared to these of olivine.

The model-predicted mean chemical composition of olivine \mbox{$\mathrm{Fa} \approx \resnum{23}$} is about \resnum{5}~pp lower than the mean of Itokawa particle analytical result of \mbox{$\mathrm{Fa} \approx 28$} or the value derived by \citet{Dunn_2013} and corresponds to L ordinary chondrite values \citep{McCoy_2001}. In contrast, the composition of orthopyroxene \mbox{$\mathrm{Fs} \approx \resnum{23}$} is \resnum{almost identical} to the analytical result \citep{Nakamura_2011, Nakamura_2014} as well as the value reported by \citet{Dunn_2013} and corresponds to LL ordinary chondrite \citep{McCoy_2001}. The reduction of the predicted mean Fa number in our model (which is closer to typical L-chondrite values rather than expected LL) can also be attributed to space weathering as a similar reduction was observed within a series of Q--Sq--S asteroids analysed in \citet{Korda_2023} and may be an indicator of iron reduction from Fe$^{2+}$ into npFe$^0$ in the topmost optically active surface layers of olivine grains. However, this is just a tentative explanation and based on available data we cannot confirm the reason behind lower predicted Fa values.

The spatial variations of the Fa number  typically exhibit low variations (see Fig.~\ref{fig:Itokawa_Fa}) with a distinguishable correlation to olivine content or Q-type score (Figs.~\ref{fig:Itokawa_OL}, \ref{fig:Itokawa_Q}, and \ref{fig:corr_mat}) supporting the possible above-mentioned space weathering effect. A spatial scatter of orthopyroxene Fs value (see Fig.~\ref{fig:Itokawa_Fs_OPX}) is also low and is obviously not  correlated with other parameters possibly due to the increased resistance of orthopyroxene to space-weathering-induced surficial reduction \citep{Quadery_2015}. Model-predicted composition of clinopyroxene shows a substantial Wo component (see Table~\ref{tab:E_I_predictions}) in line with particle analysis results. Fs and En numbers show larger deviations from analytical results. We attribute this to the lower accuracy of our model in the evaluation of clinopyroxene mineral composition (Tables~\ref{tab:metrics_comp} and \ref{tab:quantile}) and the overall low predicted amount of clinopyroxene (\resnum{3--4~vol\%}) on Itokawa. Surprisingly, our model-predicted Wo values spatially correlate quite well with space weathering (compare Figs.~\ref{fig:Itokawa_Wo_CPX}, \ref{fig:Itokawa_S}, and \ref{fig:corr_mat}). We interpret this again as intrinsic model behaviour rather than real surface composition changes.

To summarise, Itokawa shows a more refined pattern with well-distinguished local variations that are often well-correlated with geological units, while Eros shows a coarser pattern, often with sharp boundaries and an unclear geological correlation. We interpret this as an artefact related to the much larger area the individual Eros spectral observations do cover (${\approx} 200~\mathrm{deg}^2$ compared to ${\approx} 6~\mathrm{deg}^2$ in the case of Itokawa) and a smaller amount of observed spectra (${\approx} 1700$ compared to ${\approx} 8400$ spectra in the case of Itokawa). The combination of these two factors results in coarser averaging and, therefore, a featureless Eros appearance with few sharp boundaries. We found an anti-correlation between space weathering (S-match score) and the olivine content and its Fa number, as well as a correlation with the Wo number (see Fig.~\ref{fig:corr_mat}). Orthopyroxene Fs number seems to be most robust against space weathering.


\section{Conclusions}

We used convolutional neural network (CNN) models to estimate taxonomy class and silicate properties on the resolved surface of the Eros and Itokawa asteroids. The classification and composition models were composed of one and two hidden layers, respectively. The classification model utilised real asteroid spectra with known taxonomy classes. The composition model was trained using reflectance spectra of real silicate samples and silicate mixtures with known composition.

The accuracy metrics of classification models indicate a worsening of predictions with decreasing resolution. With 5-nm resolution, the metrics show over \resnum{90\%} confidence. Models trained on 20-nm resolution native to Eros NIS and Itokawa NIRS datasets result in decreased confidence between \resnum{72\%} and \resnum{80\%}. This may be due to unsharp boundaries between the taxonomy classes that are less reliably resolved with a decrease in spectrum resolution. As demonstrated in the Itokawa model, the full coverage of the 2-\textmu{}m band is not strictly important for the taxonomy classification. We further tested the classification model on meteorites, minerals, and Sq-type asteroids. In most of the cases, the model predicted the expected taxonomy class, namely, Q-type for ordinary chondrites, V-type for HED meteorites and pyroxene, A and K-types for olivine, and S and Q-types for Sq-type asteroids.

The typical RMSEs of the composition models are 6~pp for olivine modal abundance and olivine and orthopyroxene chemical compositions, 8~pp for the clinopyroxene chemical composition, and 10~pp for orthopyroxene and clinopyroxene abundances. The higher error in pyroxene metrics is due to an occasional mismatch between orthopyroxene and clinopyroxene. The RMSE does not significantly depend on the resolution of reflectance spectra, and as in the case of the classification models, the complete 2-\textmu{}m band coverage is not necessary to get precise mineral or chemical estimates.

The predicted dominant taxonomy class of Eros and Itokawa is S-type with \resnum{80\%} match score. While the surface of Eros is, to a large extent, homogeneous -- the surface of Itokawa shows more pronounced local variations in space weathering that are often correlated with surface roughness variations, craters, or total (gravity) potential maxima and minima. The composition model predictions for the Eros spectra correspond to the L/LL-type ordinary chondrite composition while the predictions for the Itokawa spectra are more consistent with the LL composition. The observed surface of both asteroids is compositionally homogeneous with most detected variations attributed to model artefacts rather than to real changes in mineralogy or chemistry.


\begin{acknowledgements}
This research is supported by Academy of Finland projects 325805 and 335595, NASA SSERVI Center for Asteroid and Lunar Surface Science~(CLASS), and within institutional support RVO~67985831 of the Institute of Geology of the Czech Academy of Sciences. 
We would like to thank Francesca DeMeo and Richard Binzel for providing us with the dataset on the asteroid spectra. We utilised data stored in the RELAB Spectral Database managed by Brown University and the C-Tape database managed by The University of Winnipeg. This research has made use of NASA's Astrophysics Data System Bibliographic Services.
The authors would like to thank the anonymous referee for the valuable comments and suggestions, which greatly improved the quality of the paper.
\end{acknowledgements}

\bibliographystyle{aa}
\bibliography{BIBL}

\appendix
\onecolumn


\appsection{Correction of Eros NIS spectra}

\wfigure[!ht]{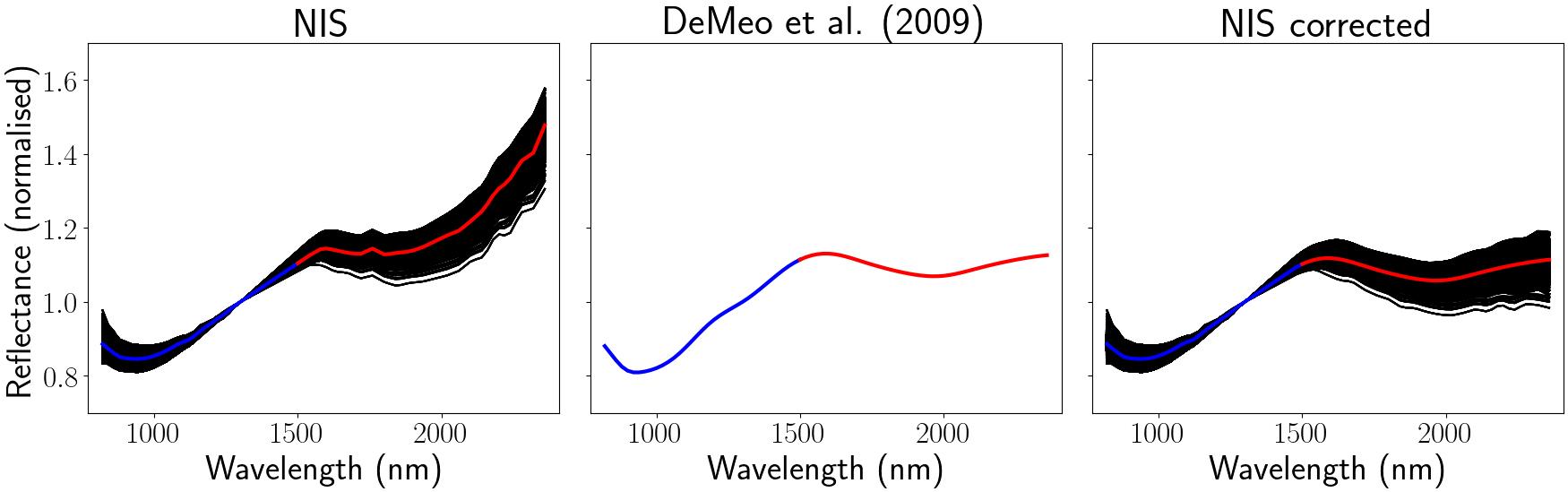}
{Correction procedure of Eros NIS spectra. \textit{Left}: NIS Eros spectra used in this study before correction (black) and their mean (blue-red curve). \textit{Middle}: Ground-based Eros spectrum. \textit{Right}: Corrected NIS spectra (black) with their mean (blue-red curve). The blue curve corresponds to wavelengths where no correction was applied while the red curve is in the corrected InGaAs spectral part.}
{fig:Eros_correction}


\section{Model hyperparameters}

\begin{table*}[!ht]
    \caption{Hyperparameters, search ranges for the hyperparameters, and other properties of the searched models.}
    \label{tab:hp}
    \centering
    \begin{tabular}{l l l l l l l l}
        \hline\hline
        & \multicolumn{2}{c}{Full res.} & \multicolumn{2}{c}{Eros res.} & \multicolumn{2}{c}{Itokawa res.} & \\
        & Class & Comp & Class & Comp & Class & Comp & \\
        \hline
        \mc{Property} & 
        \multicolumn{6}{c}{Value} & 
        \mc{Range} \\
        \hline
        Network type & conv. & conv. & conv. & conv. & conv. & conv. & conv., dense\\
        Number of hidden layers & 1 & 2 & 1 & 2 & 1 & 2 & 1--3\\
        Nodes in the input l. & 401 & 401 & 78 & 78 & 64 & 64 & fixed\\
        Nodes/filters in hid. l. & 24 & 24 and 8 & 24 & 24 and 8 & 24 & 24 and 8 & 4--32\\
        Nodes in the output l. & 16 & 10 & 16 & 10 & 16 & 10 & fixed\\
        Conv. kernel size & 5 & 5 and 5 & 5 & 5 and 5 & 5 & 5 and 5 & 3--7\\
        Hid. l. activation & ELU & ReLU & ELU & ReLU & ELU & ReLU & ReLU, tanh, sigmoid, ELU\\
        Out. l. activation & softmax & sigmoid & softmax & sigmoid & softmax & sigmoid & sigmoid, softmax\\
        Dropout rate (in.-hid.) & 0.0 & 0.0 & 0.1 & 0.0 & 0.005 & 0.0 & 0.0--0.3\\
        Dropout rate (hid.-hid.) & N/A & 0.3 & N/A & 0.3 & N/A & 0.3 & 0.0--0.5\\
        Dropout rate (hid.-out.) & 0.3 & 0.4 & 0.3 & 0.4 & 0.3 & 0.4 & 0.0--0.5\\
        $L_1$ trade-off parameter & 0.1 & 0.005 & 0.01 & 0.005 & 0.01 & 0.005 & 0.00001--1.0\\
        $L_2$ trade-off parameter & 0.1 & 0.00001 & 0.01 & 0.00001 & 0.01 & 0.00001 & 0.00001--1.0\\
        Training algorithm & Adam & Adam & Adam & Adam & Adam & Adam & Adam, SGD\\
        Learning rate & 0.0032 & 0.0005 & 0.001 & 0.0005 & 0.0013 & 0.0005 & 0.0001--1.0\\
        Batch size & 144 & 8 & 128 & 8 & 56 & 8 & 1--128\\
        Batch norm. before activation & False & False & False & False & False & False & True, False\\
        Max. number of epochs & 1500 & 2000 & 1500 & 2000 & 1500 & 2000 & fixed\\
        \hline
    \end{tabular}
    \tablefoot{Class and Comp stand for classification and composition models, respectively. N/A stands for `not applicable'.}
\end{table*}


\appsection{Accuracy metrics}

\begin{table*}[!ht]
    \sisetup{table-format = 1.2}
    \caption{Accuracy metrics computed for each class of the reduced taxonomy system and the number of spectra in the classes for the classification models.}
    \label{tab:metrics_tax}
    \centering
    \begin{tabular}{l S[table-format=3] S S S | S S S | S S S}
        \hline\hline
        \multicolumn{2}{c}{} & 
        \multicolumn{3}{c}{Full} & 
        \multicolumn{3}{c}{Eros} & 
        \multicolumn{3}{c}{Itokawa} \\
        \hline
        \mc{Class} & \mc{Counts} & \mc{Precision} & \mc{Recall} & \mc{$F_1$} & \mc{Precision} & \mc{Recall} & \mc{$F_1$} & \mc{Precision} & \mc{Recall} & \mc{$F_1$}\\
        \hline
        A & 10 & 1.00 & 1.00 & 1.00 & 1.00 & 1.00 & 1.00 & 1.00 & 1.00 & 1.00 \\
        B & 12 & 1.00 & 1.00 & 1.00 & 0.92 & 1.00 & 0.96 & 0.92 & 0.92 & 0.92 \\
        C & 30 & 0.90 & 0.93 & 0.92 & 0.78 & 0.83 & 0.81 & 0.69 & 0.83 & 0.76 \\
        Cgh & 12 & 0.91 & 0.83 & 0.87 & 0.67 & 0.17 & 0.27 & 0.00 & 0.00 & 0.00 \\
        Ch & 19 & 0.94 & 0.89 & 0.92 & 0.40 & 0.74 & 0.52 & 0.39 & 0.68 & 0.50 \\
        D & 22 & 1.00 & 0.91 & 0.95 & 0.91 & 0.95 & 0.93 & 0.88 & 1.00 & 0.94 \\
        K & 15 & 0.93 & 0.93 & 0.93 & 0.81 & 0.87 & 0.84 & 0.86 & 0.80 & 0.83 \\
        L & 33 & 0.86 & 0.91 & 0.88 & 0.81 & 0.64 & 0.71 & 0.76 & 0.76 & 0.76 \\
        Q & 42 & 1.00 & 1.00 & 1.00 & 0.85 & 0.67 & 0.75 & 0.84 & 0.62 & 0.71 \\
        S & 221 & 0.94 & 0.97 & 0.96 & 0.82 & 0.96 & 0.89 & 0.83 & 0.96 & 0.89 \\
        Sr & 40 & 0.85 & 0.72 & 0.78 & 0.80 & 0.30 & 0.44 & 0.86 & 0.47 & 0.61 \\
        T & 4& 1.00 & 0.75 & 0.86 & 0.00 & 0.00 & 0.00 & 0.00 & 0.00 & 0.00 \\
        V & 28 & 1.00 & 1.00 & 1.00 & 1.00 & 1.00 & 1.00 & 1.00 & 1.00 & 1.00  \\
        X & 19 & 0.70 & 0.84 & 0.76 & 0.46 & 0.58 & 0.51 & 0.44 & 0.42 & 0.43 \\
        Xe & 10 & 0.82 & 0.90 & 0.86 & 0.50 & 0.10 & 0.17 & 0.67 & 0.20 & 0.31 \\
        Xk & 21 & 0.72 & 0.62 & 0.67 & 0.30 & 0.29 & 0.29 & 0.50 & 0.43 & 0.46 \\
        \hdashline
        Total & 538 & 0.92 & 0.92 & 0.92 & 0.77 & 0.77 & 0.75 & 0.77 & 0.78 & 0.77 \\
        $CA$ & 538 & & & 0.92 & & & 0.77 & & & 0.78 \\
        $\kappa_C$ & 538 & & & 0.90 & & & 0.71 & & & 0.72 \\
        \hline
    \end{tabular}
\end{table*}

\begin{table*}[!ht]
    \sisetup{table-format = 2.2}
    \caption{Accuracy metric values of the composition models.}
    \label{tab:metrics_comp}
    \centering
    \begin{tabular}{l S S S | S S S | S S S}
        \hline\hline
        \mc{} & 
        \multicolumn{3}{c}{Full} & 
        \multicolumn{3}{c}{Eros} & 
        \multicolumn{3}{c}{Itokawa} \\
        \hline
        \mc{} & OL & OPX & CPX &
        OL & OPX & CPX &
        OL & OPX & CPX \\
        \hline
        \multicolumn{10}{c}{RMSE (pp)}\\
        \hline
        Modal & 7.2 & 13.9 & 13.7 & 7.2 & 12.3 & 12.7 & 7.2 & 11.2 & 12.3 \\
        \hdashline
        Fa & 7.1 & N/A & N/A & 5.9 & N/A & N/A & 6.0 & N/A & N/A \\
        Fo & 7.1 & N/A & N/A & 5.9 & N/A & N/A & 6.0 & N/A & N/A \\
        \hdashline
        Fs & N/A & 5.8 & 8.6 & N/A & 6.0 & 8.4 & N/A & 6.3 & 8.9 \\
        En & N/A & 5.8 & 7.9 & N/A & 6.0 & 7.6 & N/A & 6.3 & 8.1 \\
        Wo & N/A & N/A & 8.4 & N/A & N/A & 8.7 & N/A & N/A & 8.4 \\
        \hline
        \multicolumn{10}{c}{$R^2$}\\
        \hline
        Modal & 0.97 & 0.88 & 0.89 & 0.97 & 0.91 & 0.90 & 0.97 & 0.92 & 0.91 \\
        \hdashline
        Fa & 0.89 & N/A & N/A & 0.92 & N/A & N/A & 0.92 & N/A & N/A \\
        Fo & 0.89 & N/A & N/A & 0.92 & N/A & N/A & 0.92 & N/A & N/A \\
        \hdashline
        Fs & N/A & 0.91 & 0.77 & N/A & 0.90 & 0.78 & N/A & 0.90 & 0.75 \\
        En & N/A & 0.91 & 0.62 & N/A & 0.90 & 0.64 & N/A & 0.90 & 0.60 \\
        Wo & N/A & N/A & 0.51 & N/A & N/A & 0.48 & N/A & N/A & 0.51 \\
        \hline
        \multicolumn{10}{c}{SAM (deg)}\\
        \hline
        Modal & 7.5 & 14.8 & 16.1 & 7.6 & 13.0 & 14.8 & 7.6 & 11.9 & 14.4 \\
        \hdashline
        Fa & 12.8 & N/A & N/A & 10.7 & N/A & N/A & 10.7 & N/A & N/A \\
        Fo & 5.1 & N/A & N/A & 4.3 & N/A & N/A & 4.3 & N/A & N/A\\
        \hdashline
        Fs & N/A & 10.1 & 19.0 & N/A & 10.6 & 18.5 & N/A & 10.9 & 19.4 \\
        En & N/A & 4.4 & 9.9 & N/A & 4.5 & 9.4 & N/A & 4.7 & 10.2 \\
        Wo & N/A & N/A & 12.6 & N/A & N/A & 12.8 & N/A & N/A & 12.2 \\
        \hline
    \end{tabular}
\end{table*}

\begin{table*}[!ht]
    \sisetup{table-format = 2}
    \caption{Fraction of composition predictions that fall within the given error interval.}
    \label{tab:quantile}
    \centering
    \begin{tabular}{l S<{\%} S<{\%} S<{\%} S<{\%} | 
    S<{\%} S<{\%} S<{\%} S<{\%} | S<{\%} S<{\%} S<{\%} S}
        \hline\hline
        & 
        \multicolumn{4}{c}{Full} & 
        \multicolumn{4}{c}{Eros} & 
        \multicolumn{4}{c}{Itokawa} \\
        \hline
        &  \multicolumn{12}{c}{Absolute error} \\
        \hline
        \mc{Value} & \mc{5~pp} & \mc{10~pp} & \mc{15~pp} & \mc{20~pp} &
        \mc{5~pp} & \mc{10~pp} & \mc{15~pp} & \mc{20~pp} &
        \mc{5~pp} & \mc{10~pp} & \mc{15~pp} & \mc{20~pp} \\
        \hline
        All data & 68 & 86 & 93 & 96 & 65 & 86 & 93 & 96 & 67 & 86 & 93 & 96\% \\
        \hdashline
        OL (vol\%) & 74 & 87 & 95 & 97 & 75 & 88 & 93 & 97 & 80 & 88 & 94 & 98\% \\
        OPX (vol\%) & 65 & 79 & 89 & 93 & 63 & 78 & 88 & 92 & 66 & 82 & 91 & 94\% \\
        CPX (vol\%) & 74 & 86 & 91 & 93 & 72 & 82 & 89 & 92 & 70 & 84 & 90 & 93\% \\
        \hdashline
        Fa & 67 & 89 & 95 & 97 & 66 & 93 & 98 & 99 & 68 & 90 & 97 & 99\% \\
        \hdashline
        Fs (OPX) & 73 & 90 & 96 & 98 & 66 & 92 & 97 & 98 & 68 & 91 & 96 & 98\% \\
        \hdashline
        Fs (CPX) & 58 & 82 & 91 & 95 & 56 & 84 & 92 & 96 & 50 & 80 & 92 & 97\% \\
        En (CPX) & 58 & 87 & 96 & 97 & 59 & 86 & 95 & 97 & 58 & 82 & 94 & 97\% \\
        Wo (CPX) & 53 & 82 & 93 & 97 & 53 & 80 & 94 & 96 & 57 & 84 & 93 & 95\% \\
        \hline
    \end{tabular}
    \tablefoot{Fo and En~(OPX) have the same values as Fa and Fs~(OPX), respectively.}
\end{table*}

\begin{table}
    \sisetup{table-format = 1.1}
    \caption{1-$\sigma$ error estimates for the composition models.}
    \label{tab:one_sigma}
    \centering
    \begin{tabular}{l S S S}
        \hline\hline
        & \mc{Full} & \mc{Eros} & \mc{Itokawa} \\ 
        \hline
        & \multicolumn{3}{c}{1-$\sigma$ error (pp)} \\
        \hline
        All data & 5.0 & 5.3 & 5.1 \\
        \hdashline
        OL (vol\%) & 3.4 & 4.0 & 2.3 \\
        OPX (vol\%) & 5.8 & 6.3 & 5.4 \\
        CPX (vol\%) & 3.1 & 4.1 & 4.4 \\
        \hdashline
        Fa & 5.3 & 5.2 & 5.0 \\
        \hdashline
        Fs (OPX) & 4.2 & 5.2 & 4.7 \\
        \hdashline
        Fs (CPX) & 6.7 & 6.8 & 7.0 \\
        En (CPX) & 6.1 & 6.2 & 6.5 \\
        Wo (CPX) & 6.8 & 7.6 & 6.4 \\
        \hline
    \end{tabular}
    \tablefoot{Fo and En~(OPX) have the same absolute errors as Fa and Fs~(OPX), respectively.}
\end{table}


\appsection{Confusion matrices of classification models}
\label{app:taxonomical}

\tcfigure[!ht]{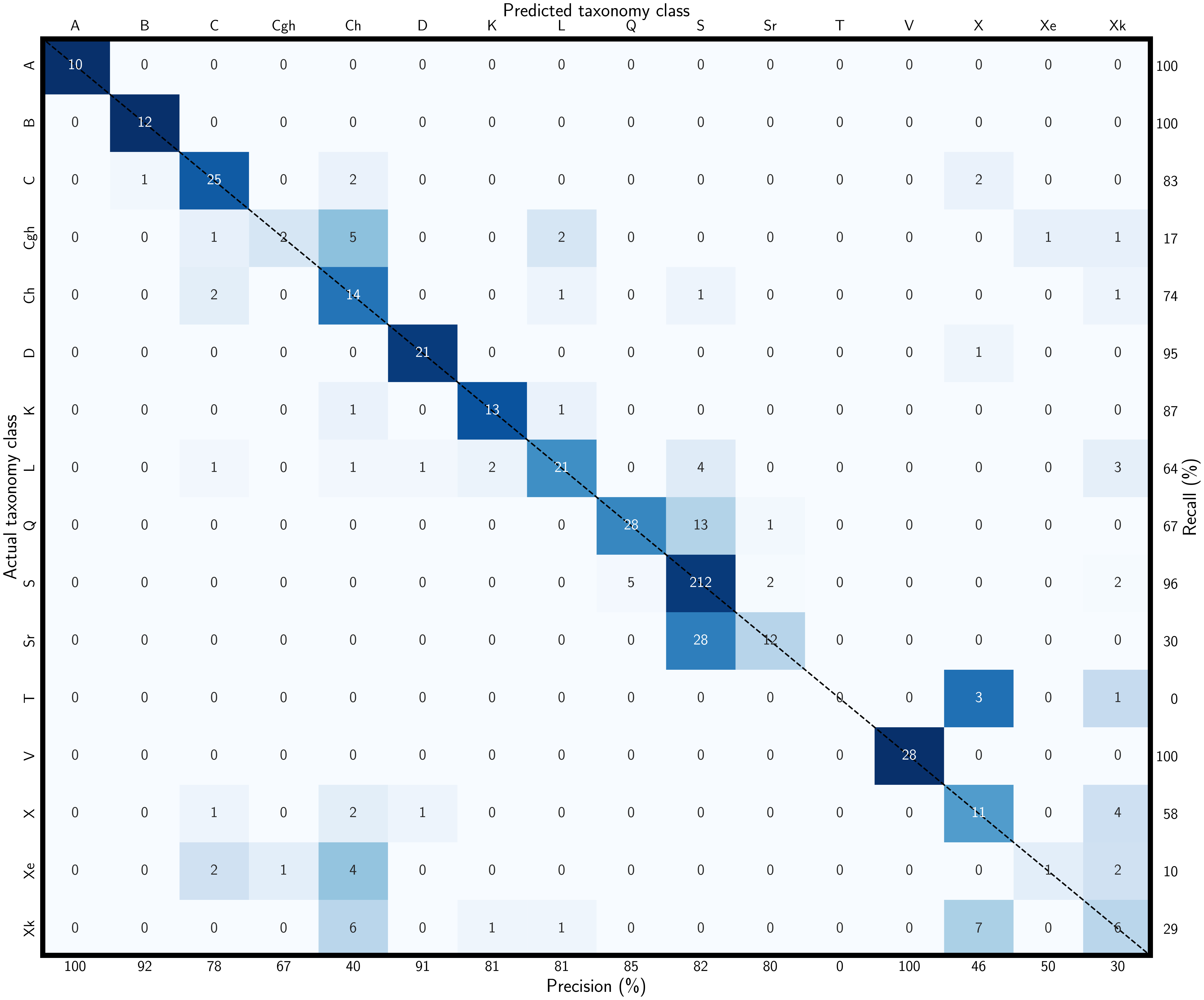}
{Confusion matrix of the Eros-resolution classification model.}
{fig:E_conf_mat}

\tcfigure[!ht]{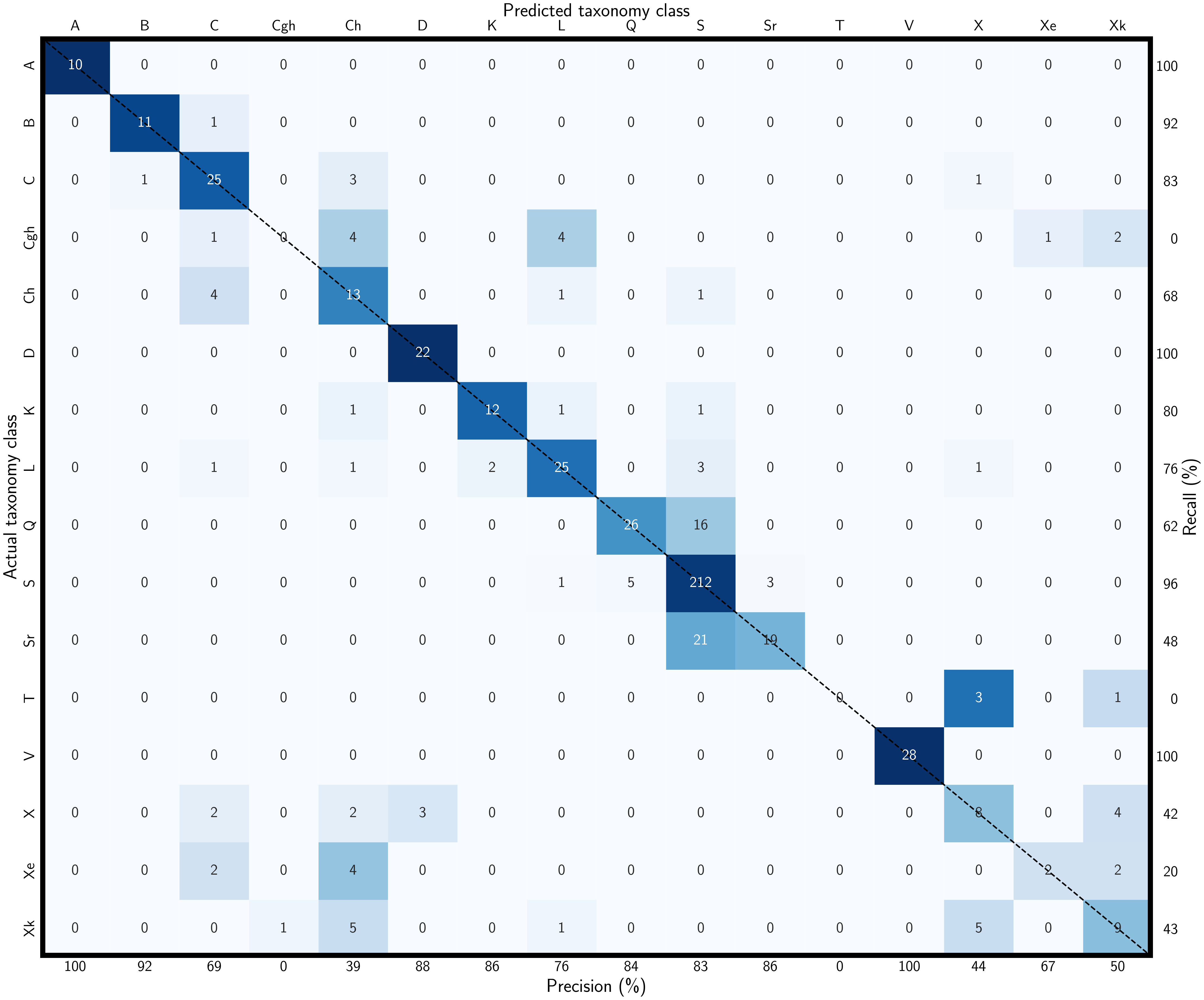}
{Confusion matrix of the Itokawa-resolution classification model.}
{fig:I_conf_mat}


\appsection{Scatter plots of composition models}
\label{app:compositional}

\wfigure[!ht]{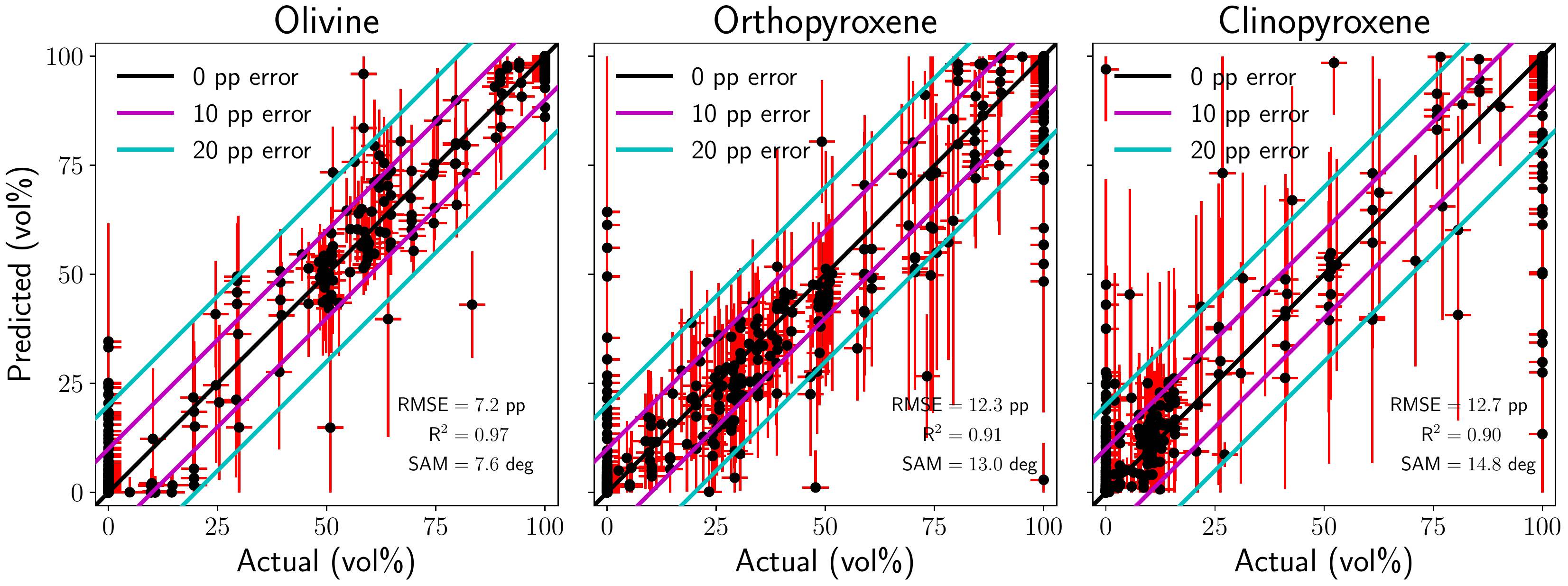}
{Comparison of the true and predicted (by the Eros-resolution model) modal abundances.}
{fig:E_mineral}

\begin{figure*}[!ht]
    \centering
    \begin{minipage}[b]{.48\textwidth}
        \resizebox{\hsize}{!}{\includegraphics{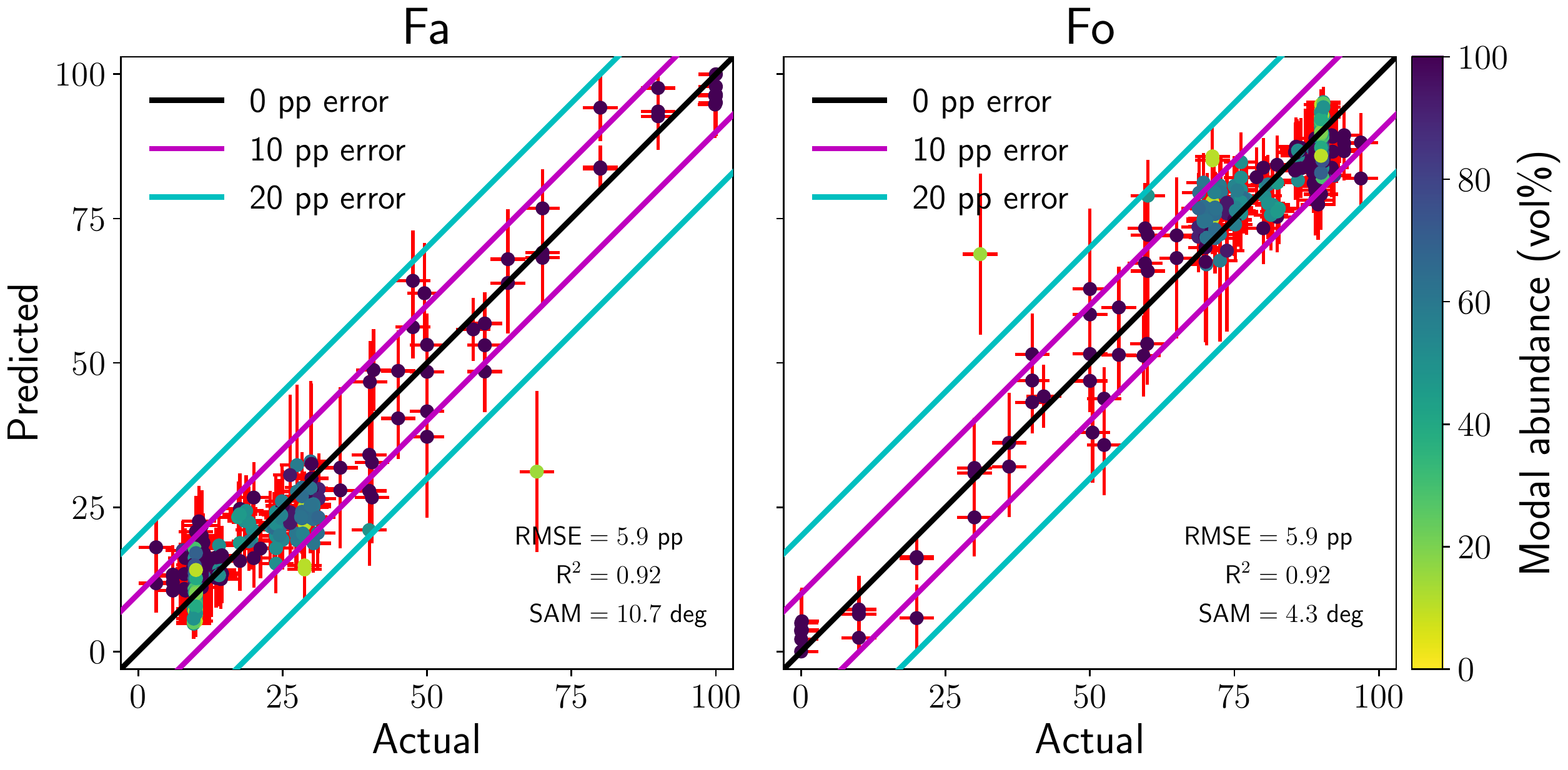}}
        \caption{Comparison of the true and predicted (by the Eros-resolution model) olivine composition. \textit{Left}: Iron content. \textit{Right}: Magnesium content. See the caption of Fig.~\ref{fig:full_ol} for details.}
        \label{fig:E_ol}
    \end{minipage}
    \hfill
    \begin{minipage}[b]{.48\textwidth}
        \resizebox{\hsize}{!}{\includegraphics{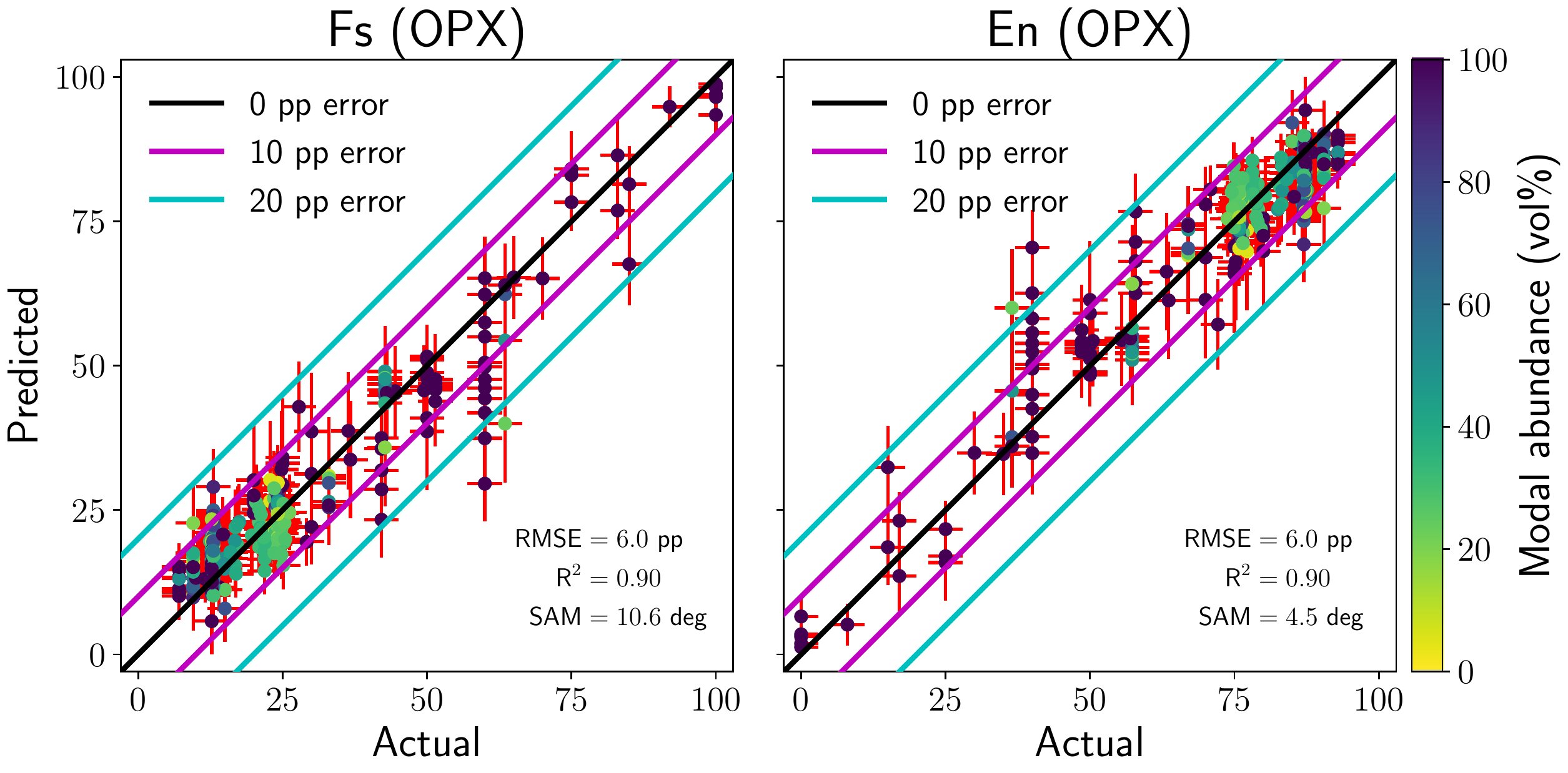}}
        \caption{Comparison of the true and predicted (by the Eros-resolution model) orthopyroxene composition. \textit{Left}: Iron content. \textit{Right}: Magnesium content. See the caption of Fig.~\ref{fig:full_opx} for details.}
        \label{fig:E_opx}
    \end{minipage}
\end{figure*}

\wfigure[!ht]{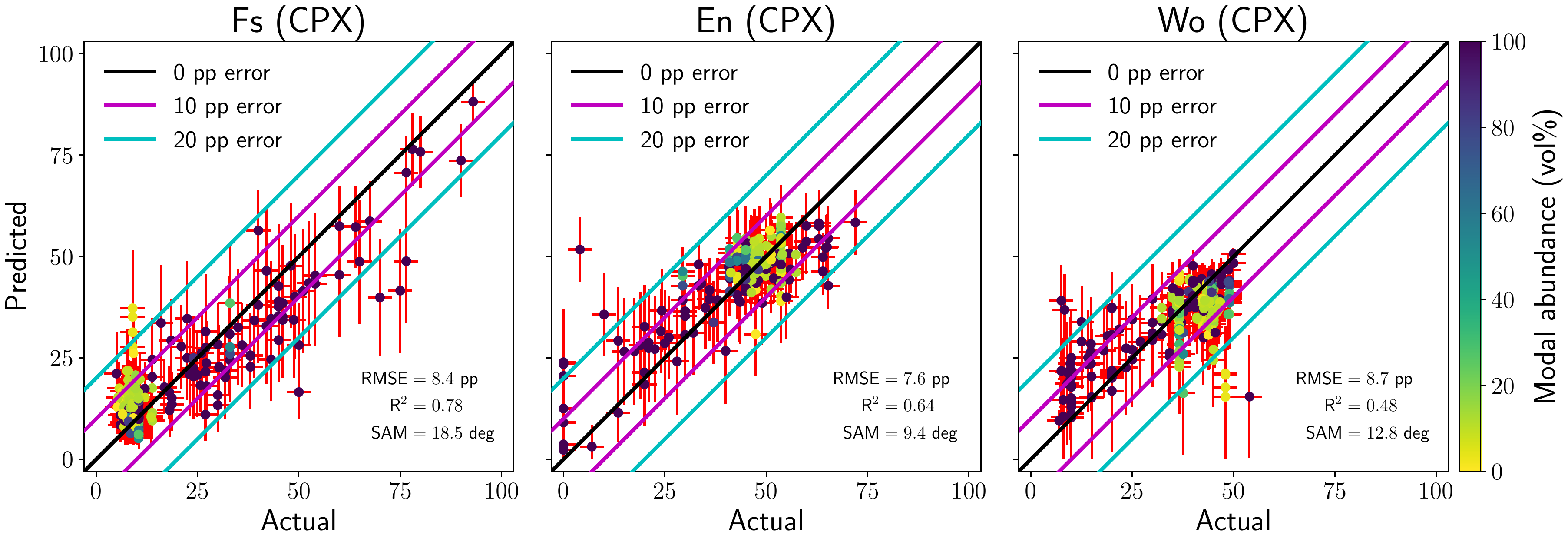}
{Comparison of the true and predicted (by the Eros-resolution model) clinopyroxene composition. \textit{Left}: Iron content. \textit{Middle}: Magnesium content. \textit{Right}: Calcium content. See the caption of Fig.~\ref{fig:full_cpx} for details.}
{fig:E_cpx}


\wfigure[!ht]{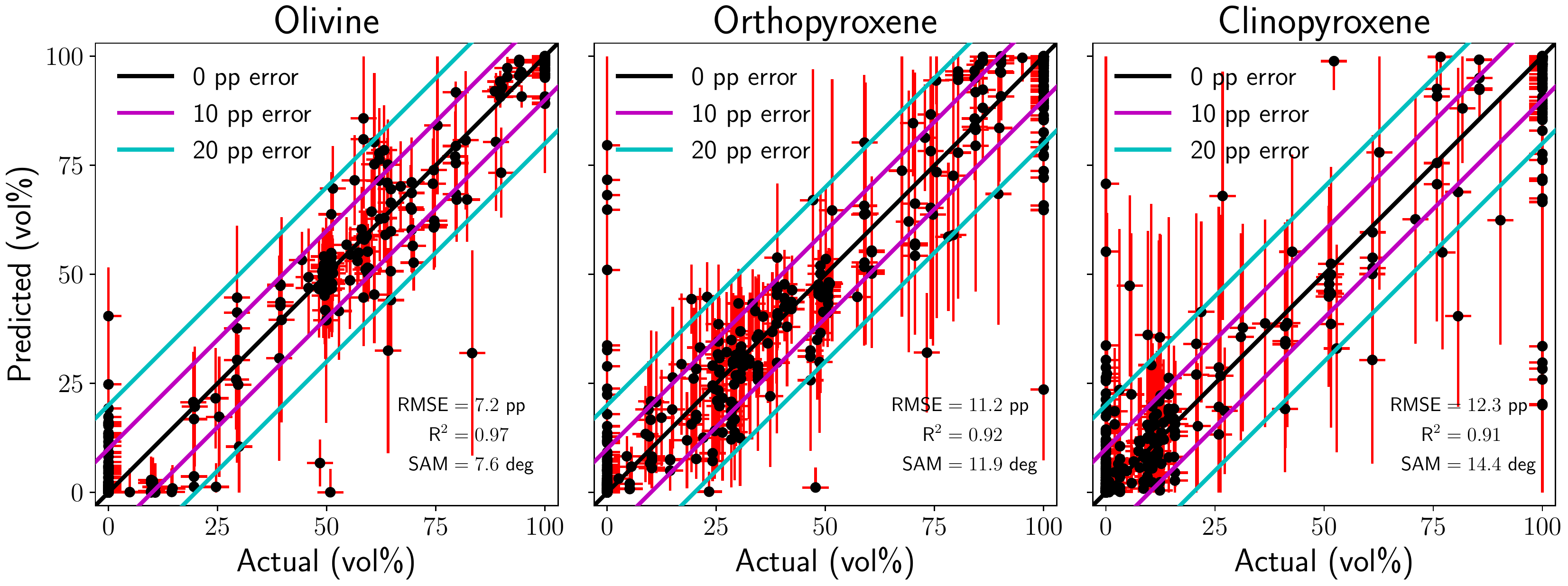}
{Comparison of the true and predicted (by the Itokawa-resolution model) modal abundances.}
{fig:I_mineral}

\begin{figure*}[!ht]
    \centering
    \begin{minipage}[b]{.48\textwidth}
        \resizebox{\hsize}{!}{\includegraphics{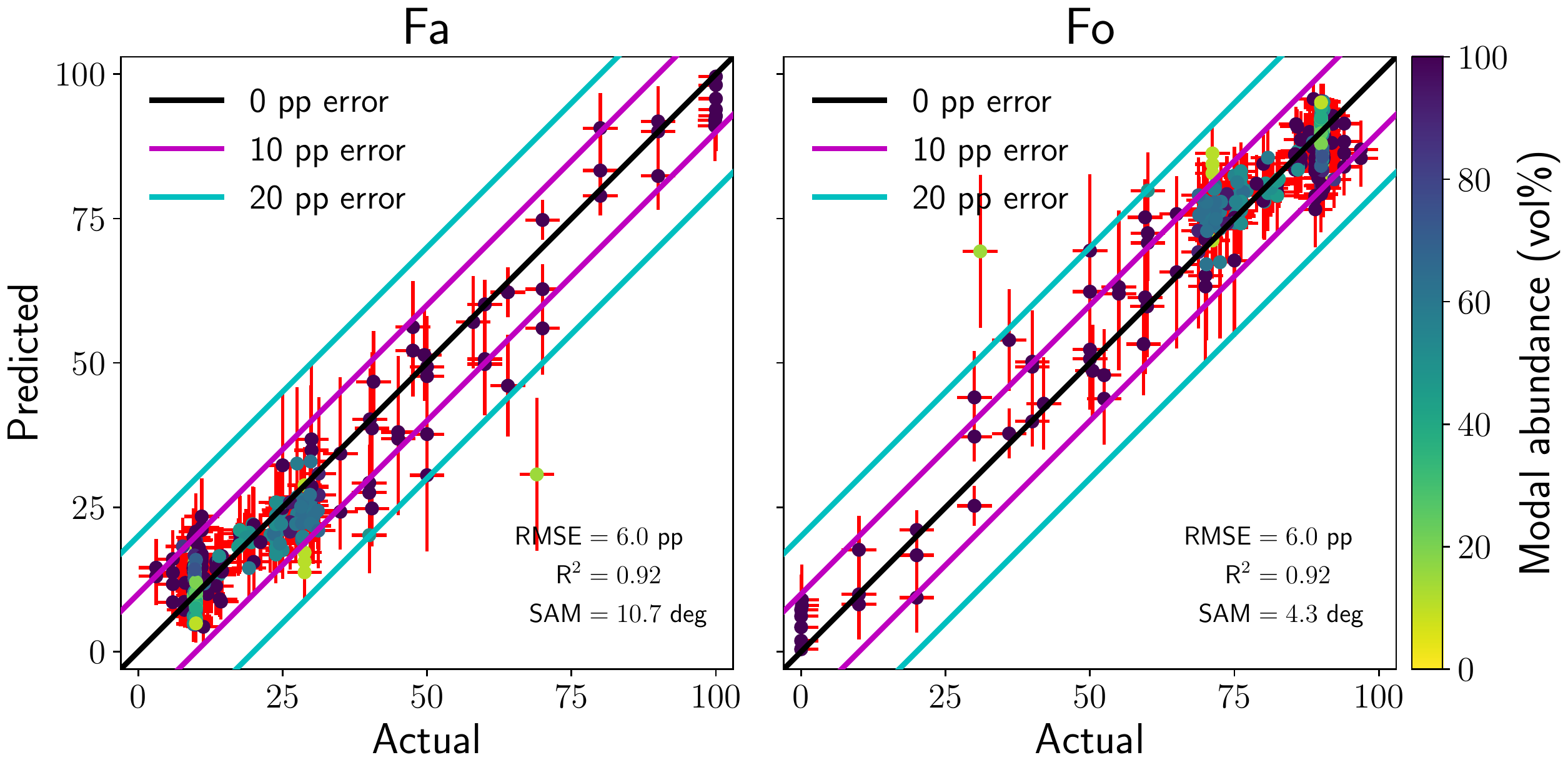}}
        \caption{Comparison of the true and predicted (by the Itokawa-resolution model) olivine composition. \textit{Left}: Iron content. \textit{Right}: Magnesium content. See the caption of Fig.~\ref{fig:full_ol} for details.}
        \label{fig:I_ol}
    \end{minipage}
    \hfill
    \begin{minipage}[b]{.48\textwidth}
        \resizebox{\hsize}{!}{\includegraphics{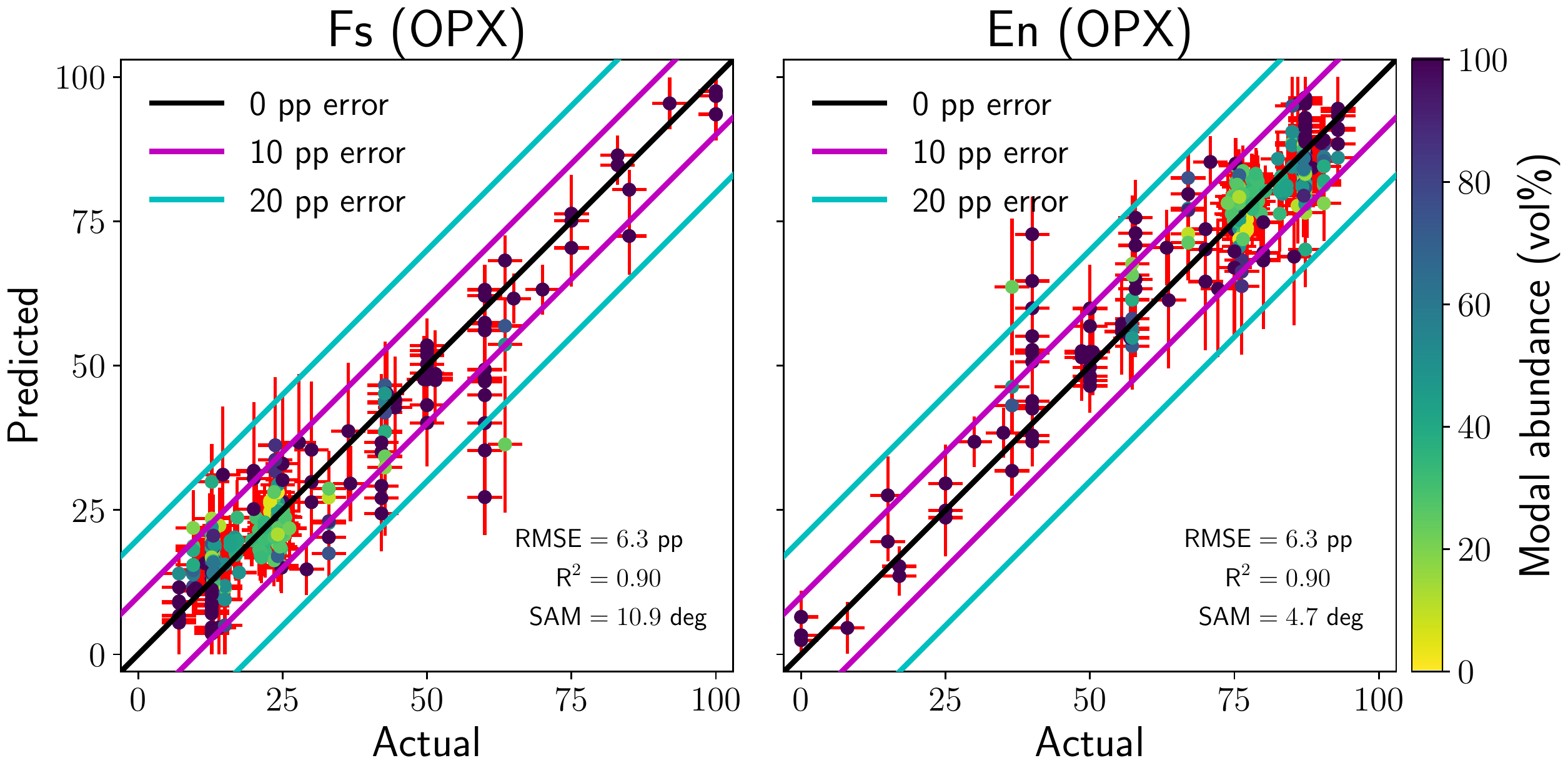}}
        \caption{Comparison of the true and predicted (by the Itokawa-resolution model) orthopyroxene composition. \textit{Left}: Iron content. \textit{Right}: Magnesium content. See the caption of Fig.~\ref{fig:full_opx} for details.}
        \label{fig:I_opx}
    \end{minipage}
\end{figure*}

\wfigure[!ht]{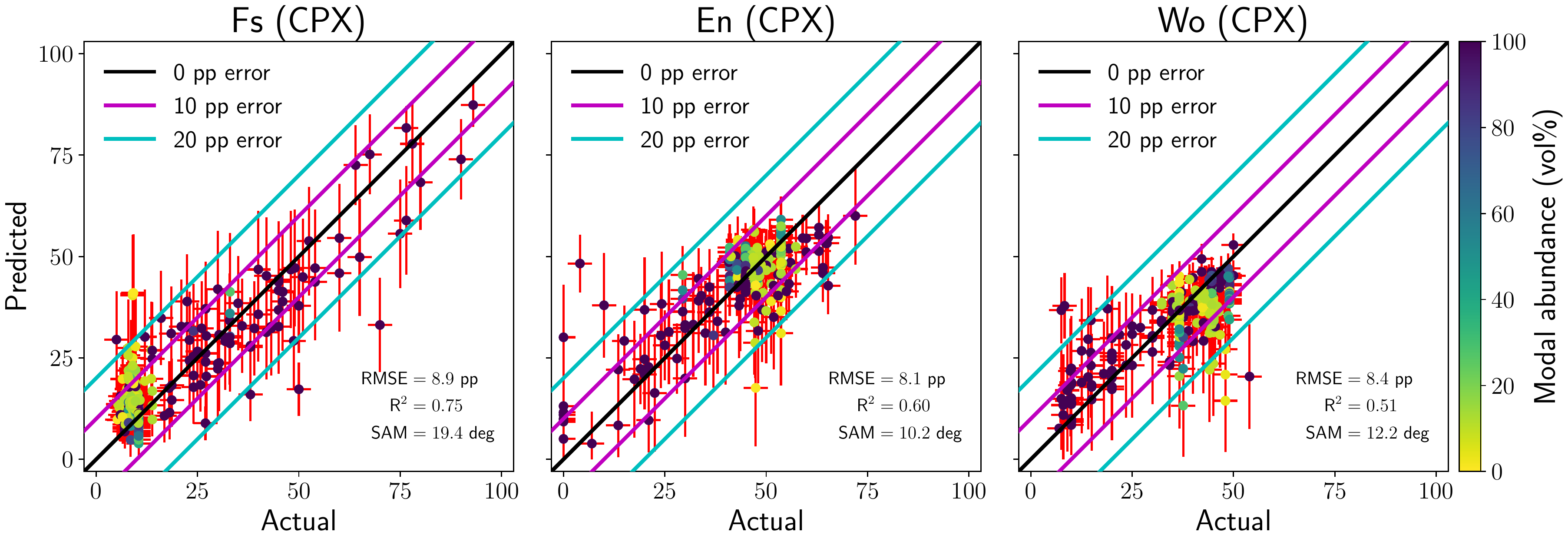}
{Comparison of the true and predicted (by the Itokawa-resolution model) clinopyroxene composition. \textit{Left}: iron content. \textit{Middle}: magnesium content. \textit{Right}: calcium content. See the caption of Fig.~\ref{fig:full_cpx} for details.}
{fig:I_cpx}

\pagebreak
\section{Error plots of classification and composition models}
\label{app:error_plots}

\begin{figure*}[!ht]
    \centering
    \resizebox{0.48\textwidth}{!}{\includegraphics{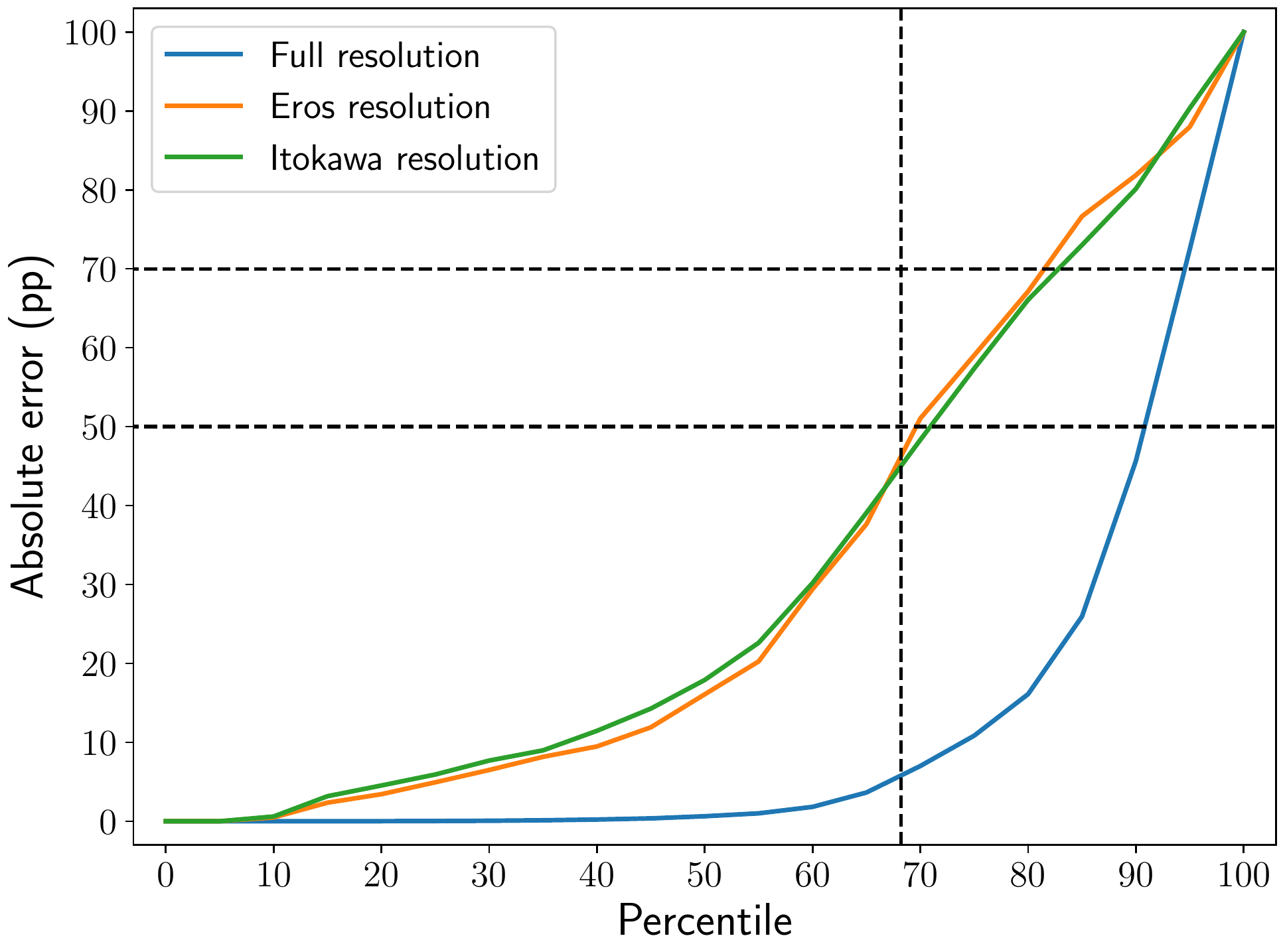}}
    \hfill
    \resizebox{0.48\textwidth}{!}
    {\includegraphics{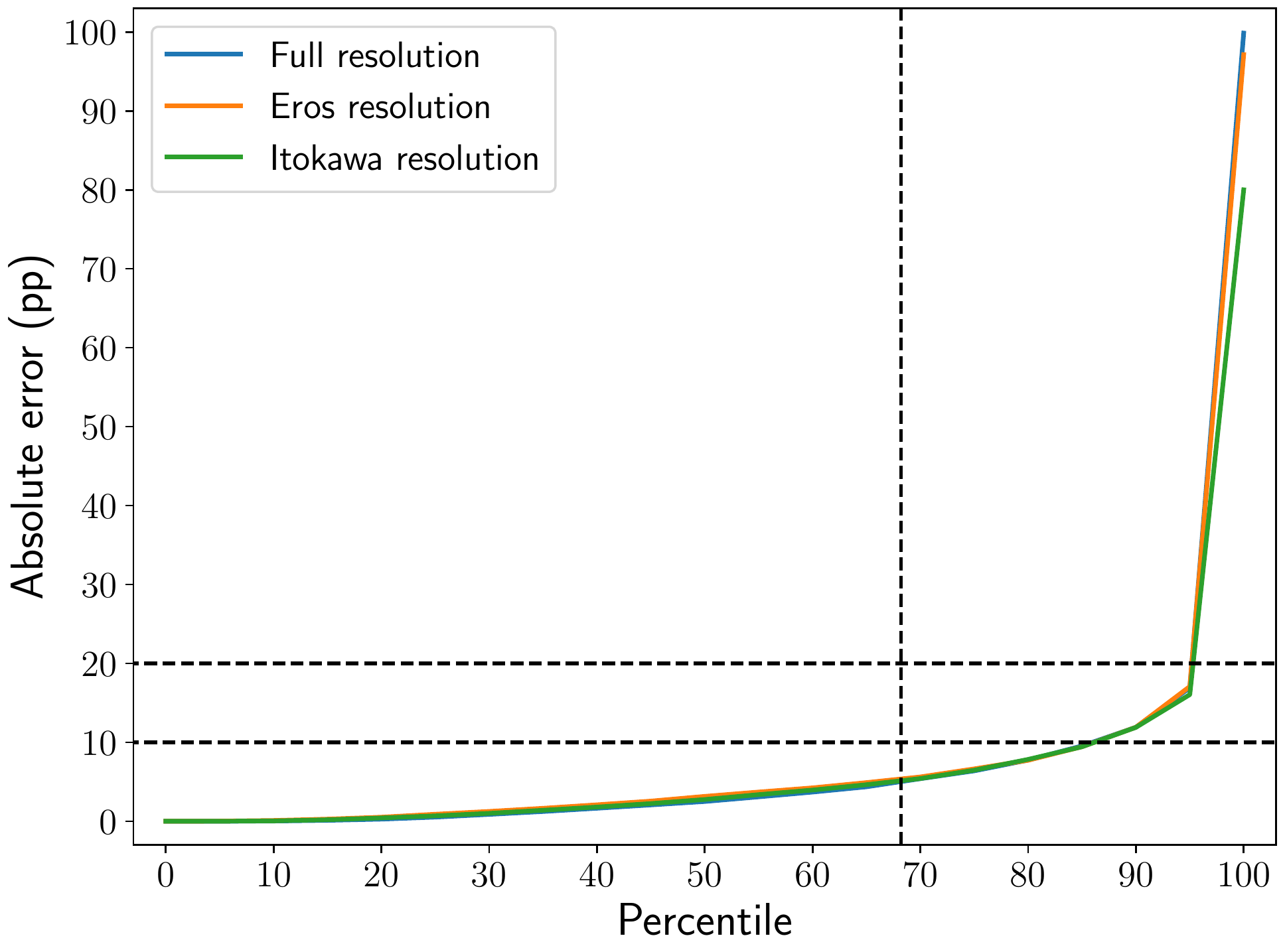}}
    \caption{Quantiles of the absolute errors computed from all predictions. \textit{Left}: classification models. \textit{Right}: composition models. The vertical dashed line delimits the 1-$\sigma$ error estimate. The horizontal dashed black lines indicate acceptable absolute errors.}
    \label{fig:error_all}
\end{figure*}


\appsection{Taxonomy and composition maps of Eros}
\label{app:Eros}

\tcfigure[!ht]{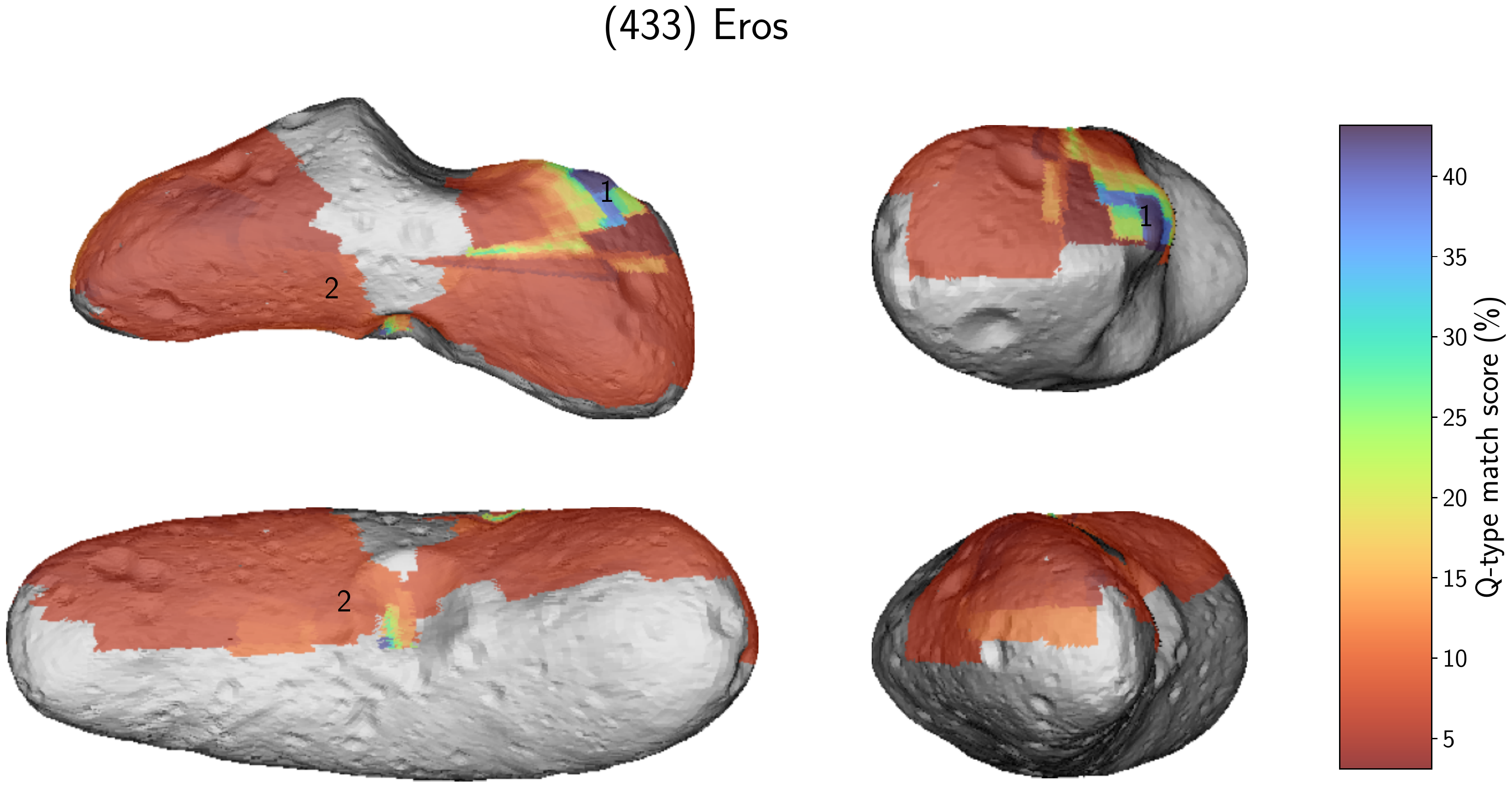}
{Predicted match score of the Q-type asteroids on the surface of Eros. The numbers designate areas discussed in the text.}
{fig:Eros_Q}


\tcfigure[!ht]{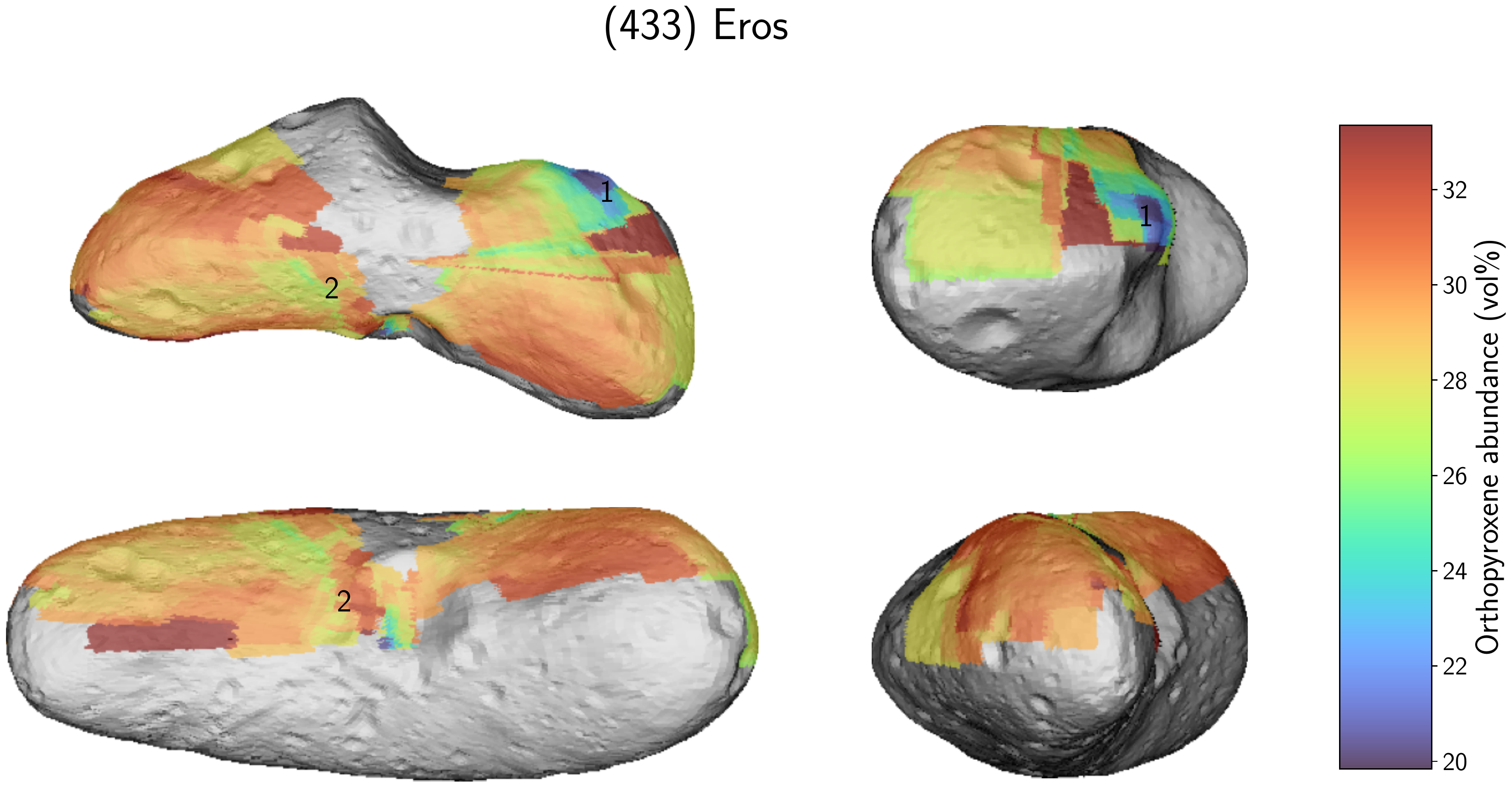}
{Predictions of orthopyroxene abundance on the surface of Eros. The numbers designate areas discussed in the text.}
{fig:Eros_OPX}

\tcfigure[!ht]{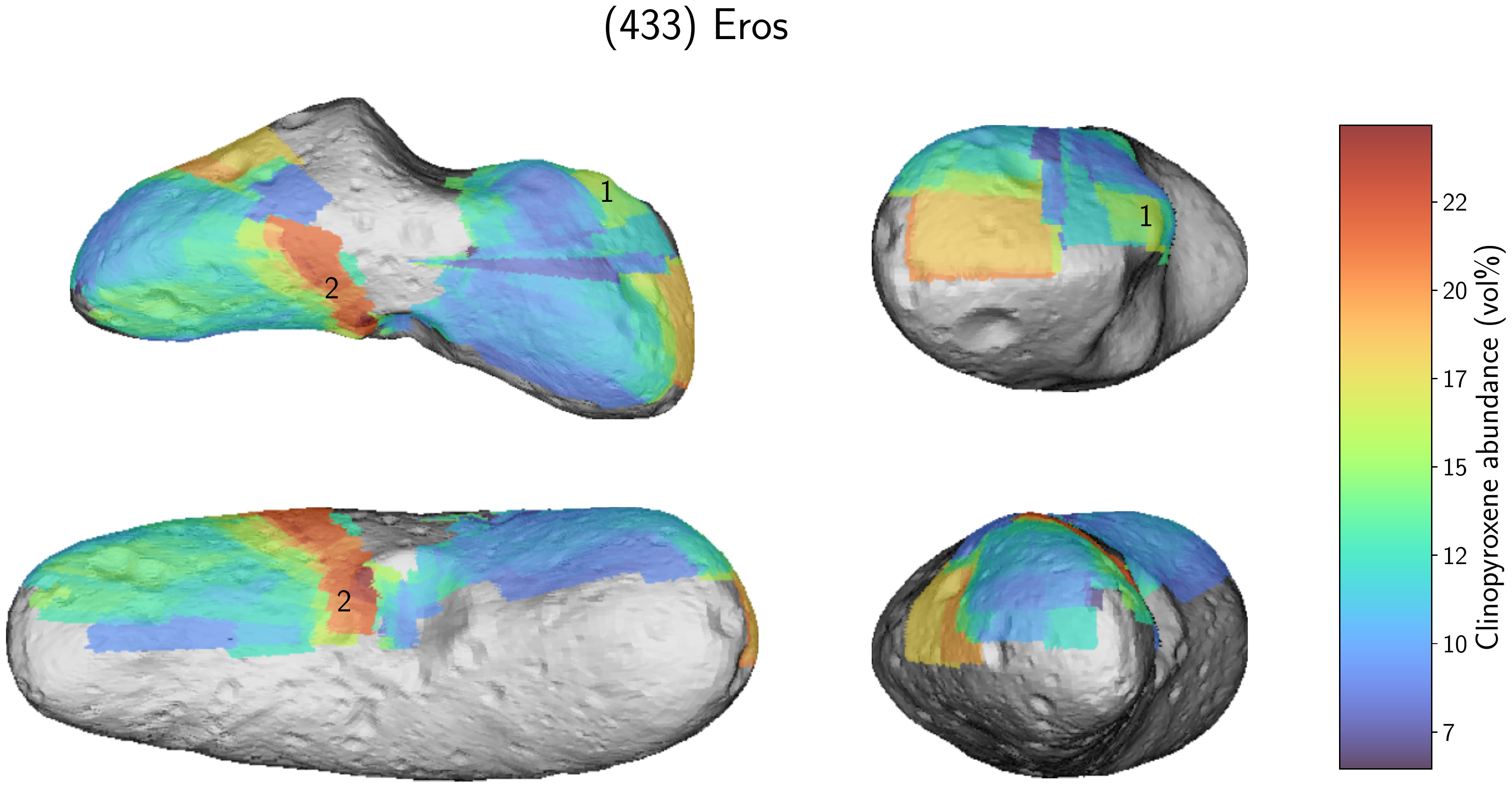}
{Predictions of clinopyroxene abundance on the surface of Eros. The numbers designate areas discussed in the text.}
{fig:Eros_CPX}

\tcfigure[!ht]{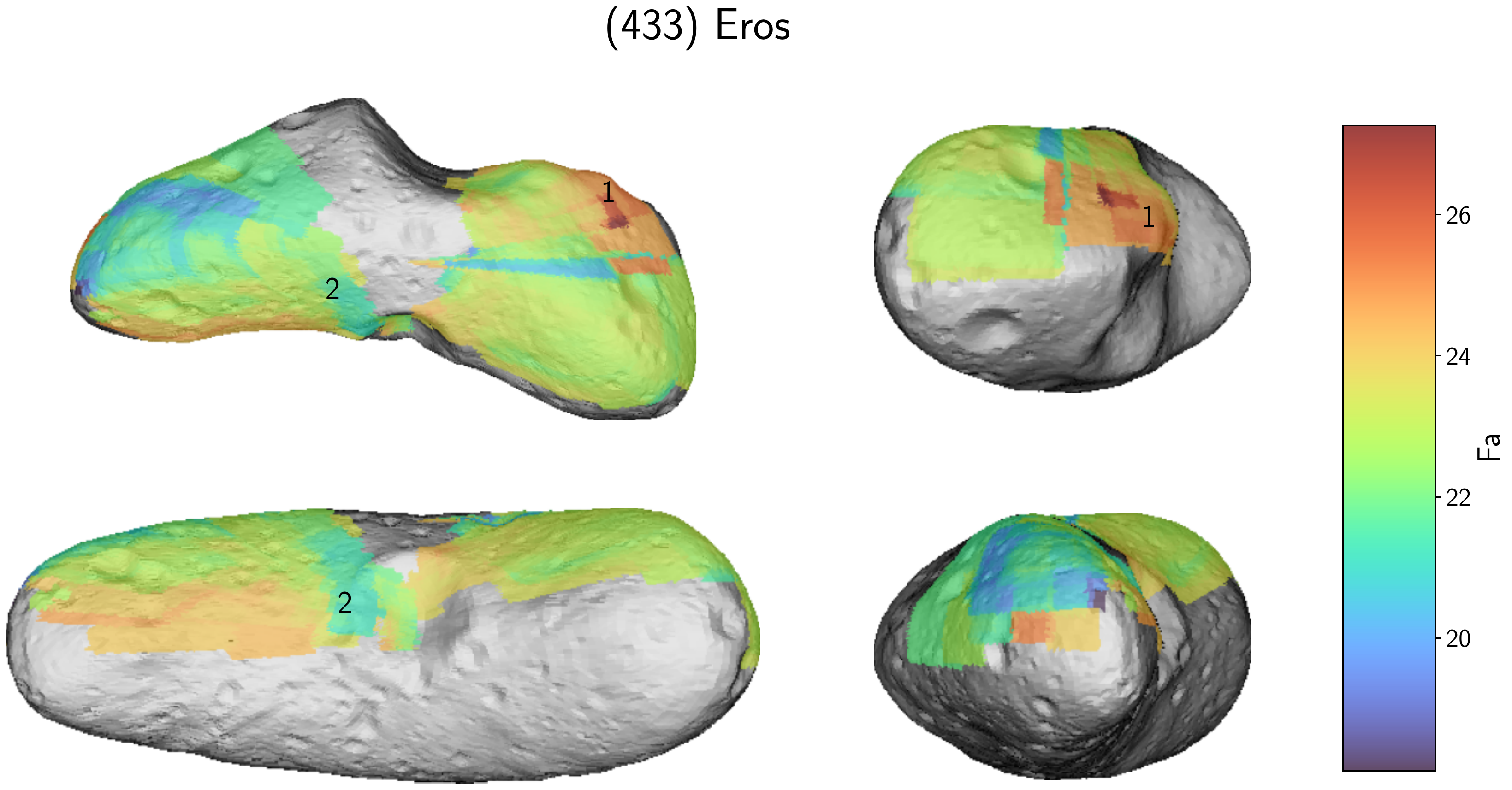}
{Predictions of iron content in olivine on the surface of Eros. The numbers designate areas discussed in the text.}
{fig:Eros_Fa}

\tcfigure[!ht]{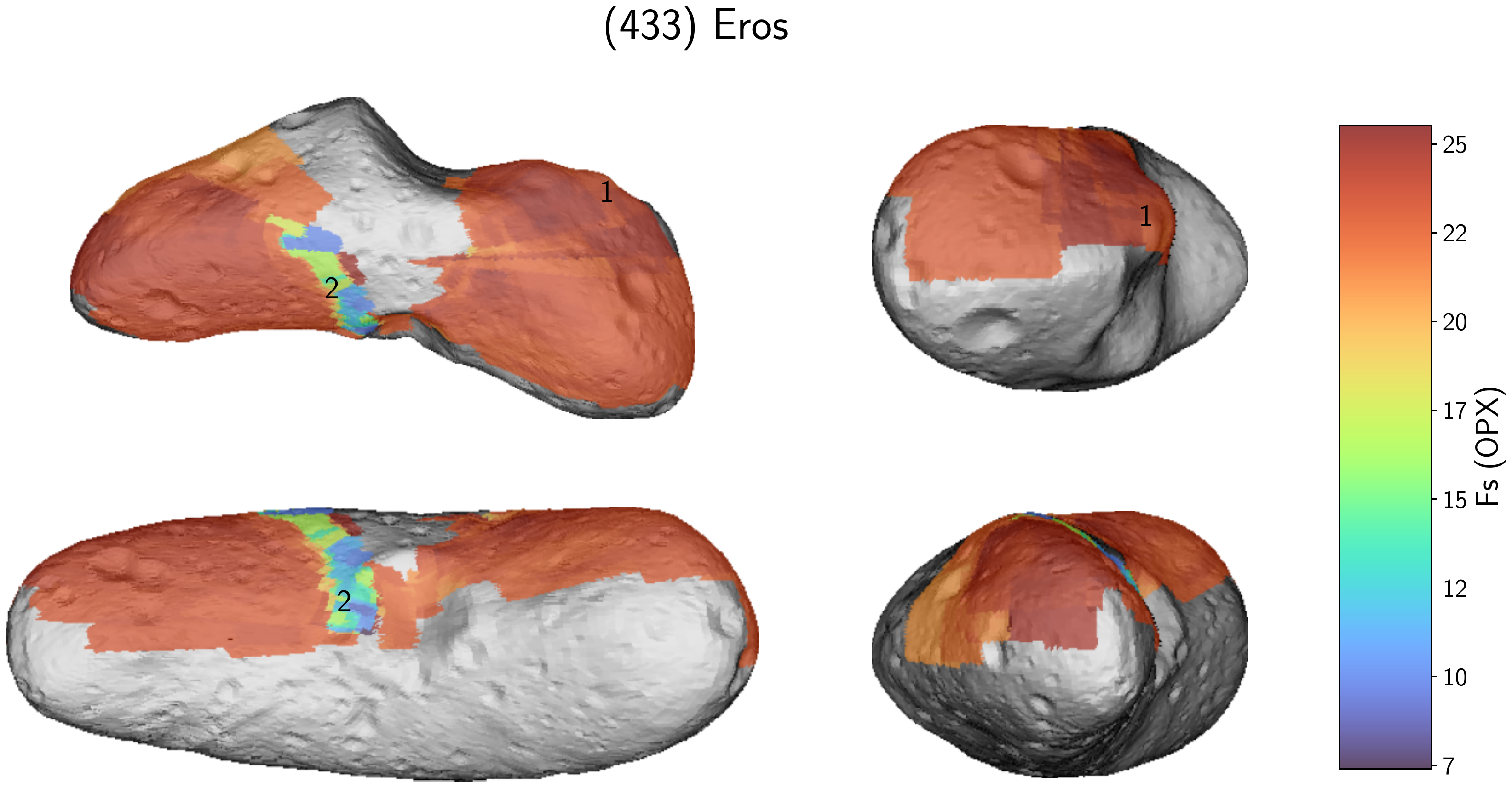}
{Predictions of iron content in orthopyroxene on the surface of Eros. The numbers designate areas discussed in the text.}
{fig:Eros_Fs_OPX}

\tcfigure[!ht]{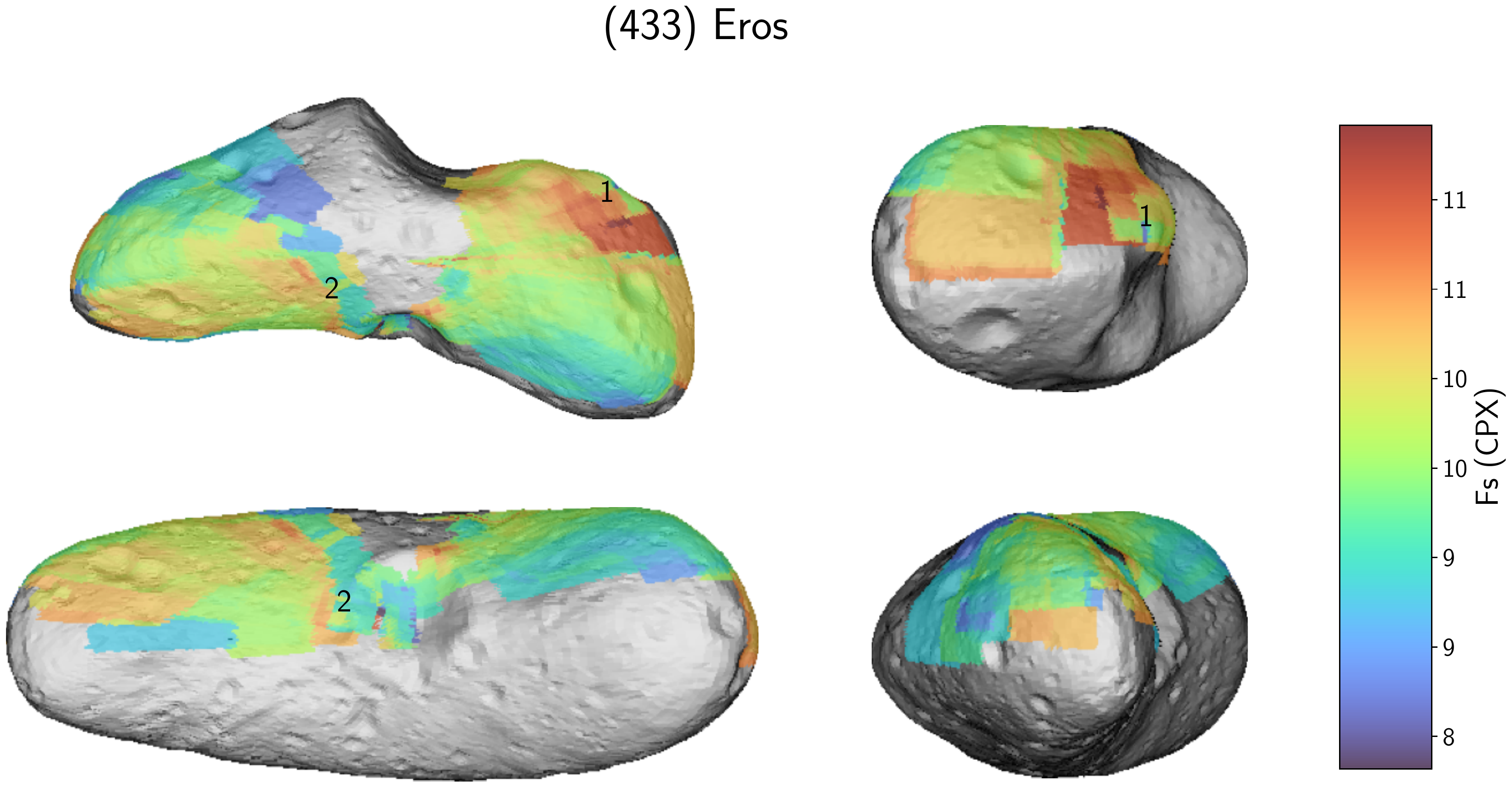}
{Predictions of iron content in clinopyroxene on the surface of Eros. The numbers designate areas discussed in the text.}
{fig:Eros_Fs_CPX}

\tcfigure[!ht]{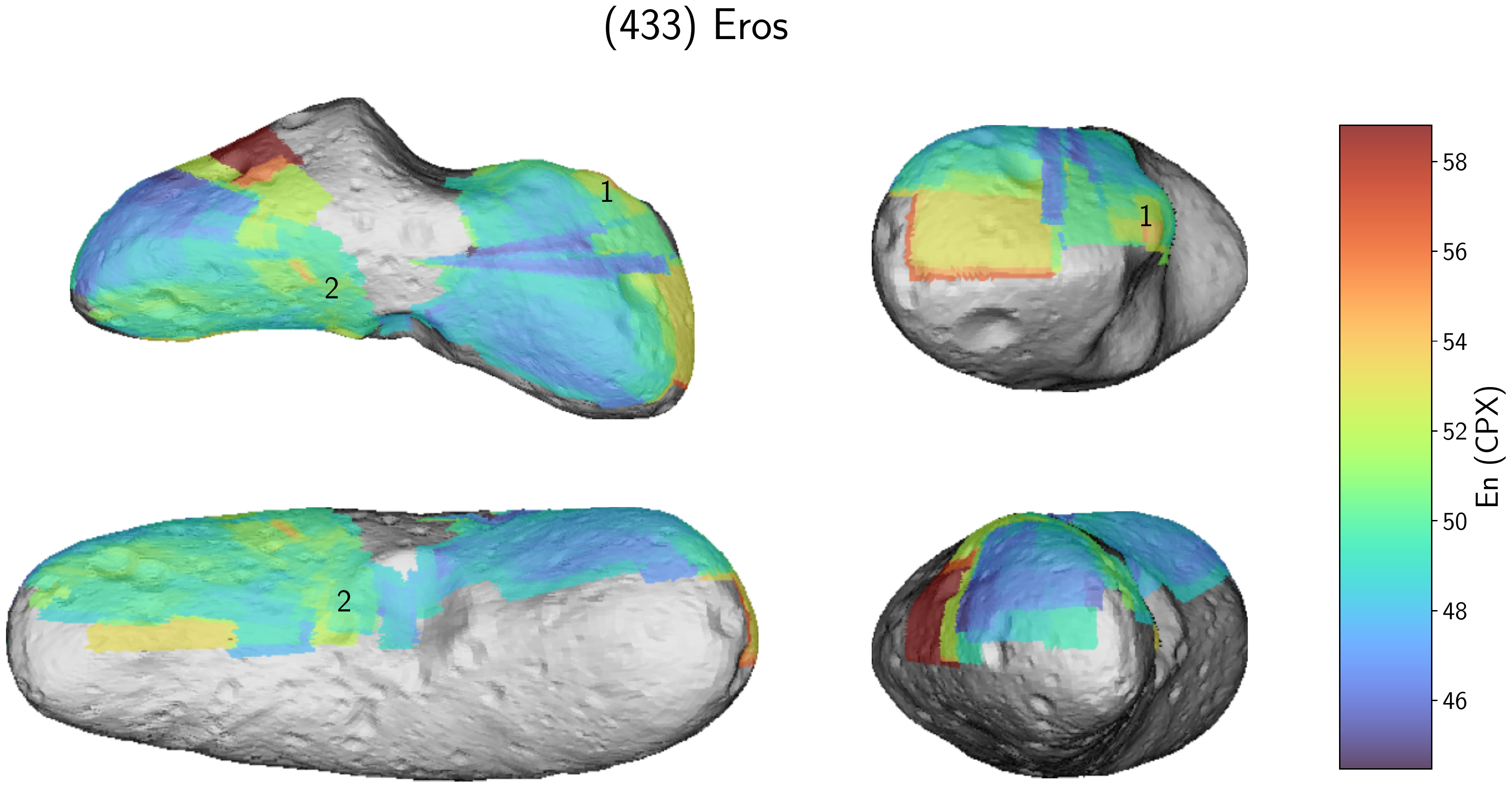}
{Predictions of magnesium content in clinopyroxene on the surface of Eros. The numbers designate areas discussed in the text.}
{fig:Eros_En_CPX}

\tcfigure[!ht]{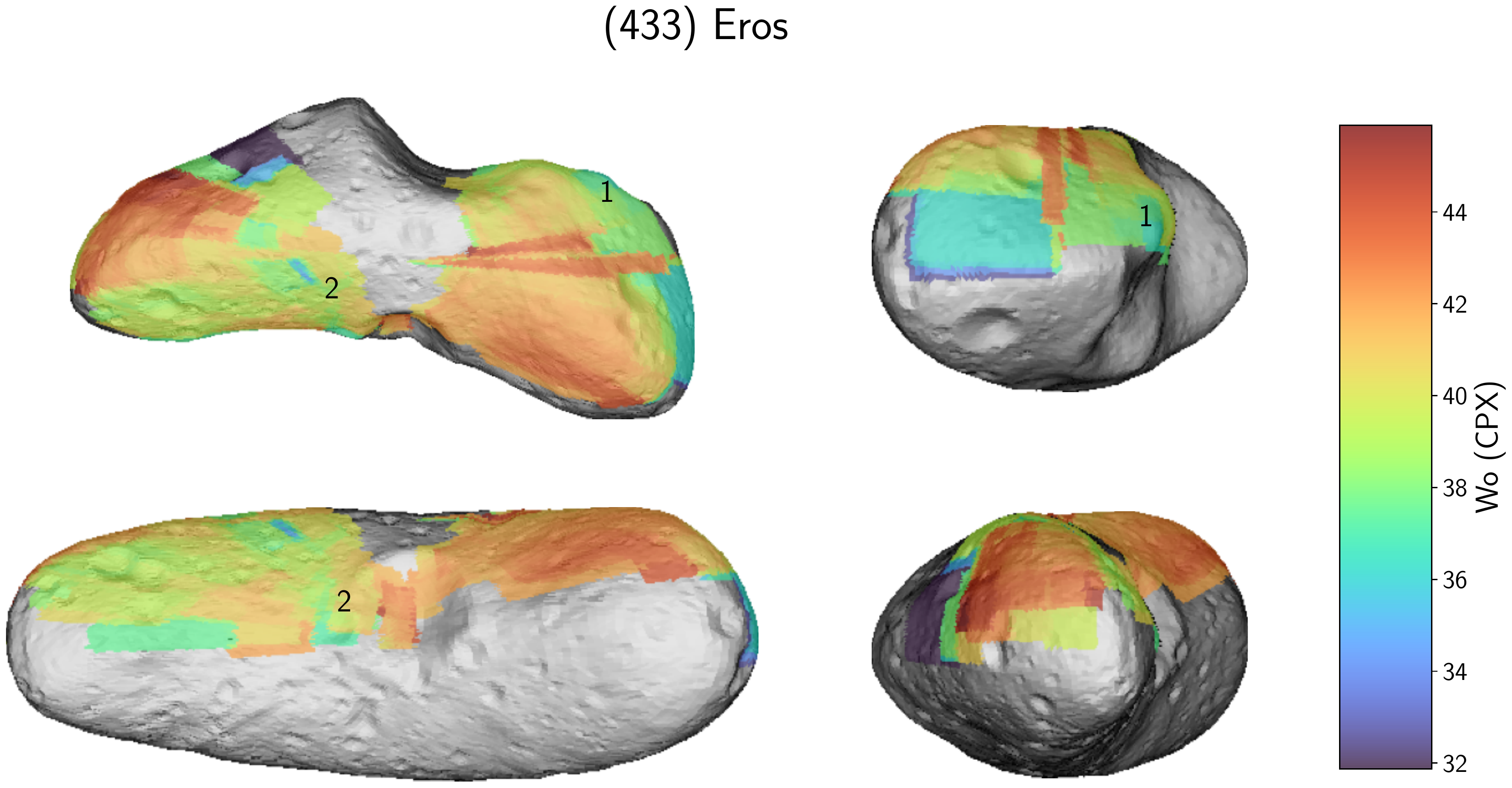}
{Predictions of calcium content in clinopyroxene on the surface of Eros. The numbers designate areas discussed in the text.}
{fig:Eros_Wo_CPX}


\appsection{Taxonomy and composition maps of Itokawa}
\label{app:Itokawa}

\tcfigure[!ht]{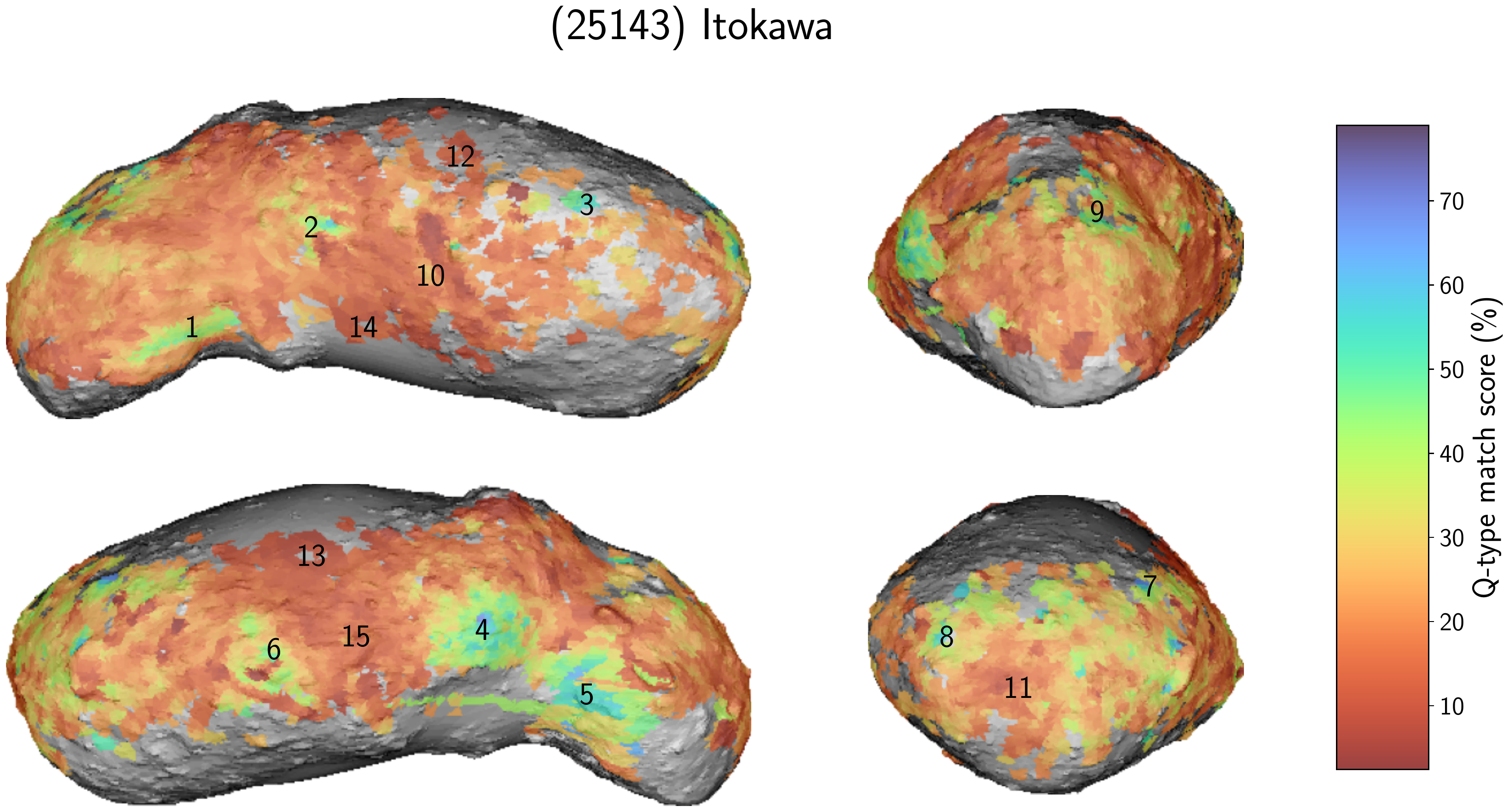}
{Predicted match score of the Q-type asteroids on the surface of Itokawa. The numbers designate fresh and mature areas.}
{fig:Itokawa_Q}


\tcfigure[!ht]{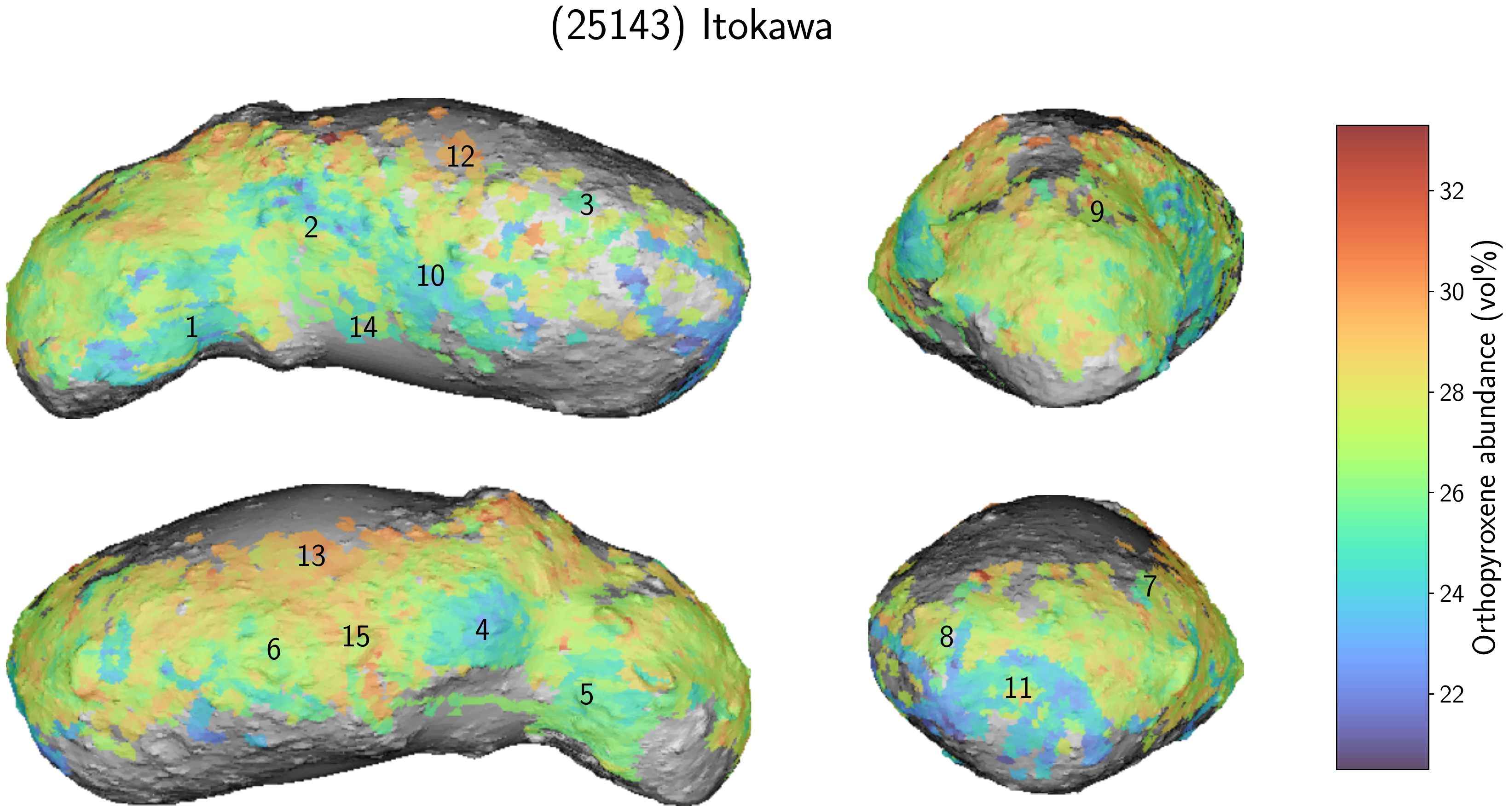}
{Predictions of orthopyroxene abundance on the surface of Itokawa. The numbers designate fresh and mature areas.}
{fig:Itokawa_OPX}

\tcfigure[!ht]{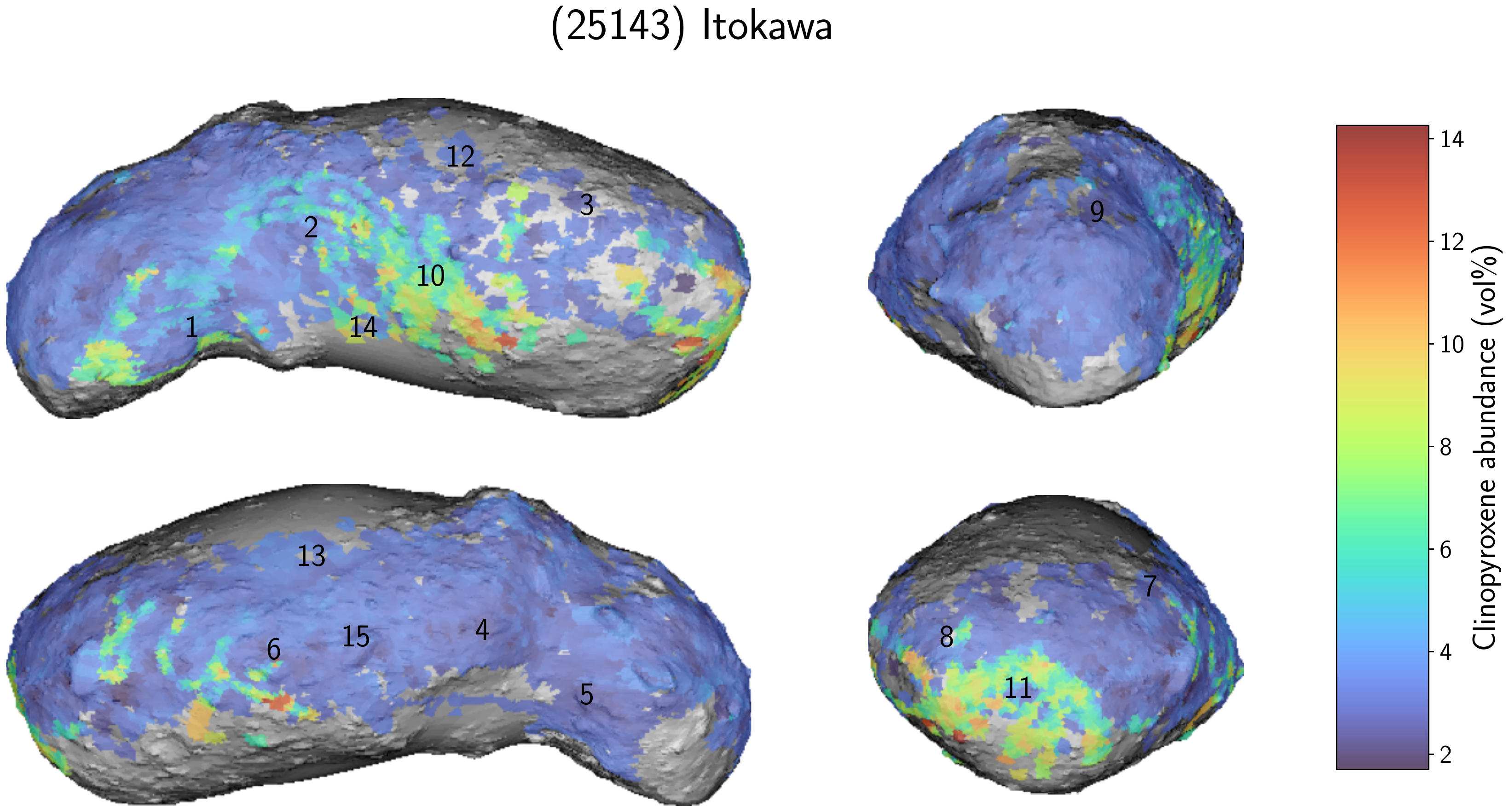}
{Predictions of clinopyroxene abundance on the surface of Itokawa. The numbers designate fresh and mature areas.}
{fig:Itokawa_CPX}

\tcfigure[!ht]{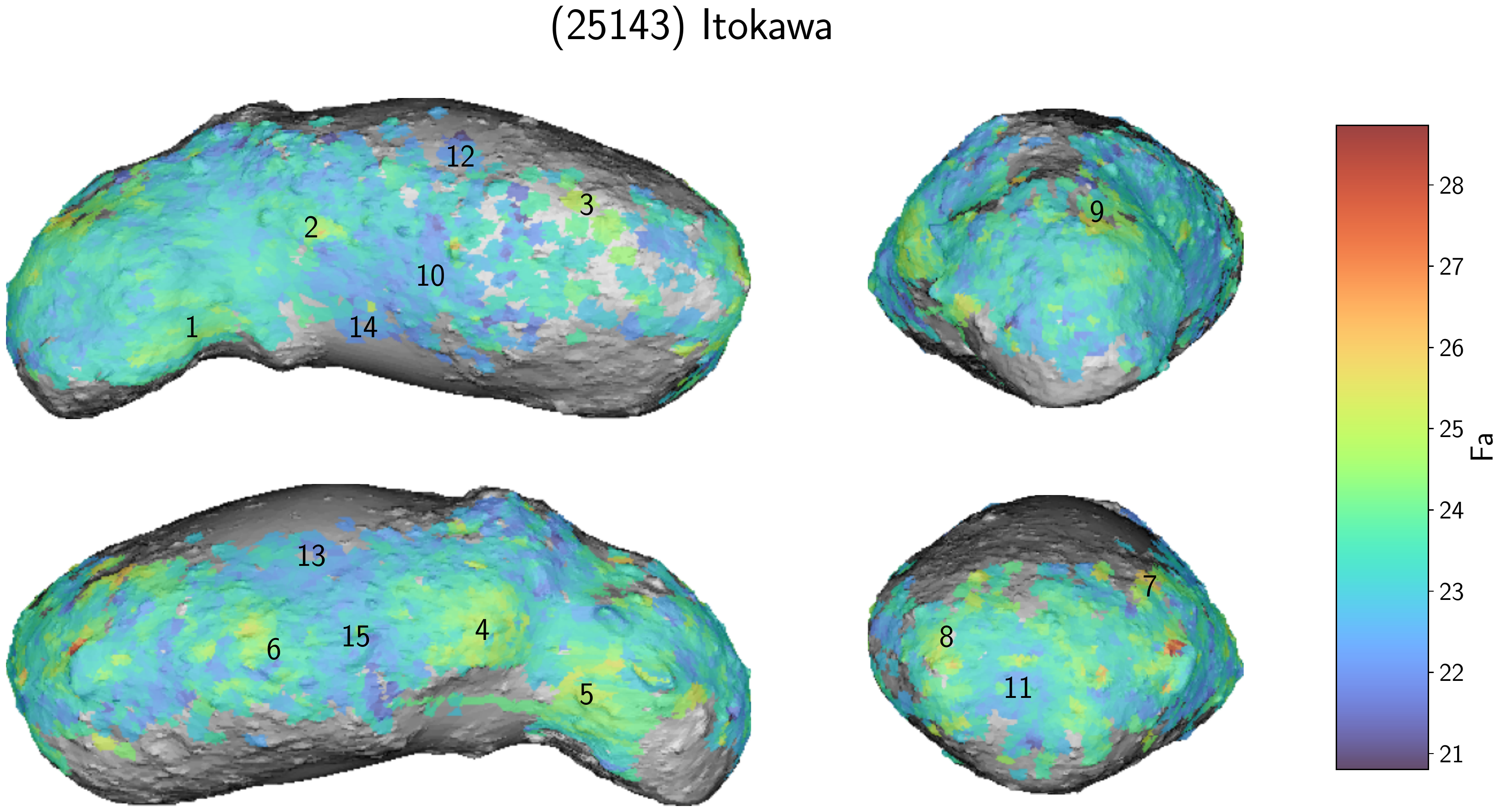}
{Predictions of iron content in olivine on the surface of Itokawa. The numbers designate fresh and mature areas.}
{fig:Itokawa_Fa}

\tcfigure[!ht]{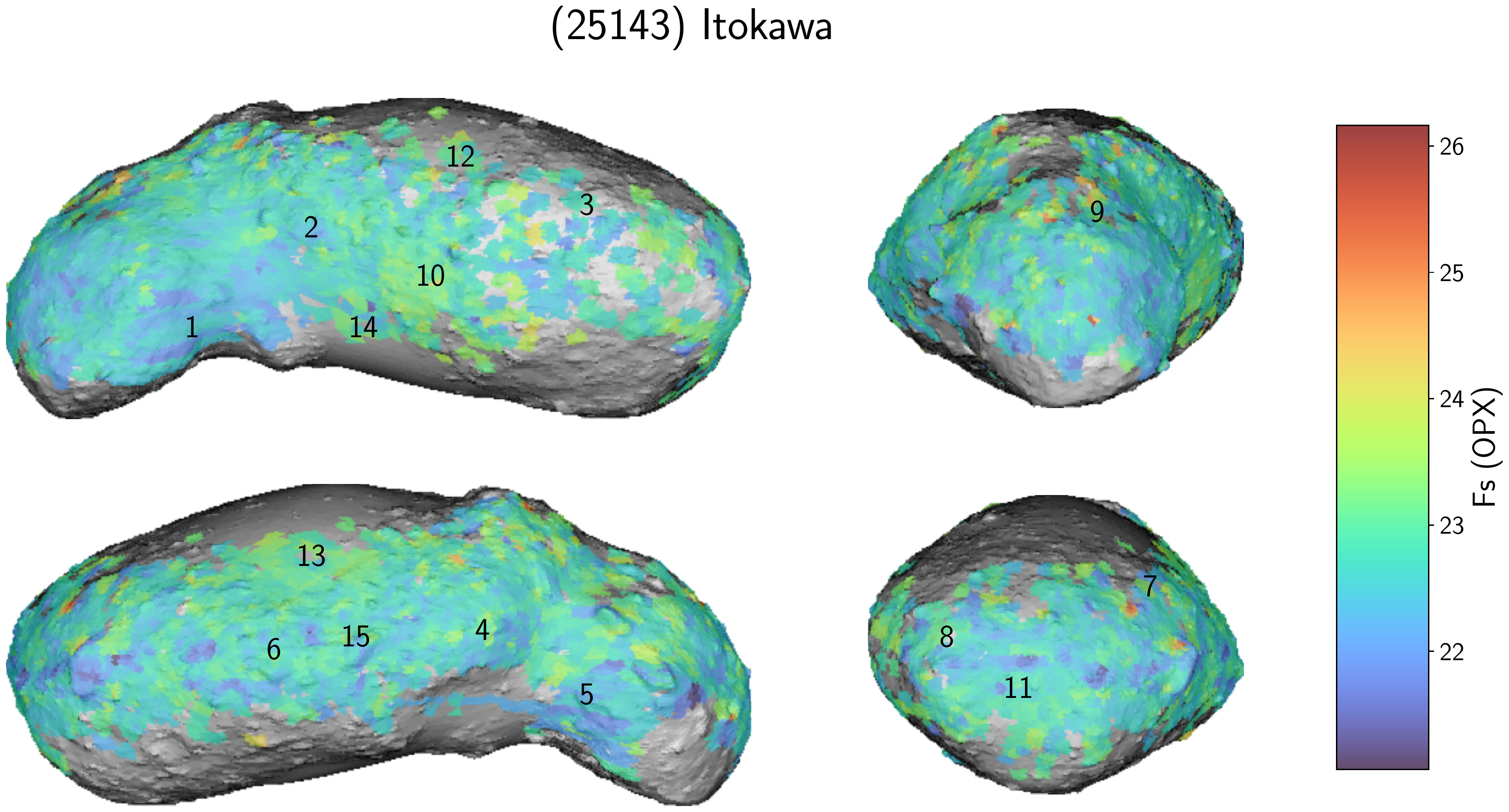}
{Predictions of iron content in orthopyroxene on the surface of Itokawa. The numbers designate fresh and mature areas.}
{fig:Itokawa_Fs_OPX}

\tcfigure[!ht]{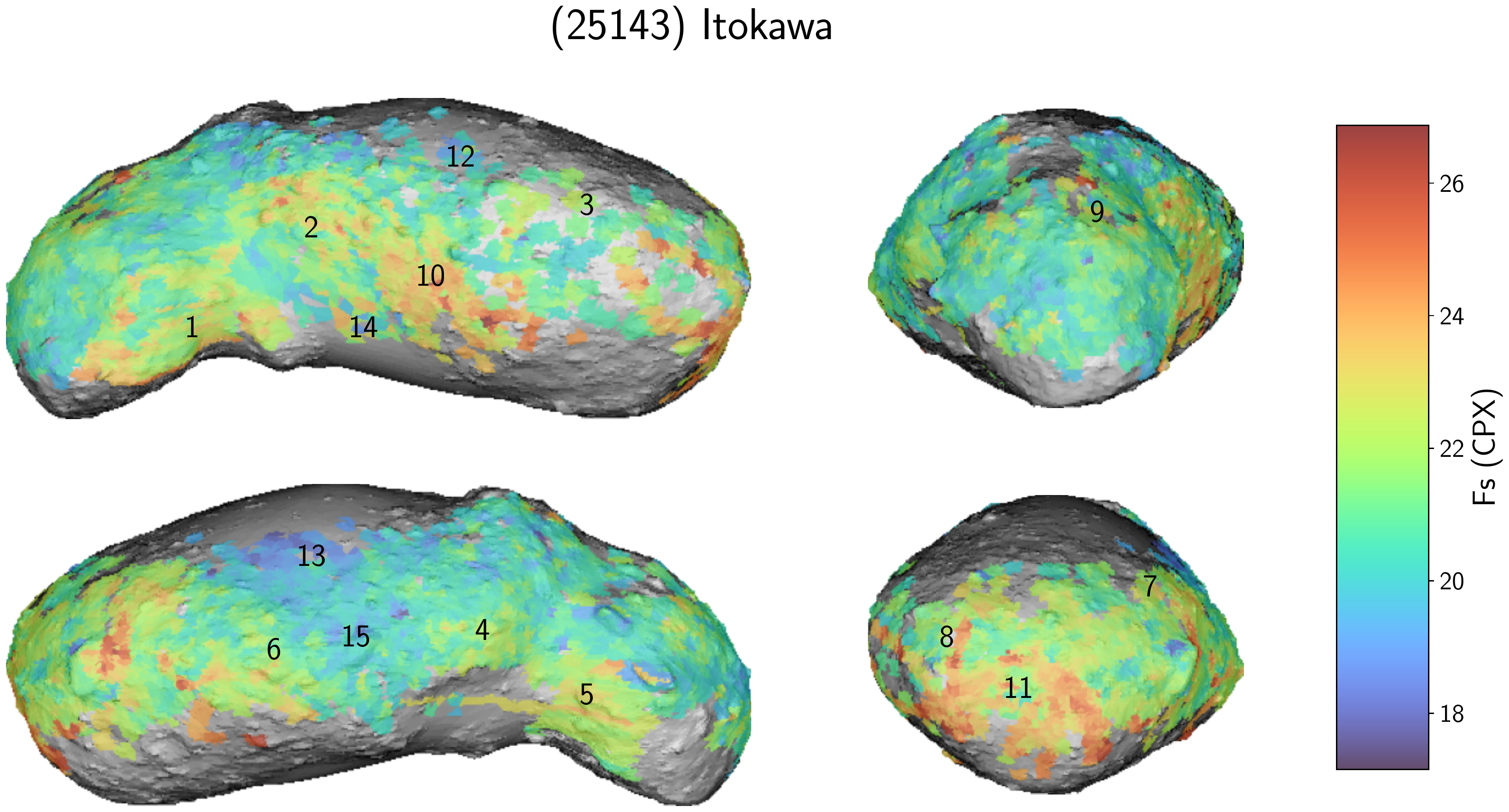}
{Predictions of iron content in clinopyroxene on the surface of Itokawa. The numbers designate fresh and mature areas.}
{fig:Itokawa_Fs_CPX}

\tcfigure[!ht]{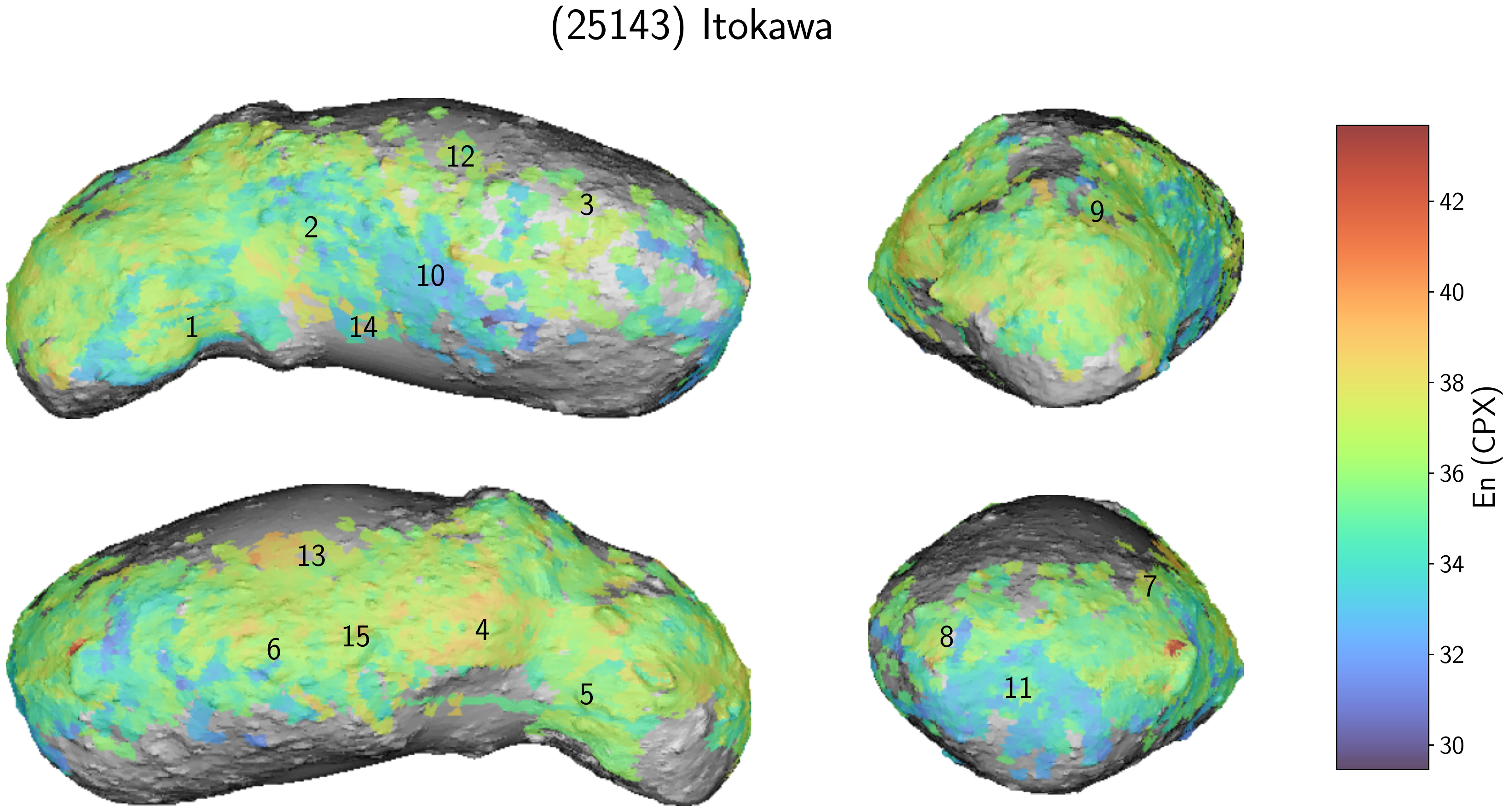}
{Predictions of magnesium content in clinopyroxene on the surface of Itokawa. The numbers designate fresh and mature areas.}
{fig:Itokawa_En_CPX}

\tcfigure[!ht]{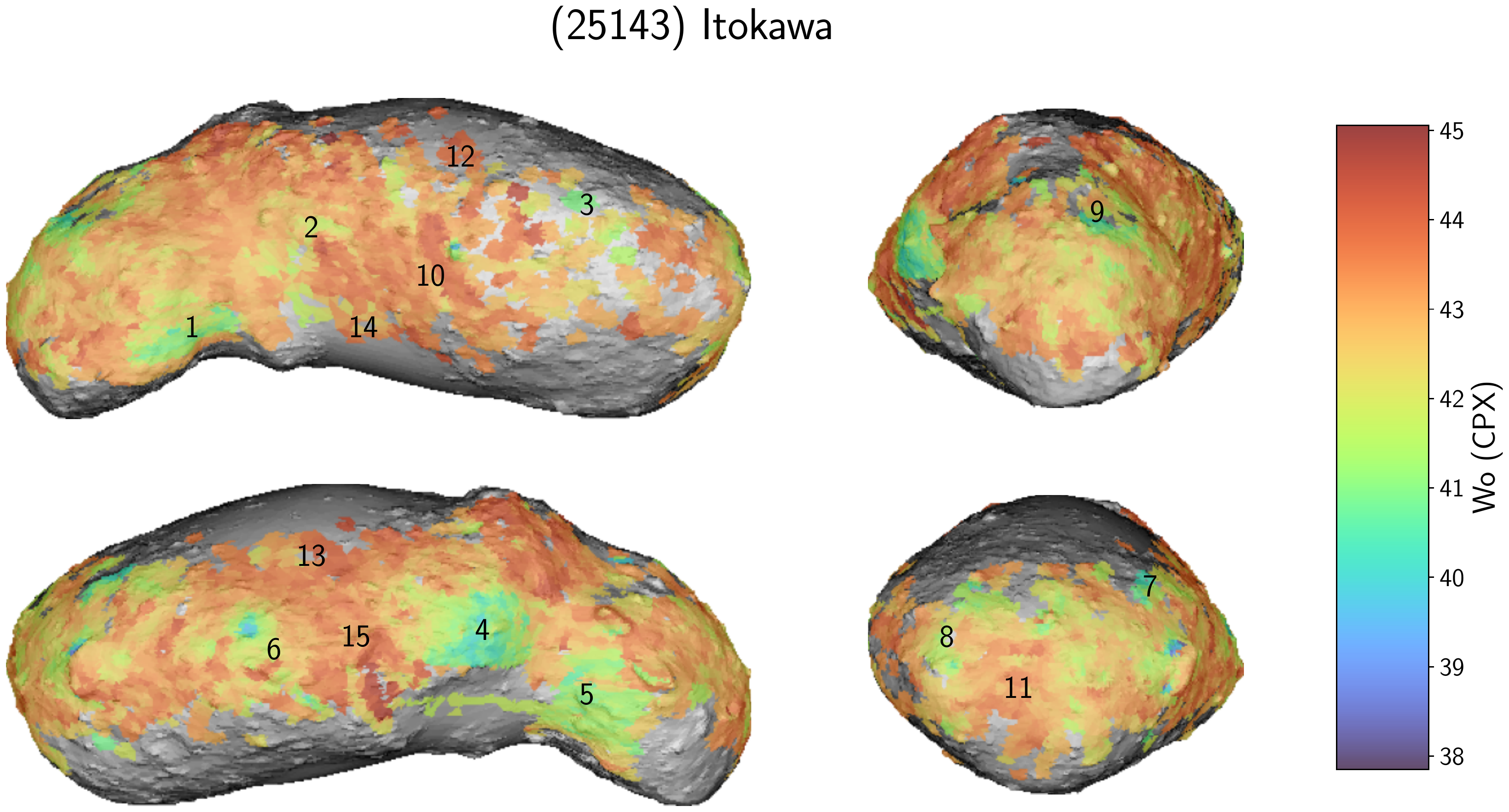}
{Predictions of calcium content in clinopyroxene on the surface of Itokawa. The numbers designate fresh and mature areas.}
{fig:Itokawa_Wo_CPX}

\clearpage

\tcfigure[!ht]{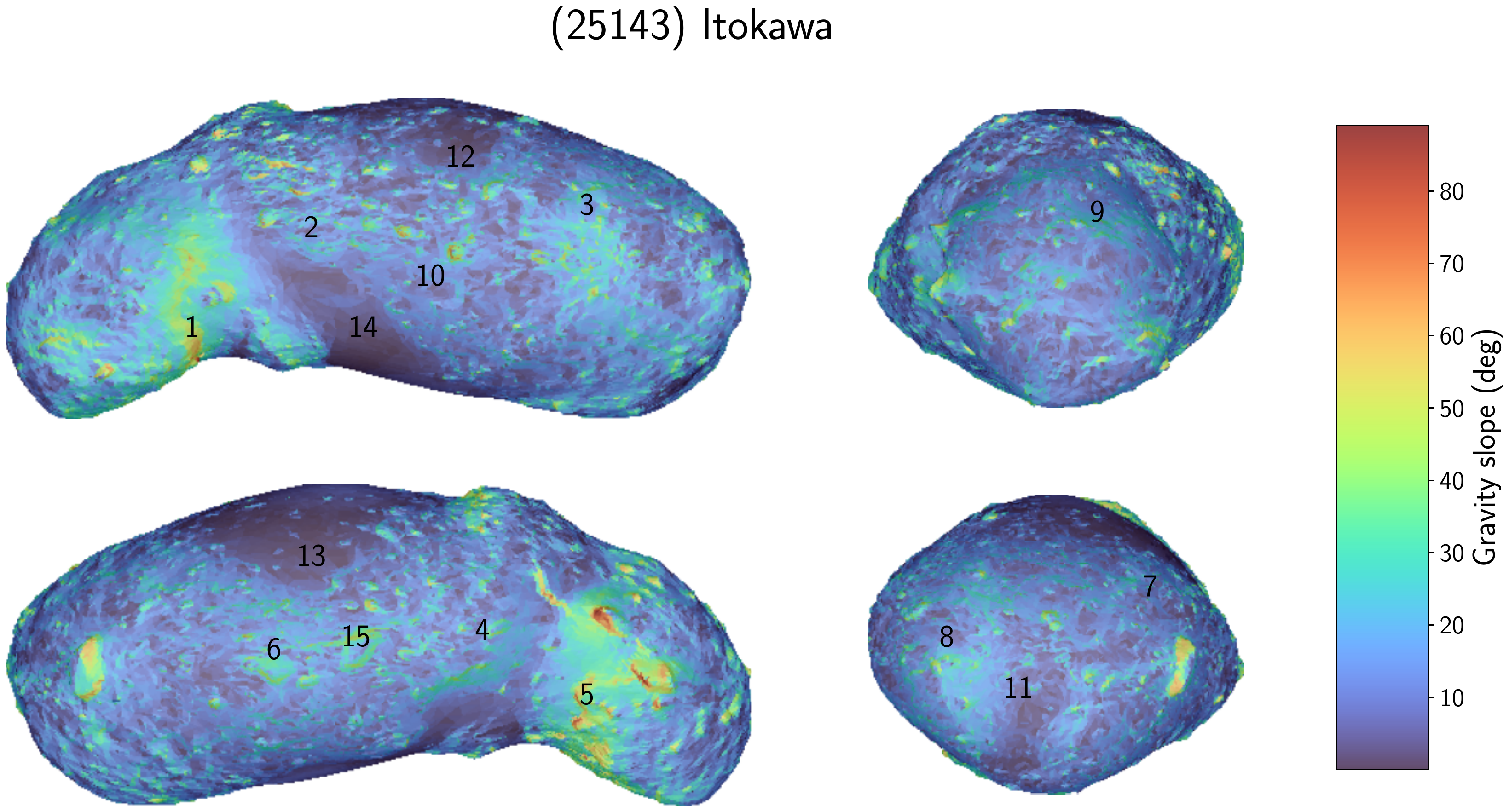}
{Slope of gravity surface acceleration on Itokawa. The numbers designate fresh and mature areas.}
{fig:Itokawa_slope}


\section{Density plots of composition model}

\tcfigure[!ht]{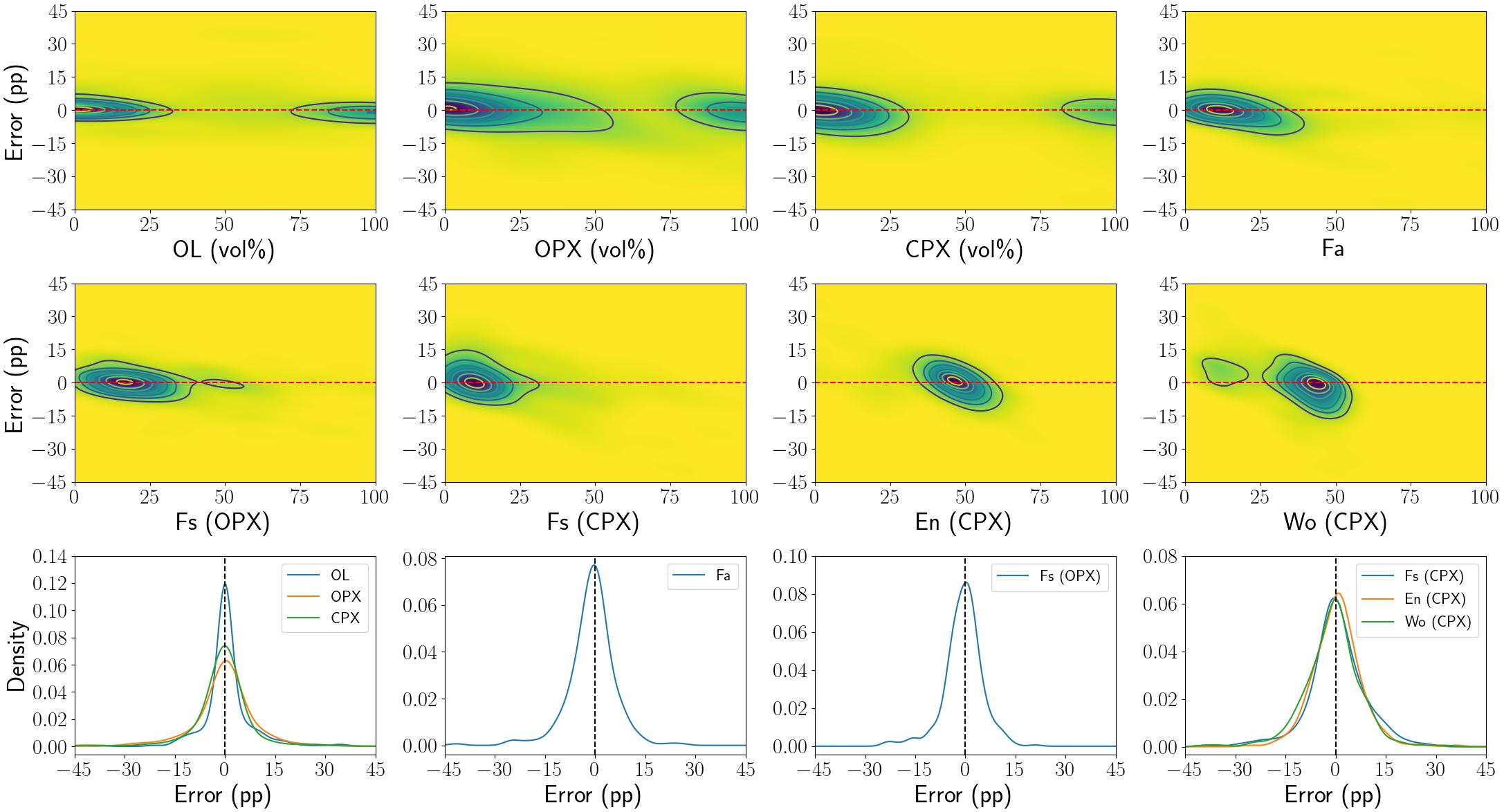}
{Density plots based on specific compositions and prediction errors. The dashed lines delimit zero error. \textit{Top} and \textit{middle rows}:  Concentrations of the data as a function of composition and prediction errors. \textit{Bottom row}: Prediction error distributions.}
{fig:full_density}


\appsection{Correlation matrices of combined predictions of classification and composition models}
\label{app:corr_mat}

\tcfigure[!ht]{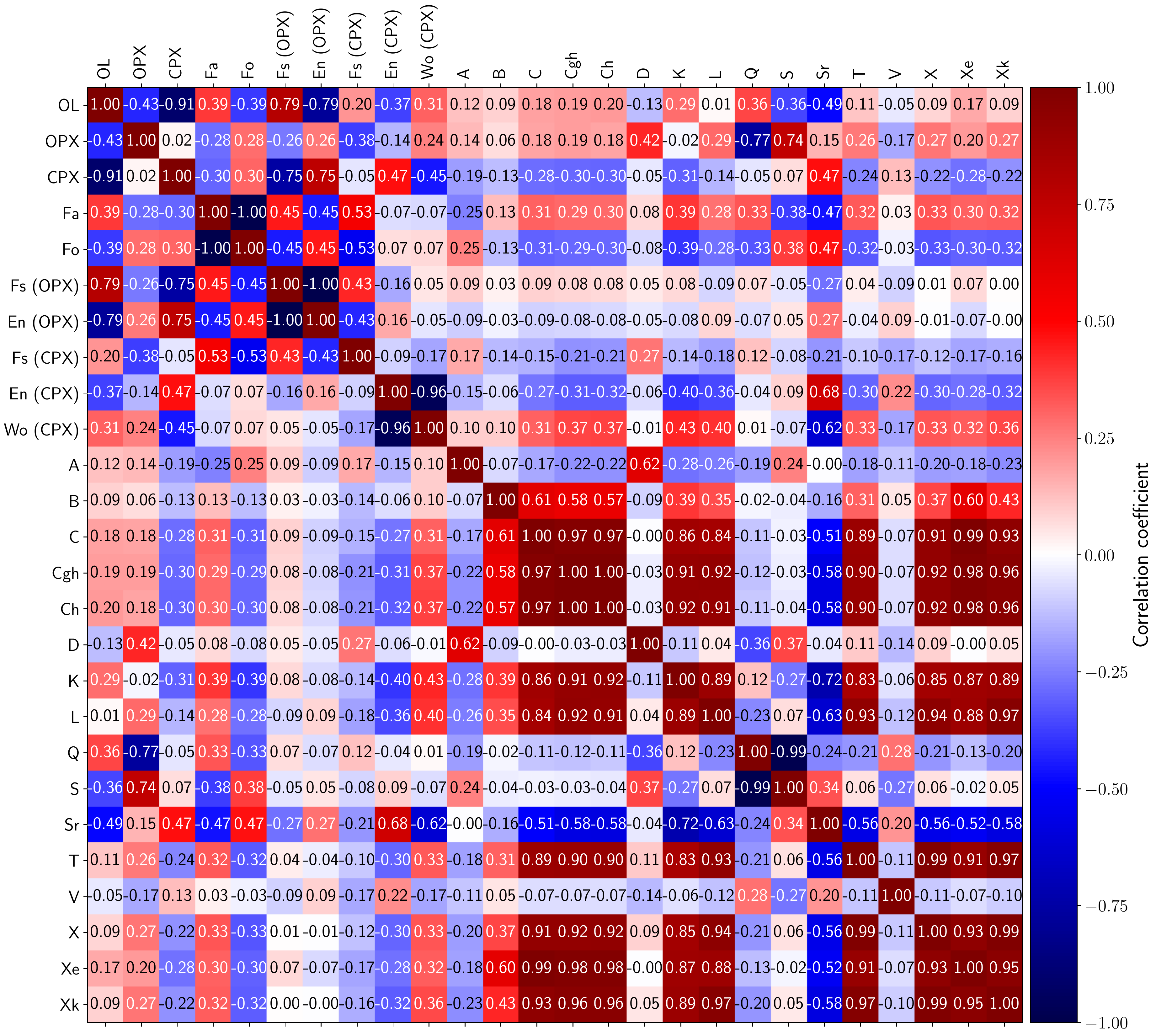}
{Correlation matrix for combined outputs of composition and classification models for the Eros asteroid.}
{fig:E_corr_full}

\tcfigure[!ht]{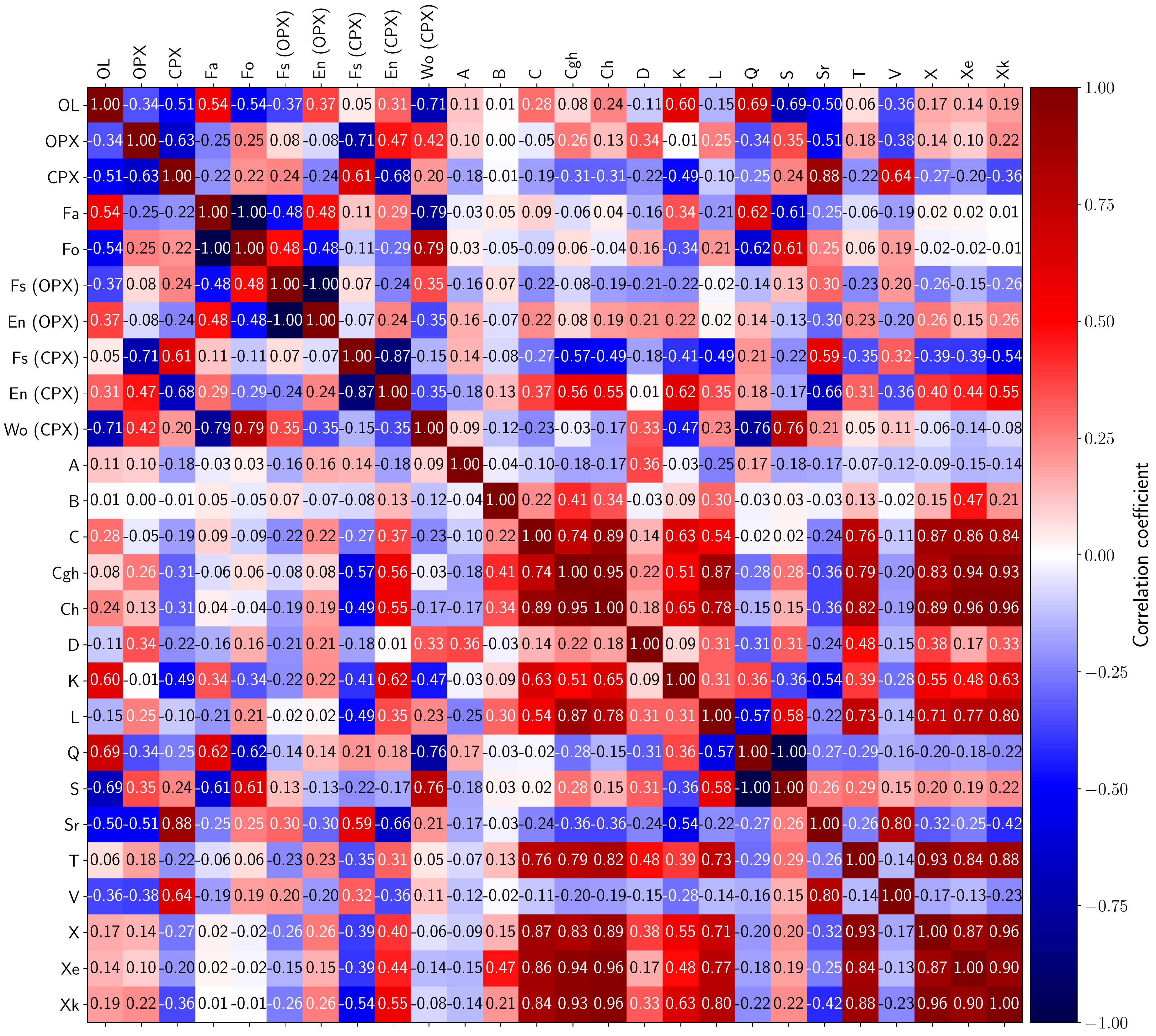}
{Correlation matrix for combined outputs of composition and classification models for the Itokawa asteroid.}
{fig:I_corr_full}

\end{document}